\documentclass[%
 reprint,
 amsmath,amssymb,
 aps,
]{revtex4-2}
\usepackage{graphicx}% Include figure files
\usepackage{dcolumn}% Align table columns on decimal point
\usepackage[colorlinks=true,
            linkcolor=blue,
            urlcolor=blue,
            citecolor=blue]{hyperref}
\usepackage{bm}% bold math
\usepackage{enumerate}
\usepackage{lipsum}
\usepackage{threeparttable}
\usepackage{epstopdf}
\usepackage{float}
\usepackage{xcolor,colortbl}
\usepackage[caption = false]{subfig}
\usepackage{tabularx}
\usepackage{chngpage}
\usepackage{csquotes}
\MakeOuterQuote{"}
\usepackage{natbib}
\usepackage[maxfloats=256]{morefloats}

\begin{document}

%\preprint{APS/123-QED}

\title{Forces on alkali Rydberg atoms due to non-linearly polarized light}
\author{A.  Bhowmik}
\email{anal.bhowmik-1@ou.edu}
\affiliation{Homer L. Dodge Department of Physics and Astronomy, The University of Oklahoma, Norman, Oklahoma 73019,USA}
\affiliation{Center for Quantum Research and Technology, The University of Oklahoma, Norman, Oklahoma 73019, USA}
\author{D. Blume}
\email{doerte.blume-1@ou.edu}
\affiliation{Homer L. Dodge Department of Physics and Astronomy, The University of Oklahoma, Norman, Oklahoma 73019,USA}
\affiliation{Center for Quantum Research and Technology, The University of Oklahoma, Norman, Oklahoma 73019, USA}

\date{\today}


%\affiliation{
%\affiliation{
 %Third institution, the second for Charlie Author
%\affiliation{%



\begin{abstract}
Trapped Rydberg atoms are   highly promising candidates for quantum science experiments.  While several approaches   have been put forward to exert (trapping) forces on isolated Rydberg atoms, a widely applicable lossless technique is lacking.  This paper proposes  a robust versatile alternative technique that avoids lifetime compromising losses. Our proposal leverages the vector polarizability, which is induced by non-linearly polarized light and is shown to be several orders of magnitude larger than the usual  scalar and tensor polarizabilities for  commonly used alkali Rydberg series such as the $nS$, $nP$, and $nD$ series with principal quantum number $n$ as low as 30. The resulting force can be used to trap isolated Rydberg atoms over long times, which constitutes a key advance that is expected to impact 
quantum 
simulation applications, as well as to generate large light--Rydberg-atom hybrid states, which possess non-trivial position-dependent forces. 

   \end{abstract}

\maketitle

\section{Introduction}

Light-matter interactions play a pivotal role   in fundamental research and a wide range of applications.  Examples include remarkable developments in quantum optics \cite{Lukin2003, Peyronel2012,Firstenberg2016, Li2019}, quantum simulations~\cite{Weimer2010,Zeiher2017,Malz2023}, quantum information protocols~\cite{ Adams2020},  quantum computing~\cite{Saffman2016}, quantum sensing~\cite{Cox2018}, high-precision spectroscopy~\cite{Knuffman2007}, precision measurement of fundamental constants~\cite{Jentschura2008},  and molecular physics~\cite{Hollerith2019}.

Electromagnetic fields  fundamentally alter the optical properties of materials~\cite{Gorlach2020} and  are nowadays routinely used to trap cold atoms~\cite{Kaufman2021,Hsu2022} as well as microparticles ~\cite{Almeida2023, Sneh2024} and microspheres~\cite{Shahabadi2021}. This work considers the force on an atom with a high principal quantum number  $n$, specifically an alkali Rydberg atom, in the presence of an external electromagnetic field.   Quite generally, the  force  introduced by the external  field  is commonly  expressed in terms of the dynamic electric dipole polarizability tensor with elements $\alpha_{jk}(\omega)$, where $\omega$ denotes the angular frequency of the electromagnetic field ($j,k=x, y,$ and $z$).  The application of an electromagnetic field along the $x$-direction, e.g., produces a polarization vector with $x$-, $y$-, and $z$-components, with the response being quantified by $\alpha_{xx}(\omega)$, $\alpha_{yx}(\omega)$, and $\alpha_{zx}(\omega)$. The dependence  of the polarizability tensor  on $\omega$ provides an exquisite tuning knob that underlies a variety of enabling tools, including magic wavelength-based traps and state-dependent optical lattices  for low-lying atomic states ~\cite{Katori1999, Ye2008,LeBlanc2007, Cooper2018, Das2020}. In Rydberg atoms, an external electromagnetic field additionally introduces a so-called ponderomotive force, which is associated with the "quiver motion" of the outermost electron~\cite{Dutta2000}. Considering far off-resonant linearly polarized light, earlier work showed that the ponderomotive force dominates over the polarizability-dependent force, i.e., the polarizability-dependent Stark shift or light shift have been shown to play a secondary role in Rydberg atoms~\cite{Zhang2011, Topcu2013_1,Topcu2013,  Barredo2020}. This work identifies a regime where the light shift, which is tunable, dominates over the energy shift due to the ponderomotive force.

Instead of expressing the light shift in terms of the components of the polarizability tensor, it is more convenient to use a decomposition into  the scalar polarizability $\alpha^{S}(\omega)$ (rank-0 tensor), the vector polarizability
 $\alpha^{V}(\omega)$ (rank-1 tensor),  and  the tensor polarizability $\alpha^{T}(\omega)$ (rank-2 tensor)~\cite{Manakov1986,Mitroy2010, Flambaum2008}.    The  effect of the vector and tensor contributions  depends on the experimental geometry, i.e., the  orientation of  the polarization vector   and the propagation vector of the light  relative to the quantization axis.  As a result, the light shift may exhibit a high degree of tunability, offering flexibility through appropriate adjustments of the experimental geometry.  Independent tunability of the vector and tensor parts has, e.g., been demonstrated for 
 lanthanide 
 atoms in the ground state~\cite{Becher2018,Tsyganok2019}.  For alkali atoms, in contrast,  
 the vector polarizability 
 tends to be small for the ground state~\cite{Arora2012}, thereby appreciably restricting the 
 tunability; note, though, that tunability of low-lying excited alkali states and alkaline earth metal states via $\alpha^{V}(\omega)$ has been demonstrated~\cite{Arora2012,Sahoo2013,Sherman2005}.

 Focusing on linearly polarized light, for which the vector contribution vanishes identically and the 
  polarizability away from resonances scales as $-e^2/(m_e \omega^2)$~\cite{Anderson2011, Wilson2022} [here, $e$ and $m_e$ denote the electron charge and electron mass,
  respectively], three distinct strategies have been proposed, namely working with 
 (i)  blue-detuned light near an atomic resonance~\cite{Bai2020_1,Bhowmik2024,Saffman2005}, (ii) blue-detuned bottle beam optical traps that utilize the interference of two Gaussian  beams with different beam waists~\cite{Zhang2011}, and (iii) beam profiles that exhibit a non-trivial spatial dependence such as optical lattices~\cite{Li2013} or Laguerre-Gauss beams~\cite{Verde2023,Bhowmik2018,Das2020}.
  Even though a very recent  study~\cite{Bhowmik2024} predicted that certain Rydberg series---such as, e.g., the $D_{3/2}$ and 
  $D_{5/2}$ series of cesium---feature unexpectedly large detunings and, correspondingly, unexpectedly low losses, 
  strategy (i) does not provide a one-fits-all solution.
  Similarly, strategies (ii) and (iii) have  short-comings~\cite{Wilson2022}, 
  that have---up to now---limited their use to niche applications.
  
  This work proposes a versatile alternative strategy, namely the use of elliptically polarized light to exert a force on an alkali Rydberg atom in the direction of the high intensity region of the intensity profile of, e.g., a Gaussian beam. The approach leads to  spatially dependent (trapping) forces for various alkali Rydberg series away from resonances, i.e., without
  introducing lifetime compromising losses. Since the force exerted has contributions from the scalar and tensor parts, which scale as $\omega^{-2}$, as well as the vector part, which scales as $\omega^{-1}$, the proposed strategy allows for the realization of forces that   depend comparatively weakly on the wavelength of the light.   It is shown that there exists a fairly broad range of experimentally accessible parameters, where the light shift dominates over the energy shift generated by   the ponderomotive force.  
  The proposed strategy works for alkali Rydberg atoms of varying size, including $n$ values  as low as $30$.
  This feature distinguishes the proposed strategy from a recent 
  alternative approach, which only works for ultra-large Rydberg atoms for which the outer electron is, on average, so far away from the 
  ionic core that it resides in the region where the Gaussian beam has vanishingly small intensity
  such that the ionic core is trapped via the scalar polarizability and the outer electron hangs on to the core~\cite{Wilson2022}.

 The remainder of this paper is organized as follows. Section~\ref{sec_theory} reviews the theoretical framework employed in this work. Section~\ref{sec_results} presents our results, focusing primarily on rubidium. Finally, Sec.~\ref{sec_conclusion} summarizes.  Technical details as well as calculations for cesium are relegated to several appendixes.

\section{Theoretical framework}
\label{sec_theory}

We start by discussing the  energy  shift $U_{\text{gr}}(X,Y,Z,\omega)$ that a ground state atom experiences in the presence of a non-resonant laser with frequency $\omega$. The  potential experienced by a ground state atom with single outer or ``active'' electron  
is given by the second-order Stark shift $U_{\text{stark}}(X, Y, Z,\omega)$~\cite{Manakov1986,Mitroy2010, Flambaum2008},
 \begin{eqnarray}\label{Eq1}
 U_{\text{gr}}&&(X, Y, Z,\omega) =U_{\text{stark}}(X, Y, Z,\omega),
 \end{eqnarray}
 where
 \begin{eqnarray}
   \label{eq_stark}
   U_{\text{stark}}(X, Y, Z,\omega)  
 =-\frac{{I(X,Y,Z)}}{2\epsilon_0c} \bigg[ \alpha^{S}(\omega)\nonumber \\
+f^V(A,\theta_k,J,M_J) \alpha^{V}(\omega)+f^T(\theta_p,J,M_J) \alpha^{T}(\omega) \bigg].
 \end{eqnarray}In Eq.~(\ref{eq_stark}), $I(X,Y,Z,\omega)$ denotes the spatially varying intensity of the laser at the position $(X,Y,Z)$ of the atom,  $\epsilon_0$ the vacuum permittivity, and $c$ the speed of light in vacuum.  Explicit expressions for the polarizabilities $\alpha^{S}(\omega)$, $\alpha^{V}(\omega)$, and $\alpha^{T}(\omega)$ are provided in Refs.~\cite{Flambaum2008,Bhowmik2020,Zhang2024} and not reproduced here. 
  Since we will be working away from resonances, our focus is on the real parts of the polarizabilities, which can be calculated    by setting the lifetimes of the intermediate states to zero. Correspondingly, $\alpha^S(\omega)$ denotes the real part of the scalar polarizability and similarly for the vector, tensor, and total polarizabilities.  To determine the dynamic scalar, vector, and tensor  polarizabilities, we employ the sum-over-states approach~\cite{Mitroy2010}.
 
\begin{figure}[t]
\vspace{-1.cm}
{\includegraphics[ scale=.25]{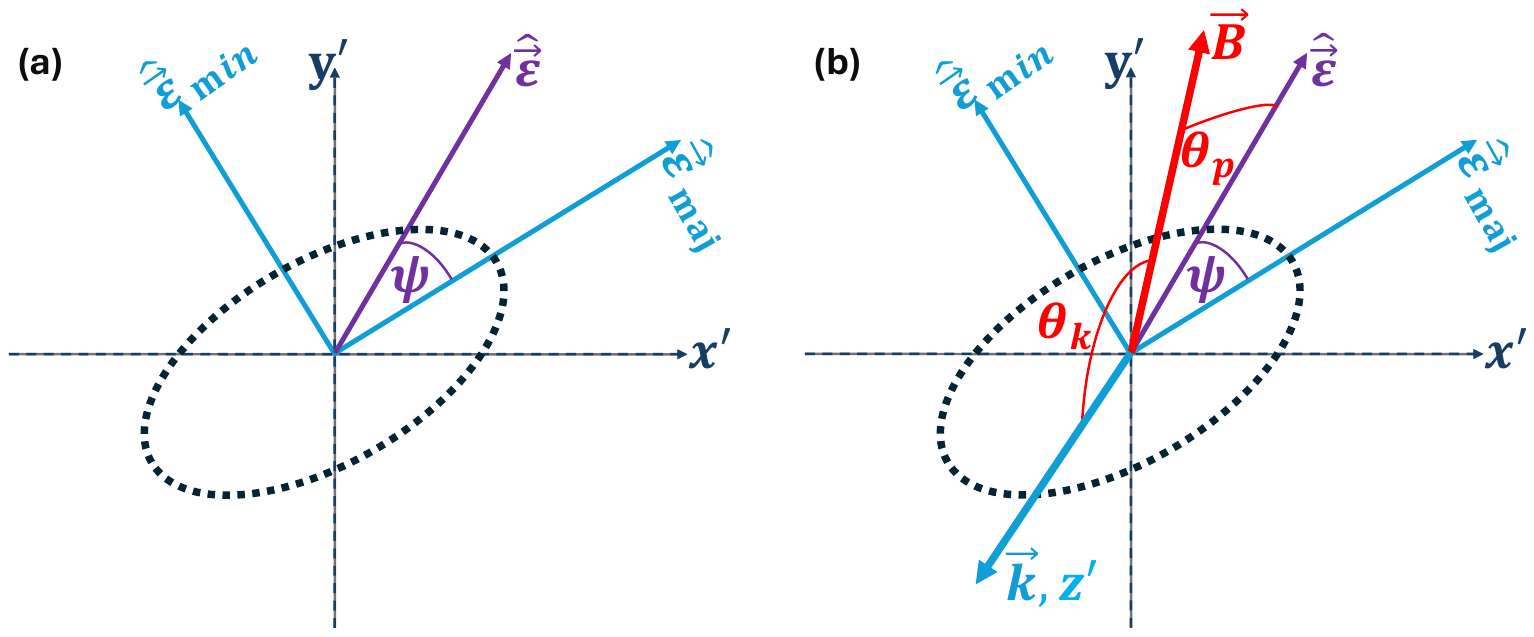}}\\
\vspace{-1.5cm}
\caption{"Geometric arrangement." (a) The polarization ellipse (dotted line) lies in the $x'y'$-plane. $\hat{\vec{\epsilon}}$, $\hat{\vec{\epsilon}}_{\text{maj}}$, and $\hat{\vec{\epsilon}}_{\text{min}}$ denote the polarization vector, major axis, and minor axis, respectively;  $A$ is given by the angle $\psi$, $A= \sin (2 \psi)$. (b) The propagation vector $\vec{k}$, which lies along the $z'$-axis, is orthogonal to $\hat{\vec{\epsilon}}_{\text{maj}}$ and $\hat{\vec{\epsilon}}_{\text{min}}$. The angles $\theta_k$ and $\theta_p$ define  the orientation of $\hat{\vec{k}}$ and $\hat{\vec{\epsilon}}$ relative to the external magnetic field vector $\vec{B}$.}
\label{Fig0}
\end{figure}

The prefactors $f^V(A,\theta_k,J,M_J)$ and $f^T(\theta_p,J,M_J)$ are coefficients that depend on the total orbital angular momentum quantum number $J$ of the electron and its associated projection quantum number $M_J$. The quantity $A$ denotes the degree of ellipticity~\cite{Zhang2024};
 $A=0$ and $A=\pm1$ correspond, respectively, to linearly and circularly polarized light.
  The angles $\theta_k$ and $\theta_p$ (Fig.~\ref{Fig0}) are defined through 
 $\cos \theta_k=\hat{\vec{k}} \cdot \hat{\vec{B}}$ and $\cos ^2 \theta_p=|\hat{\vec{\epsilon}} \cdot \hat{\vec{B}}|^2$, respectively, where 
 $\vec{k}$ denotes the propagation vector, $\vec{B}$ the external magnetic field vector ($\vec{B}$ sets the quantization axis), and 
 $\hat{\vec{\epsilon}}$ the polarization vector ($\hat{\vec{k}}=\imath \hat{\vec{\epsilon}}^* \times \hat{\vec{\epsilon}}$).
 The explicit forms of  $f^V$ and $f^T$  read~\cite{Flambaum2008,Zhang2024}
 \begin{eqnarray}\label{Eq2}
 f^V(A,\theta_k,J,M_J)=
 A \cos \theta_k \frac{M_J}{2J}
 \end{eqnarray}
 and
 \begin{eqnarray}\label{Eq3}
     f^T(\theta_p,J,M_J)=\frac{3 \cos ^2 \theta_p -1}{2} \frac{3M_J^2-J(J+1)}{J(2J-1)}.
 \end{eqnarray}
 To streamline the discussion to come, we refer to 
 $A \cos \theta_k$ and $(3 \cos^2 \theta_p-1)/2$ as "geometric factors."
 This naming scheme is motivated by the fact that their values are controlled by the relative orientation of the external magnetic field vector and the vectors that characterize the electromagnetic field (Fig.~\ref{Fig0}). The former geometric factor can take values between $-1$ and $1$ while the latter geometric factor can take values between $-1/2$ and $1$.  

\begin{figure}[t]
\vspace{-1.8cm}
{\includegraphics[ trim=0cm 0cm 0cm 0cm, clip=true, totalheight=0.30\textheight, angle=0]{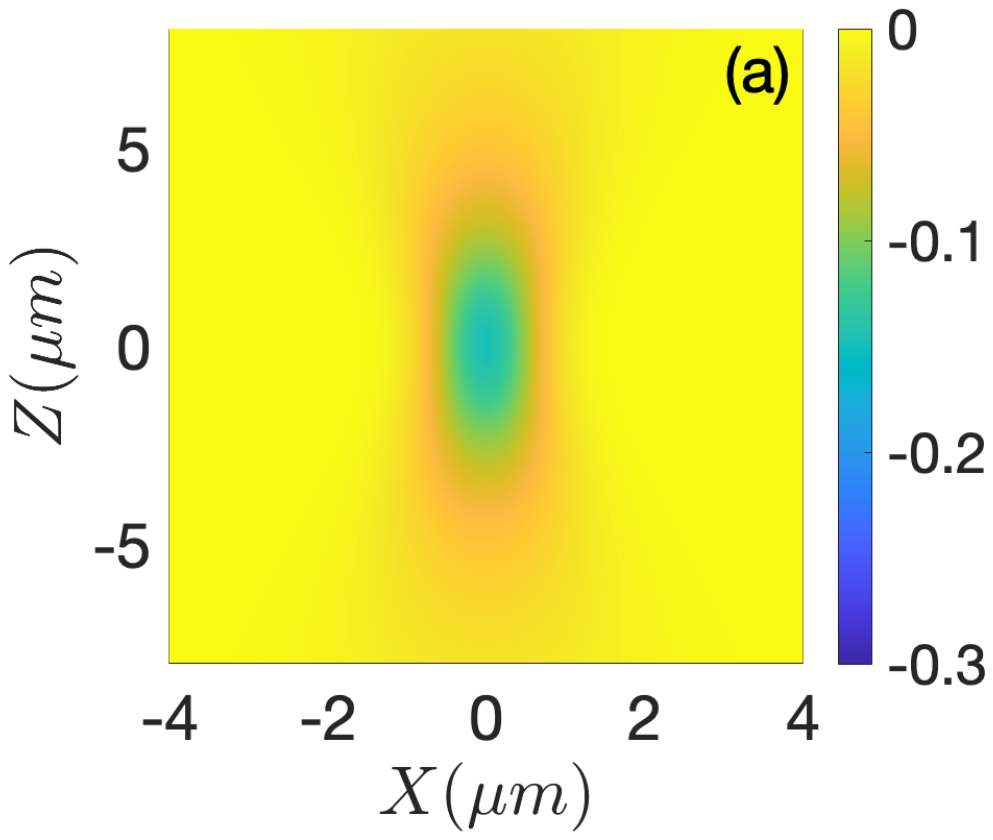}}
\hspace{-0.8cm}
{\includegraphics[trim=5.0cm 0cm 0cm 0cm, clip=true, totalheight=0.30\textheight, angle=0]{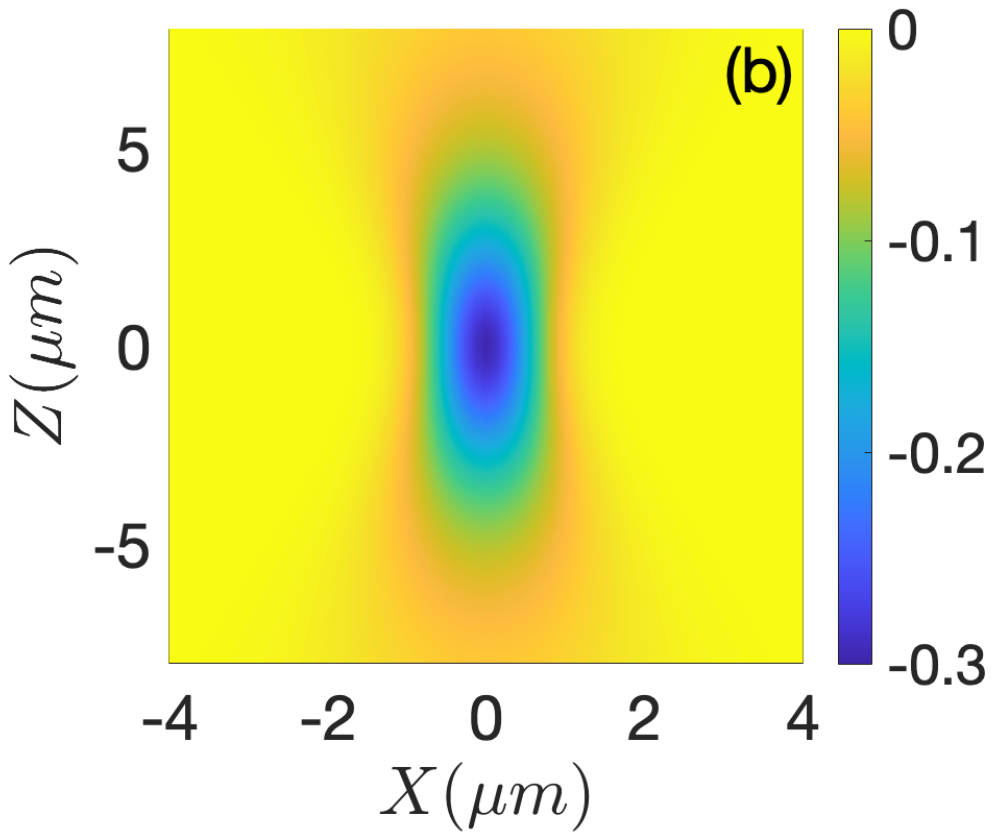}}
\vspace{-1.5cm}
\caption{Trapping potential $U_{\text{gr}}(X,0,Z,\omega)$  as functions of $X$ and $Z$ for the ground state of rubidium for $\lambda=1,000$~nm, $\alpha(\omega)=820.2$~a.u., and $w_0=1$~$ \mu$m. The laser power $P$ of the Gaussian beam is (a) $1.25$~mW and (b) $2.5$~mW. The color bars indicate the values of $U_{\text{gr}}(X,0,Z,\omega)$ in mK. }
\label{Fig1_extra}
\end{figure}

 For alkali atoms in the ground state ($S$ state), the spherical symmetry of the electronic wavefunction implies that the tensor polarizability vanishes, i.e., $\alpha^T(\omega)=0$. Additionally, the vector polarizability contribution in the ground state is typically small~\cite{Arora2012}. As a result, the combined effect of the vector polarizability and the associated  prefactor  $f^V(A,\theta_k,J,M_J)$ provides very small tunability of the total dynamic polarizability~\cite{Arora2012}. Therefore, the trapping potential experienced by an alkali atom in the ground state is essentially independent of the geometric factor $A\cos\theta_k$. As an example, Fig.~\ref{Fig1_extra} shows the trapping potential $U_{\text{gr}}(X,0,Z,\omega)$ for the ground state of the rubidium atom   for a Gaussian laser that is characterized by the following intensity pattern:
\begin{eqnarray}
\label{eq_gaussian}
    I(X,Y,Z,\omega)= I_0\frac{1}{1+\big(\frac{\lambda Z}{\pi w_0^2}\big)^2}e^{-\frac{2(X^2+Y^2)}{w_0^2\big[1+\big(\frac{\lambda Z}{\pi w_0^2}\big)^2\big]}}.
\end{eqnarray} 
In Eq.~(\ref{eq_gaussian}), $w_0$ denotes the beam waist, $\lambda$ the wavelength of the laser,  and $I_0$ the magnitude of the peak intensity, $I_0=2P/(\pi w_0^2)$ ($P$ is the power of the laser).  Figure~\ref{Fig1_extra} uses typical experimental parameters, namely $\omega_0=1$~$\mu$m,  $\lambda=1,000$~nm, and $P=1.25$~mW and $2.5$~mW. For this wavelength, we find that the polarizability $\alpha(\omega)$ of the ground state is equal to $820.2$~a.u..

\begin{figure}[t]
\vspace{-1.7cm}
{\includegraphics[ trim=0cm 0cm 1.5cm 0cm, clip=true, totalheight=0.27\textheight, angle=0]{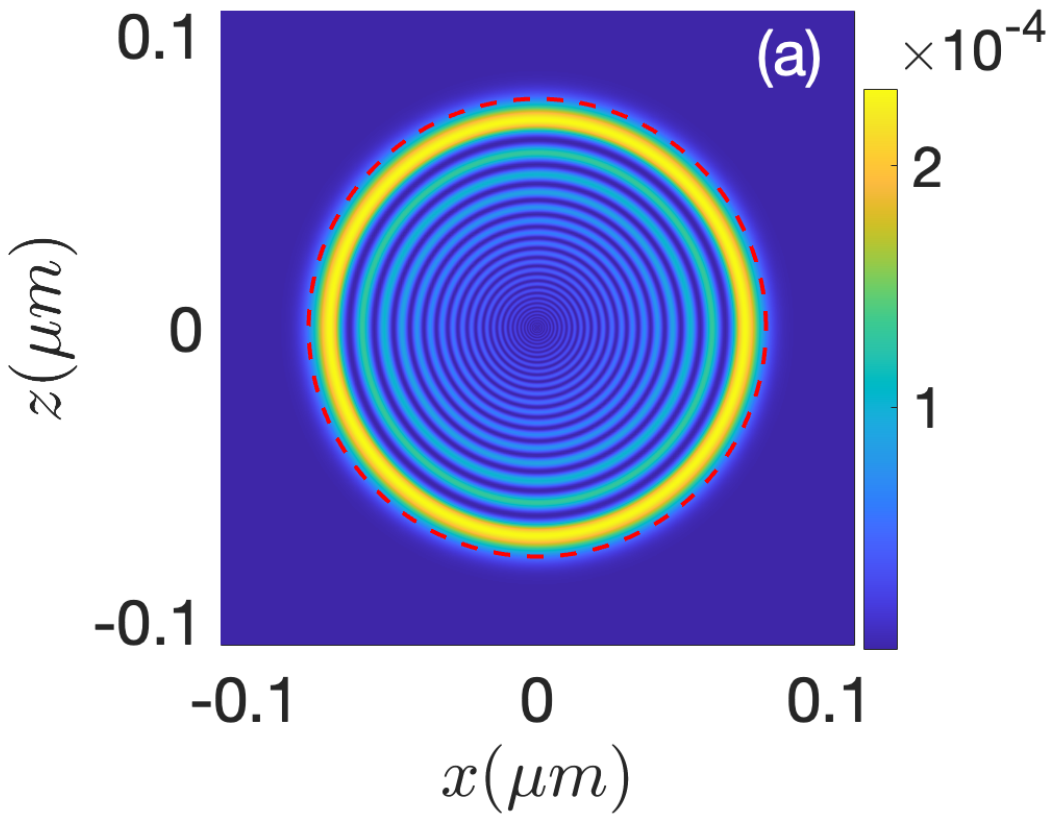}}
\hspace{-0.2cm}
{\includegraphics[trim=3.8cm 0cm 0cm 0cm, clip=true, totalheight=0.27\textheight, angle=0]{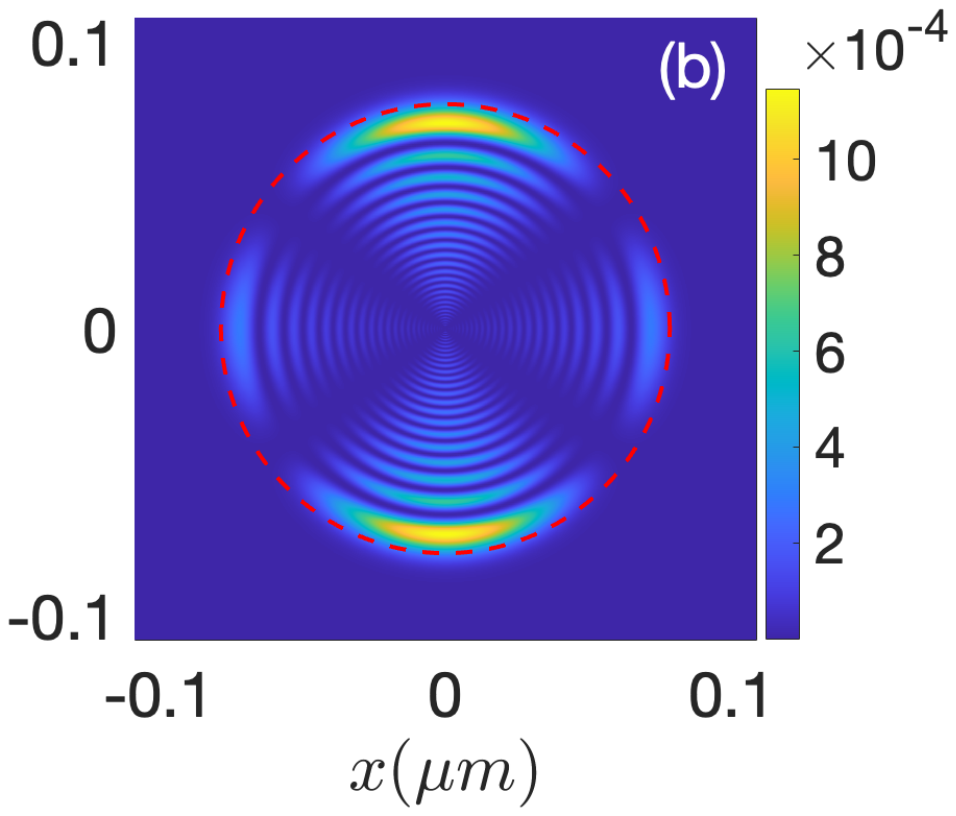}}\\
\vspace{-3.4cm}
{\includegraphics[ trim=0cm 0cm 1.5cm 0cm, clip=true, totalheight=0.27\textheight, angle=0]{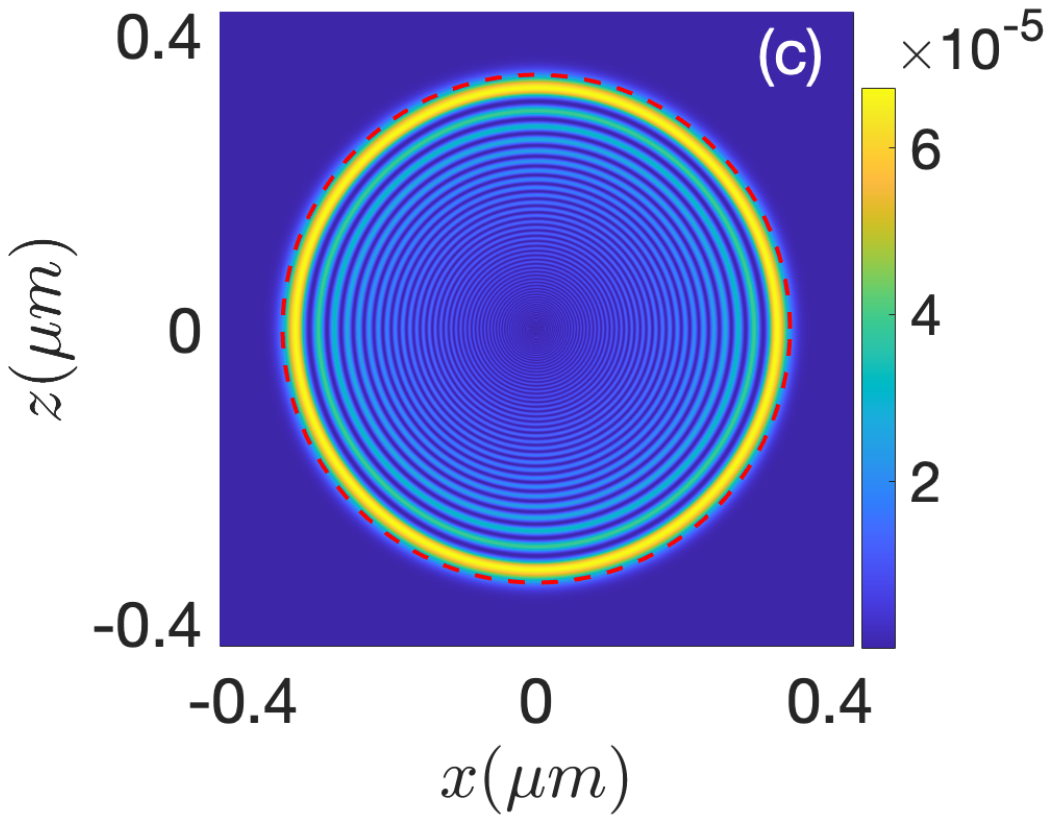}}
\hspace{-0.2cm}
{\includegraphics[trim=3.8cm 0cm 0cm 0cm, clip=true, totalheight=0.27\textheight, angle=0]{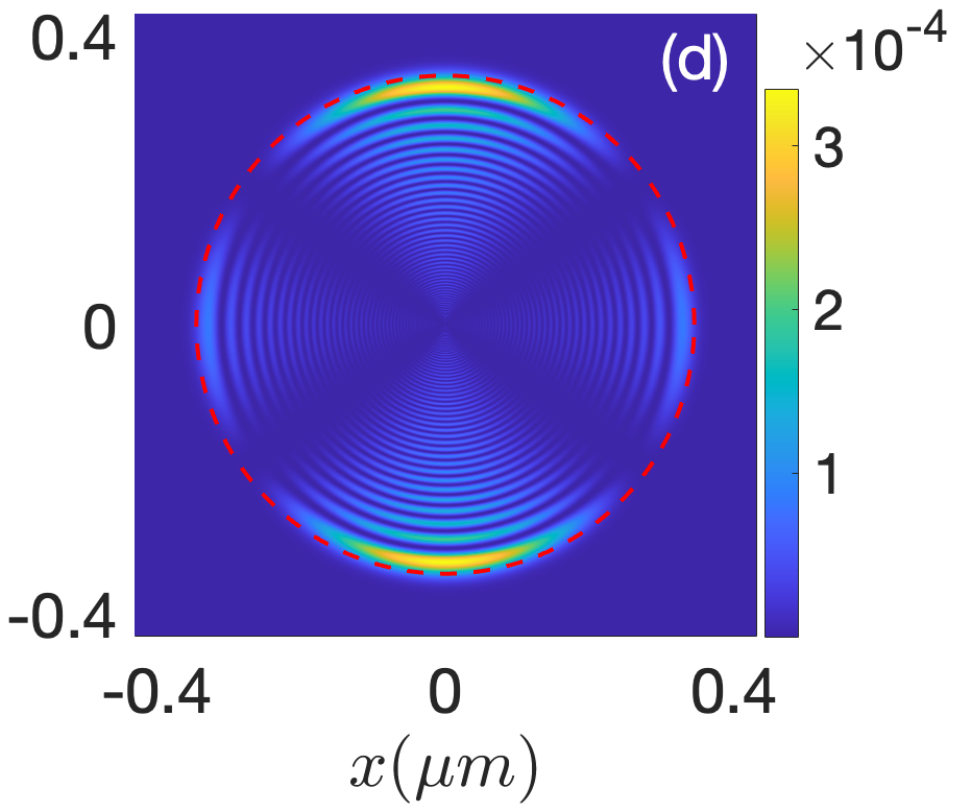}}\\
\vspace{-3.4cm}
{\includegraphics[ trim=0cm 0cm 1.5cm 0cm, clip=true, totalheight=0.27\textheight, angle=0]{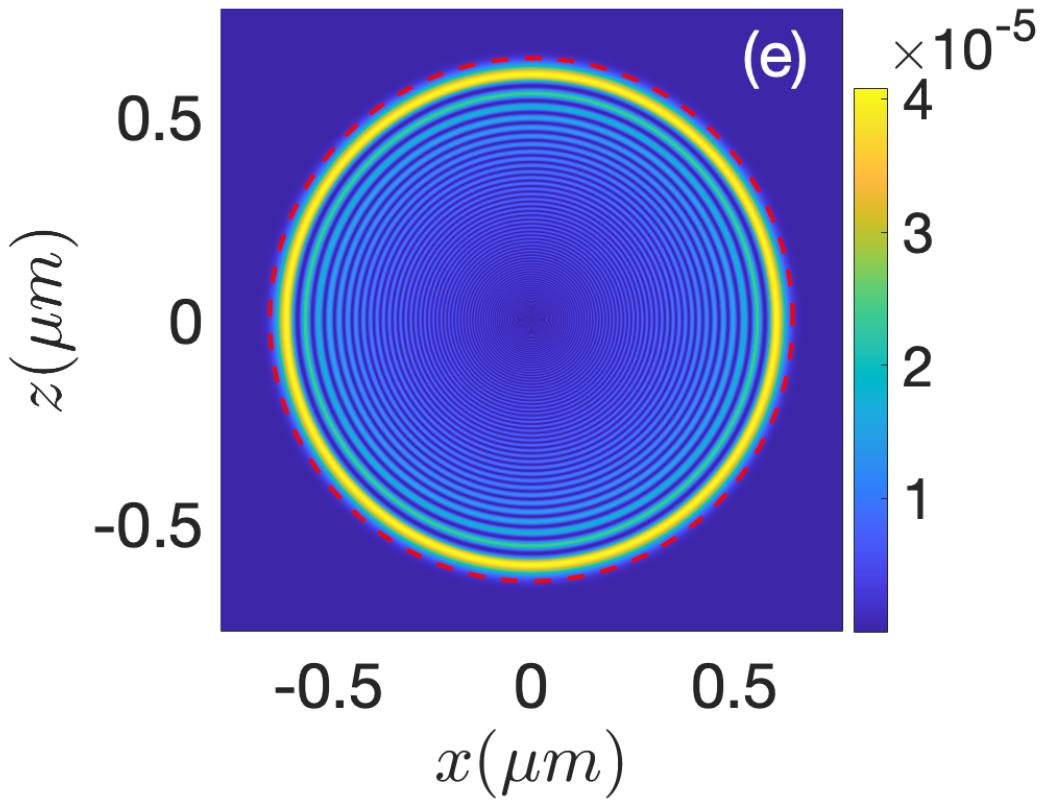}}
\hspace{-0.2cm}
{\includegraphics[trim=3.8cm 0cm 0cm 0cm, clip=true, totalheight=0.27\textheight, angle=0]{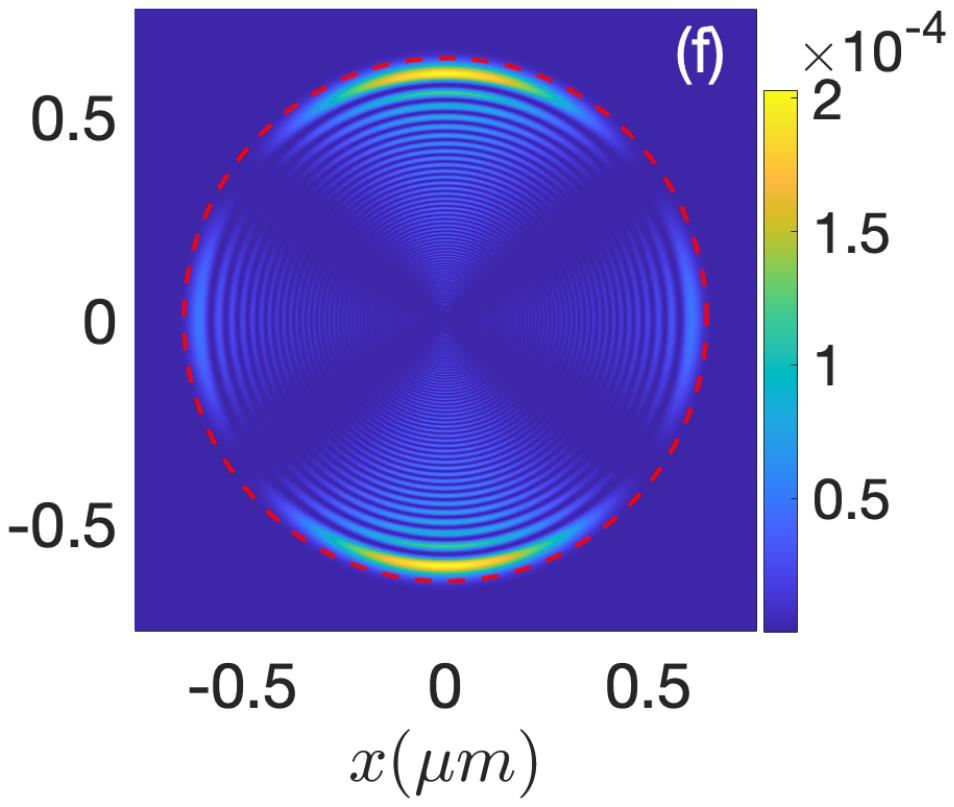}}\\
\vspace{-1.5cm}
\caption{ Probability distribution $r^2|\Psi^{(0)}_{n_{\text{eff}},L,J,M_J}(x,0,z)|^2$  of various rubidium Rydberg states as  functions of $x$ and $z$ for $y=0$.  The Rydberg states are (a) $30S_{1/2,+1/2}$, (b) $30D_{3/2,-3/2}$, (c) $60S_{1/2,+1/2}$, (d) $60D_{3/2,-3/2}$, (e) $80S_{1/2,+1/2}$, and  (f) $80D_{3/2,-3/2}$. The color bars indicate the values of $r^2|\Psi^{(0)}_{n_{\text{eff}},L,J,M_J}(x,0,z)|^2$ in ($\mu$m)$^{-1}$.   The red dashed lines show the $(x_*,z_*)$ values for which  95~\% of the $y=0$ probability of the Rydberg electron lies inside. }
\label{Fig2_extra}
\end{figure}

If the size $\Bar{r}$ of the Rydberg atom  is comparable to or larger than the beam waist $\omega_0$, then the energy shift induced by the electromagnetic field contains, in addition to the Stark shift, a diamagnetic term~\cite{Dutta2000}. In fact, the diamagnetic term has been previously found to be the dominant contribution for linearly polarized light~\cite{Zhang2011, Topcu2013_1,Topcu2013,  Barredo2020}.  To quantify the size of the Rydberg atom, Fig.~\ref{Fig2_extra}  shows the probability distribution $r^2 |\Psi^{(0)}_{n_{\text{eff}},L,J,M_J}(x,y,z)|^2$ of the rubidium Rydberg electron as functions of $x$ and $z$ for $y=0$, where  $r^2=x^2+y^2+z^2$ ($x$, $y$, and $z$ denote the Cartesian coordinates of the position vector $\vec{r}$ of the Rydberg electron measured relative to the center-of-mass of the atom). $n_{\text{eff}}$ is the effective principal quantum number, see Appendix~\ref{sec_appendix_a}.
The  unperturbed wave functions $\Psi^{(0)}_{n_{\text{eff}},L,J,M_J}(\vec{r})$, which are calculated in the absence of an external field, are normalized such that
$\int|\Psi^{(0)}_{n_{\text{eff}},L,J,M_J}(\vec{r})|^2 d\vec{r}=1$. Figure~\ref{Fig2_extra} shows probability distributions for $S$-states ($nL_{J,M_J}=nS_{1/2,+1/2}$, left column) and $D$-states ($nD_{3/2,-3/2}$, right column); throughout, the Rydberg states are labeled by the principal quantum number $n$, the orbital angular momentum quantum number $L$, the total angular momentum quantum number $J$ of the electron, and the associated projection quantum number $M_J$.  From top to bottom, $n$ increases from $n=30$ to $60$ to $80$.
The red dashed lines  demarcate the $(x_*,z_*)$ values  at which the probability of the Rydberg electron for  $y=0$ has reached 95~\%. While the $n=30$ Rydberg state is small compared to typical beam waists, the size of the $n=60$ and $80$ states approaches $w_0$.

Accounting for the  diamagnetic term, which originates from the square of the vector potential of the  oscillating electric field of the far off-resonant laser~\cite{Dutta2000}, the Rydberg atom experiences both the Stark shift [see Eq.~(\ref{eq_stark})] and  the ponderomotive potential $U_{\text{pon}}(X, Y, Z,\omega)$~\cite{Dutta2000,Zhang2011, Topcu2013_1,Topcu2013,  Barredo2020, Lampen2018}, 
\begin{equation}\label{Eq4}
    U_{\text{ryd}}(X, Y, Z,\omega)=U_{\text{stark}}(X, Y, Z,\omega)+U_{\text{pon}}(X, Y, Z,\omega).
\end{equation}
It is important to note that the polarizabilities that enter into the Stark shift depend on the state under consideration.  For the Rydberg states we employ, just as for the ground states,  the sum-over-states approach~\cite{Mitroy2010}. Reference~\cite{Bhowmik2024}  carefully benchmarked  our implementation for cesium. Additional benchmarks are presented in Appendix~\ref{sec_appendix_a}. Earlier work~\cite{Bhowmik2024} showed that the polarizabilities of Rydberg states are essentially unchanged,  for linearly polarized light, when one transforms from the $(J,M_J,I_N,M_I)$
 to the $(F,M_F,J,I_N)$ basis, where $I_N$ denotes the nuclear spin, $M_I$ the associated projection quantum number, $F$ the total angular momentum quantum number of the atom, and $M_F$ its associated projection quantum number.  We have verified that  the same holds  for non-linearly polarized light considered in this work. Consequently, this work uses the $(J,M_J,I_N,M_I)$ basis.  

The ponderomotive potential is calculated using first-order
perturbation theory. For a Rydberg atom with unperturbed state $\Psi_{n_{\text{eff}},L,J,M_J}^{(0)}(\vec{r})$, the  ponderomotive potential reads~\cite{Dutta2000}
\begin{eqnarray}\label{Eq5}
    U_{\text{pon}}(X, Y, Z,\omega)=
    \nonumber \\
    \frac{e^2}{2\epsilon_0cm_e\omega^2}\int d^3r I(\vec{R}+\vec{r})|\psi_{n_{\text{eff}},L,J,M_J}^{(0)}(\vec{r})|^2,
\end{eqnarray} 
where $\vec{R}$ denotes the center-of-mass vector of the atom and $\vec{r}$ the position vector of the Rydberg electron relative to the center of mass of the atom. The next section shows that $U_{\text{pon}}(X,Y,Z,\omega)$ varies slowly compared to the highly oscillatory density of the Rydberg electron. It is important to note that the ponderomotive potential is independent of the geometric factor, i.e., it is the same for linearly and non-linearly polarized light. It can be read off Eq.~(\ref{Eq5}) that the ponderomotive potential is repulsive.

Equation~(\ref{Eq4}) shows that the energy shift that the Rydberg atom experiences is spatially dependent and governed by an interplay between the Stark shift and the ponderomotive potential. The next section demonstrates that    far off-resonant,  non-linearly polarized laser light, i.e., light with $A\ne 0$, can lead---owing to  the induced vector polarizability---to a highly-tunable $U_{\text{stark}}(X, Y, Z,\omega)$.  
Importantly, this is distinct from the 
 ground-state manifold of alkali atoms, which does not possess this tunability.
 As a consequence, one can realize a regime in which the Stark shift and the ponderomotive potential have opposite signs, with the Stark shift dominating over the ponderomotive potential. As such, the use of non-linearly polarized light opens the door for simultaneously trapping atoms away from resonance in a Rydberg state  ($S$-series, $P$-series, $D$-series, etc.) and the ground state   using Gaussian beams at readily available wavelengths.

\section{Results}
\label{sec_results}

\begin{figure}[t]
\vspace{-1.8cm}
{\includegraphics[ trim=0cm 0cm 0cm 0cm, clip=true, totalheight=0.30\textheight, angle=0]{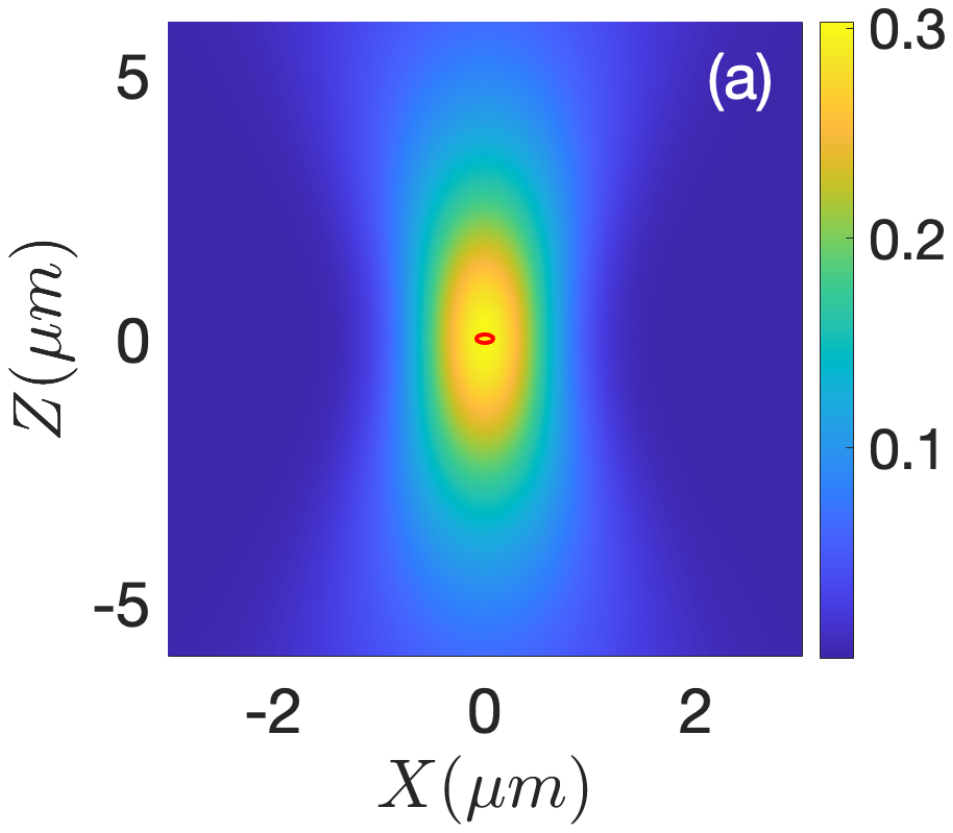}}
\hspace{-0.8cm}
{\includegraphics[trim=5.0cm 0cm 0cm 0cm, clip=true, totalheight=0.30\textheight, angle=0]{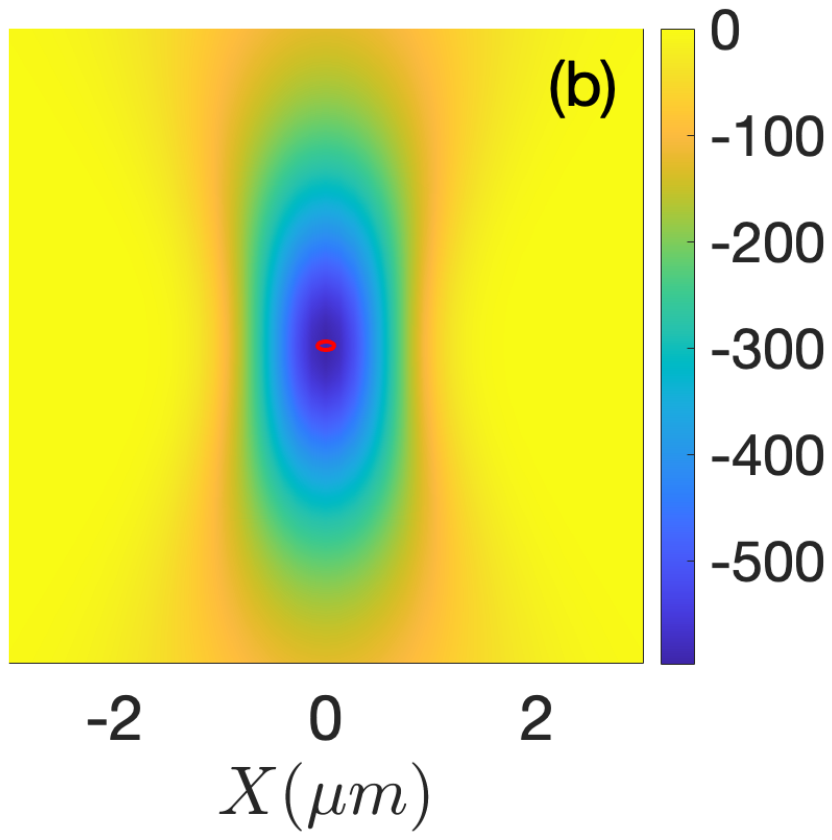}}\\
\vspace{-4.2cm}
{\includegraphics[ trim=0cm 0cm 0cm 0cm, clip=true, totalheight=0.30\textheight, angle=0]{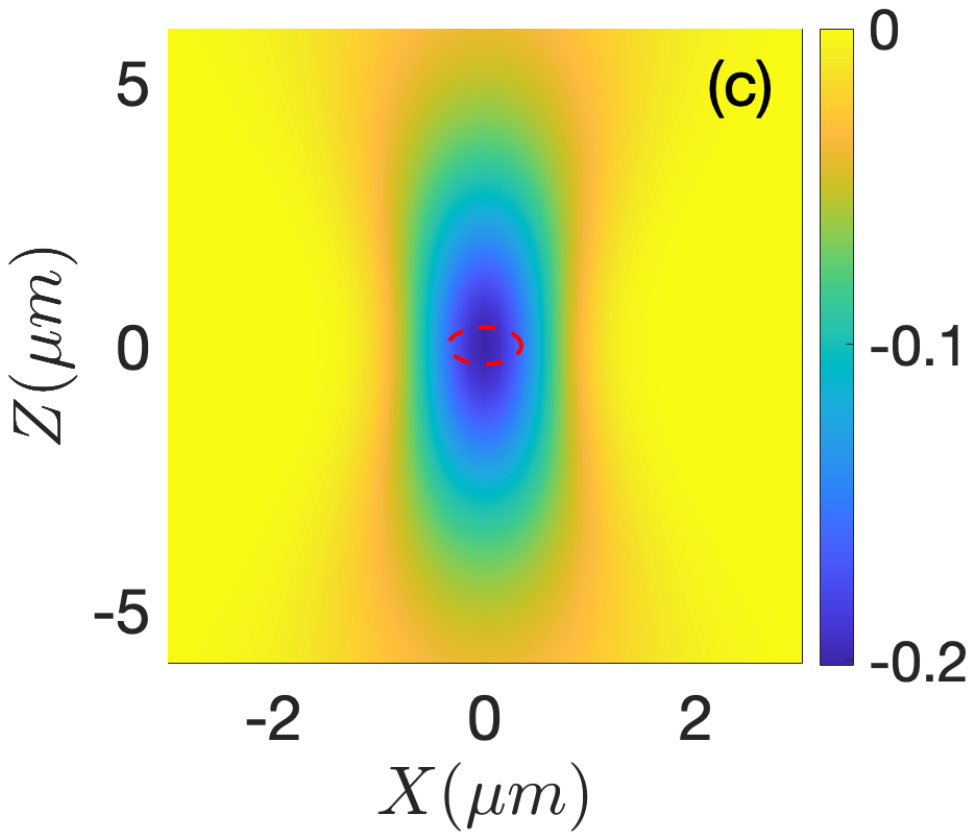}}
\hspace{-0.8cm}
{\includegraphics[trim=5.0cm 0cm 0cm 0cm, clip=true, totalheight=0.30\textheight, angle=0]{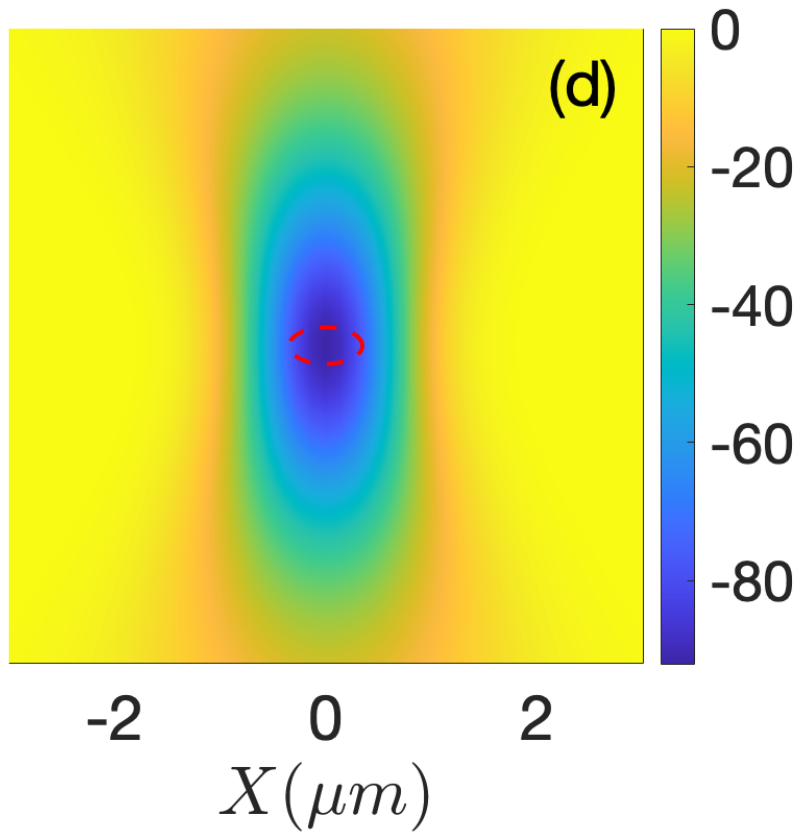}}\\
\vspace{-4.2cm}
{\includegraphics[ trim=0cm 0cm 0cm 0cm, clip=true, totalheight=0.30\textheight, angle=0]{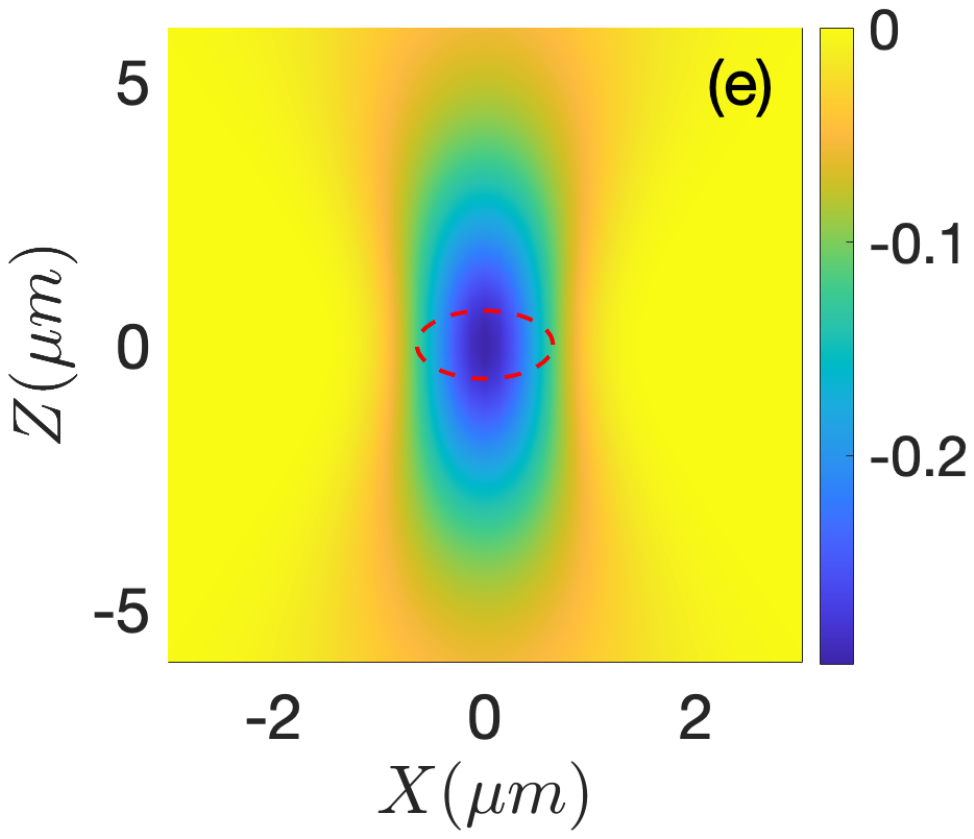}}
\hspace{-0.8cm}
{\includegraphics[trim=5.0cm 0cm 0cm 0cm, clip=true, totalheight=0.30\textheight, angle=0]{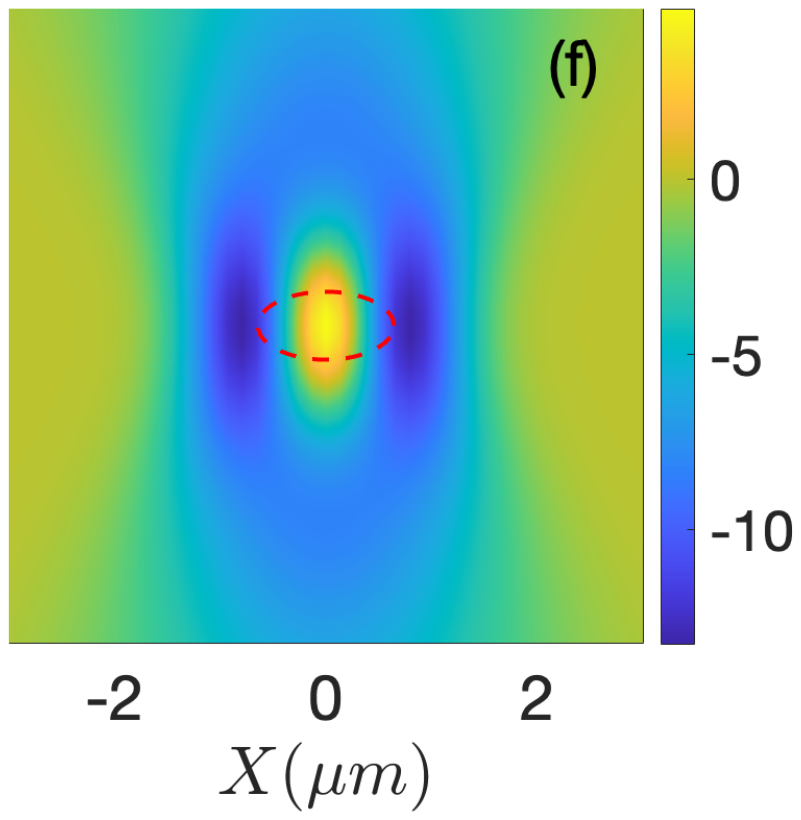}}\\
\vspace{-1.5cm}
\caption{ Trapping of $nS_{1/2,+1/2}$ series of the rubidium Rydberg atom 
for $\lambda=1,000$~nm, $w_0=1$~$\mu$m,  $P=2.5$~mW, and $\theta_p=\pi/2$. 
Left column: Rydberg trapping potential $U_{\text{ryd}}(X,0,Z,\omega)$ as functions of $X$ and $Z$ for   (a) $n=30$ and $A \cos \theta_k=1$, (c) $n=60$ and $A \cos \theta_k=1$, and (e) $n=80$ and $A \cos \theta_k=0.3923$. The color bars indicate the values of $U_{\text{ryd}}(X,0,Z,\omega)$ in mK. 
 Right column: Differential trapping potential $\Delta U(X,0,Z,\omega)$  as functions of $X$ and $Z$. The $n$ and $A \cos \theta_k$ values in (b), (d), and (f) are the same as those used in (a), (c), and (e), respectively. The color bars indicate the values of $\Delta U(X,0,Z,\omega)$ in $\mu$K (it is important to note that the color bars in the left and right columns use different scales, namely mK and $\mu$K, respectively).
 The red dashed lines demarcate the $(X=x_*,Z=z_*)$ values for which 95~\% of the probability of the Rydberg electron, for $Y=0$, lies inside the red dashed line (see  the left column of Fig.~\ref{Fig2_extra} for reference). }
\label{Fig3_extra}
\end{figure}

\begin{figure}[t]
 \vspace{-1.8cm}
{\includegraphics[ trim=0cm 0cm 0cm 0cm, clip=true, totalheight=0.30\textheight, angle=0]{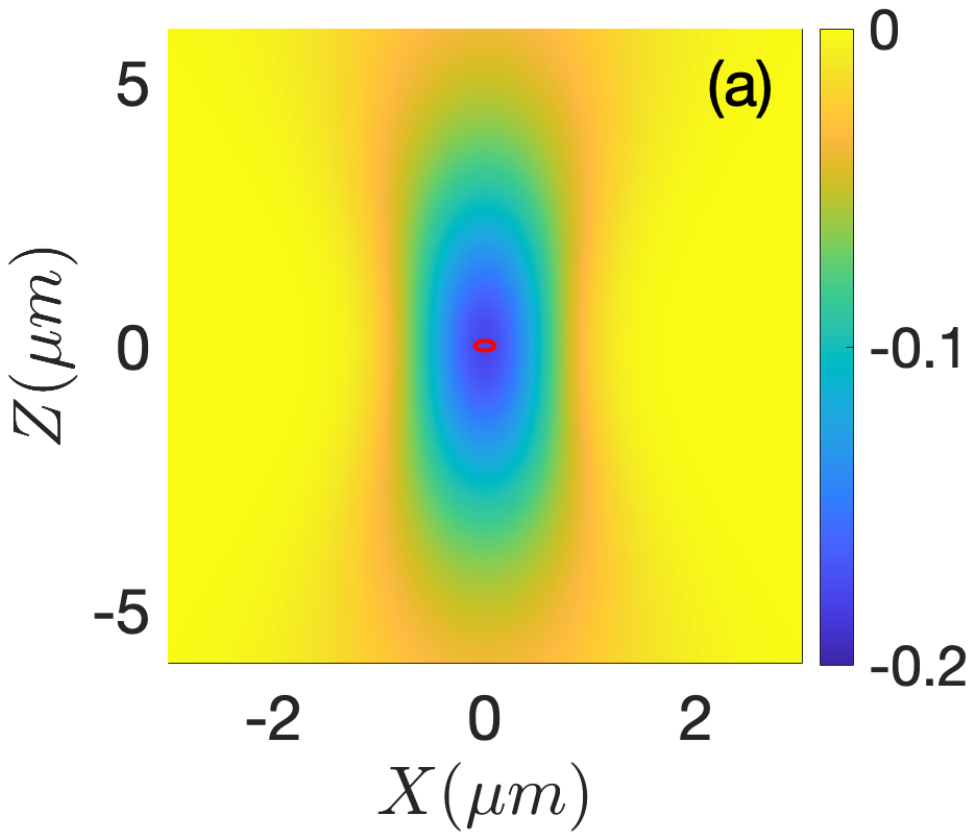}}
\hspace{-0.8cm}
{\includegraphics[trim=5.0cm 0cm 0cm 0cm, clip=true, totalheight=0.30\textheight, angle=0]{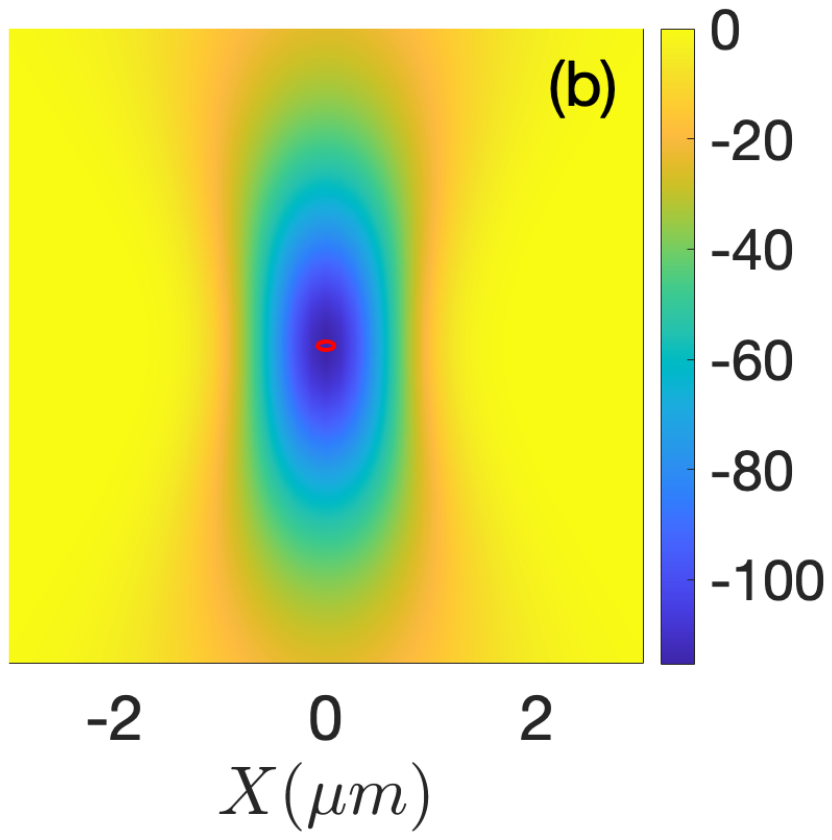}}\\
\vspace{-4.2cm}
{\includegraphics[ trim=0cm 0cm 0cm 0cm, clip=true, totalheight=0.30\textheight, angle=0]{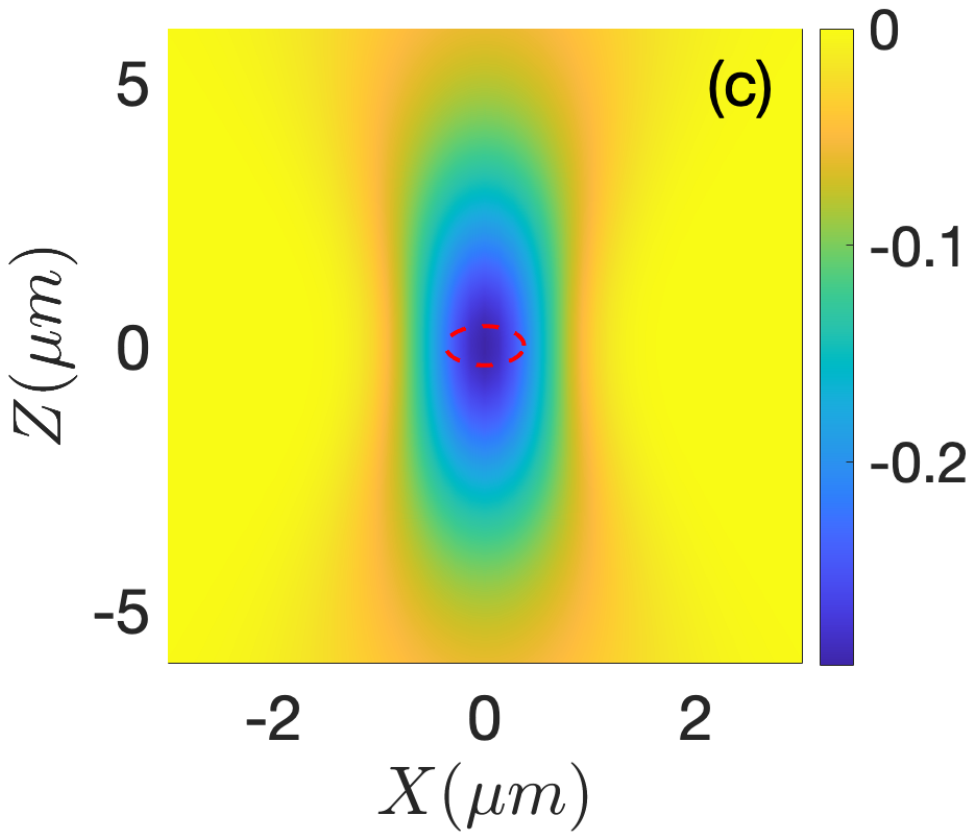}}
\hspace{-0.8cm}
{\includegraphics[trim=5.0cm 0cm 0cm 0cm, clip=true, totalheight=0.30\textheight, angle=0]{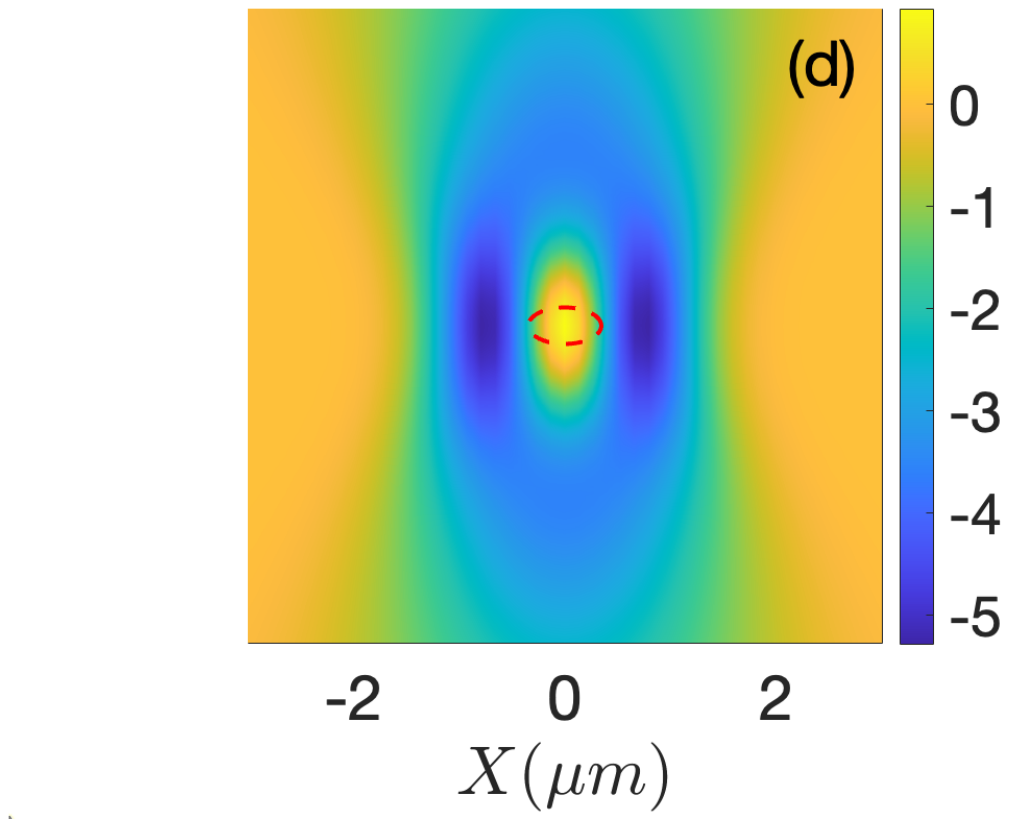}}\\
\vspace{-4.2cm}
{\includegraphics[ trim=0cm 0cm 0cm 0cm, clip=true, totalheight=0.30\textheight, angle=0]{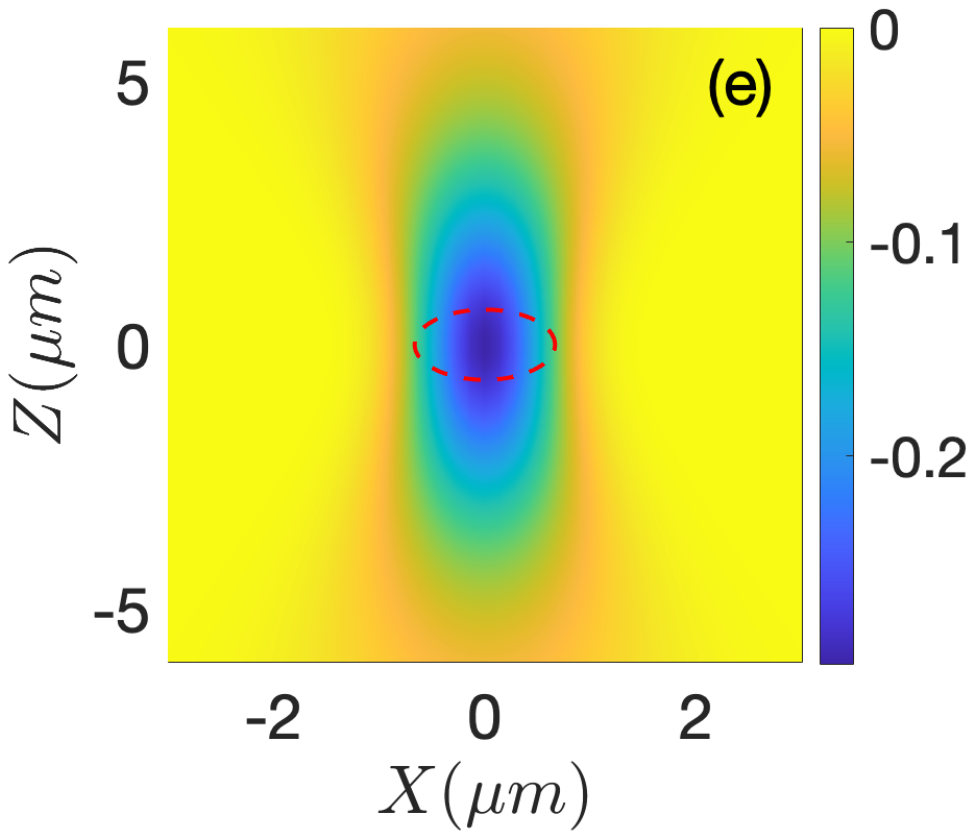}}
\hspace{-0.8cm}
{\includegraphics[trim=5.0cm 0cm 0cm 0cm, clip=true, totalheight=0.30\textheight, angle=0]{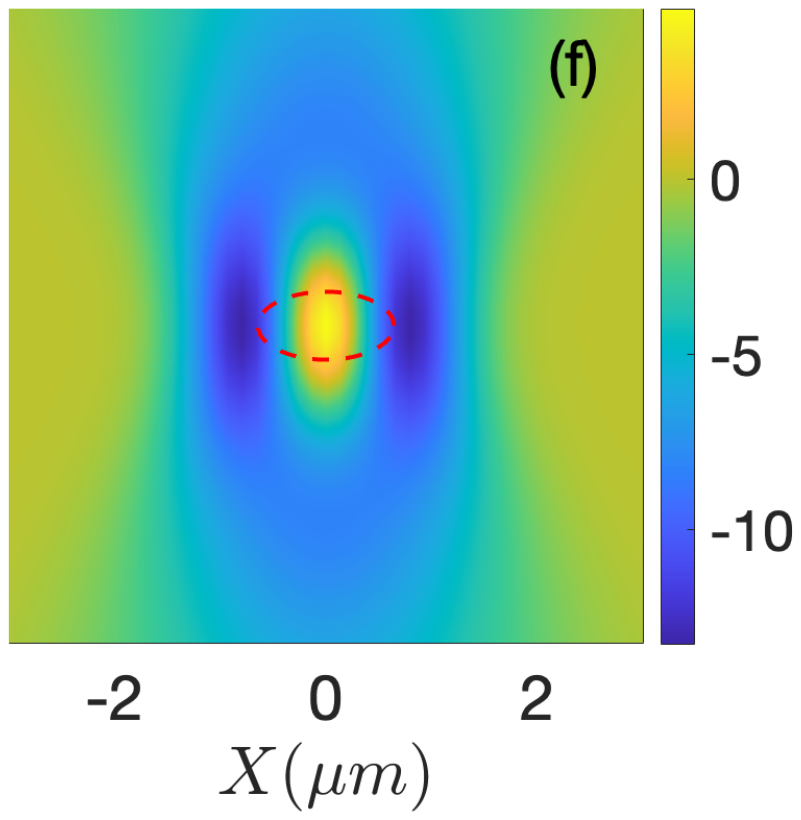}}\\
\vspace{-1.5cm}
\caption{ Trapping of $nD_{3/2,-3/2}$ series of the rubidium Rydberg atom 
for $\lambda=1,000$~nm, $w_0=1$~$\mu$m,  $P=2.5$~mW, and $\theta_p=\pi/2$. 
Left column: Rydberg trapping potential $U_{\text{ryd}}(X,0,Z,\omega)$ as functions of $X$ and $Z$ for   (a) $n=30$ and $A \cos \theta_k=1$, (c) $n=60$ and $A \cos \theta_k=0.1915$, and (e) $n=80$ and $A \cos \theta_k=0.0308$. The color bars indicate the values of $U_{\text{ryd}}(X,0,Z,\omega)$ in mK. 
 Right column: Differential trapping potential $\Delta U(X,0,Z,\omega)$  as functions of $X$ and $Z$. The $n$ and $A \cos \theta_k$ values in (b), (d), and (f) are the same as those used in (a), (c), and (e), respectively. The color bars indicate the values of $\Delta U(X,0,Z,\omega)$ in $\mu$K (it is important to note that the color bars in the left and right columns use different scales, namely mK and $\mu$K, respectively).
 The red dashed lines demarcate the $(X=x_*,Z=z_*)$ values for which 95~\% of the probability of the Rydberg electron, for $Y=0$, lies inside the red dashed line (see  the right column of Fig.~\ref{Fig2_extra} for reference).  }
\label{Fig4_extra}
\end{figure}

To start with, we fix the power, beam waist, and wavelength  at $P=2.5~$mW, $w_0=1$~$\mu$m, and $\lambda=1,000$~nm, respectively (other wavelengths will be considered later).  For these parameters, the Stark shift experienced by the ground state of rubidium is, as can be seen in Fig.~\ref{Fig1_extra}(b), negative; the light is  red-detuned. The left columns of Figs.~\ref{Fig3_extra} and
\ref{Fig4_extra} show the total trapping potential $U_{\text{ryd}}(X,0,Z,\omega)$ for selected rubidium Rydberg states  of the $nS_{1/2,+1/2}$ and  $nD_{3/2,-3/2}$ series, respectively.
In both figures, $n$ is equal to $30$, $60$, and $80$ in the top, middle, and bottom rows, respectively.
In Fig.~\ref{Fig3_extra}, the geometric factor $A\cos\theta_k$ is set to $1$ for the $30S_{1/2,+1/2}$ and   $60S_{1/2,+1/2}$ states and to  $0.3923$ for the $80S_{1/2,+1/2}$ state.
The resulting potential is positive (indicating anti-trapping) for $n=30$ but negative (indicating trapping) for $n=60$ and $80$, with trap depths around a tenth of a milli-Kelvin. The trapping potentials shown 
in Figs.~\ref{Fig3_extra}(c) and  \ref{Fig3_extra}(e) should comfortably hold Rydberg states.
In Fig.~\ref{Fig4_extra}, the geometric factor $A\cos\theta_k$ is set to $1$ for the $30D_{3/2,-3/2}$ state,
to  
$0.1915$ for the $60D_{3/2,-3/2}$ state, and to $0.0308$ for the $80D_{3/2,-3/2}$ state. 
For the $D$-states, the resulting potential is  negative for all three $n$ considered, with trap depths comparable to those for the $n=60$ and $n=80$ $S$-states.

The ponderomotive potential  $U_{\text{pon}}(X,Y,Z,\omega)$ experienced by a Rydberg state  is positive for all $(X,Y,Z)$ and wavelengths, provided one works with far off-resonant laser light.  As a result, a Rydberg state  cannot be confined by a red-detuned optical trap if the only light-induced energy shift is $U_{\text{pon}}(X,Y,Z,\omega)$. However, the Stark shift  $U_{\text{stark}}(X,Y,Z,\omega)$  for Rydberg states  possesses---as discussed in detail below---an appreciable dependence on $A \cos \theta_k$. This tunability is exploited in Figs.~\ref{Fig3_extra} 
and \ref{Fig4_extra}. Specifically,  the value of 
$A \cos \theta_k$ in Figs.~\ref{Fig3_extra}(e)-\ref{Fig3_extra}(f) and \ref{Fig4_extra}(c)-\ref{Fig4_extra}(f) are adjusted such that
(i) $U_{\text{ryd}}(X,Y,Z,\omega)$ is negative and (ii) the differential potential $\Delta U(X,0,Z,\omega)$, where $\Delta U(X,0,Z,\omega)$ is defined through
\begin{eqnarray}
    \Delta U(X,Y,Z,\omega)= U_{\text{gr}}(X,Y,Z,\omega)- U_{\text{ryd}}(X,Y,Z,\omega),\nonumber \\
\end{eqnarray}
is
minimized.
The metric for the minimization is 
\begin{eqnarray}\label{eq_metric}
\min \int_{-x_*}^{x_*} \int_{-z_*}^{z_*} |\Delta U(X,0,Z,\omega)|dXdZ.
\end{eqnarray}
For the $30S_{1/2,+1/2}$ state considered in Fig.~\ref{Fig3_extra}(a), a value of $A \cos \theta_k$  that yields a negative $U_{\text{ryd}}(X,Y,Z,\omega)$ was not found. This means that this state cannot be trapped using red-detuned light.
For the  $60S_{1/2,+1/2}$ state considered in Fig.~\ref{Fig3_extra}(c) and the   $30D_{3/2,-3/2}$ state considered in Fig.~\ref{Fig4_extra}(a), a value of $A \cos \theta_k=1$ yields the best result, with 
$\Delta U(X,0,Z,\omega)$ being comparable to or just a bit smaller than, in magnitude, the ground state and Rydberg state trapping potentials themselves.
For the $80S_{1/2,+1/2}$,
$60D_{3/2,-3/2}$, and 
$80D_{3/2,-3/2}$ states, in contrast, the magnitude of $\Delta U(X,0,Z,\omega)$ is of the order of a few $\mu$K in the regime where the Rydberg atom resides (i.e., inside the region demarcated by the red dashed line), suggesting that the chosen parameters lead to approximately equal trapping potentials for the ground and Rydberg states across the entire Rydberg atom. We can think of  this as a magic wavelength-like condition. In our case, the geometric factor $A \cos \theta_k$ is tuned to realize  approximately equal trapping potentials. A change of $A \cos \theta_k$ by about $0.01$ for $80S_{1/2,+1/2}$ and $60D_{3/2,-3/2}$  and by about $0.003$ for $80D_{3/2,-3/2}$ leads to a change of the metric defined in Eq.~(\ref{eq_metric}) by  a factor of 2.  This shows that the sensitivity of the optimal trapping conditions to the geometric factor depends 
on the Rydberg state under consideration.   While Figs.~\ref{Fig3_extra} and \ref{Fig4_extra} are for rubidium, we find analogous magic wavelength-like conditions for cesium (see Appendix~\ref{sec_appendix_b} and Figs.~\ref{FigS1_extra} and \ref{FigS2_extra}).

The discussion above demonstrates that achieving simultaneous trapping of both the ground state and a Rydberg state via a red-detuned laser  depends critically on the contribution from the AC Stark shift to the Rydberg potential. In what follows, we provide a detailed analysis of the  polarizabilities $\alpha^S(\omega)$, $\alpha^V(\omega)$, and $\alpha^T(\omega)$, and examine how they---together with the geometric factors---enable  tuning of the dynamic AC Stark shift and correspondingly of $U_{\text{ryd}}(X,Y,Z,\omega)$. A detailed understanding of the polarizabilities of the Rydberg states for non-linearly polarized light  offers valuable insights into how to arrive at optimal   trapping conditions for the simultaneous trapping of the ground state and Rydberg states.

 \begin{figure}[t]
\vspace*{-0.1cm}
{\includegraphics[ scale=.35]{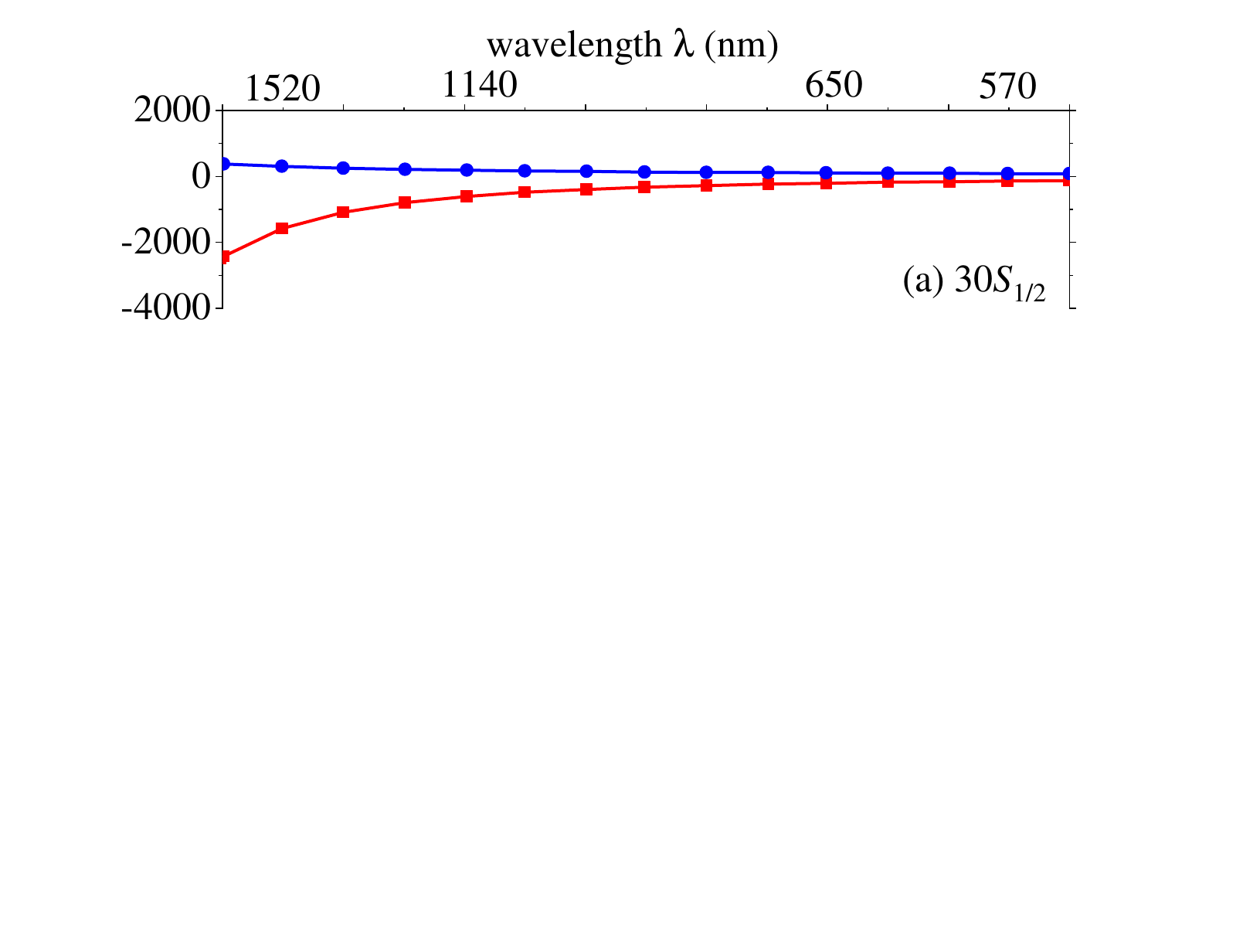}}\\
\vspace{-5.76cm}
{\includegraphics[ scale=.35]{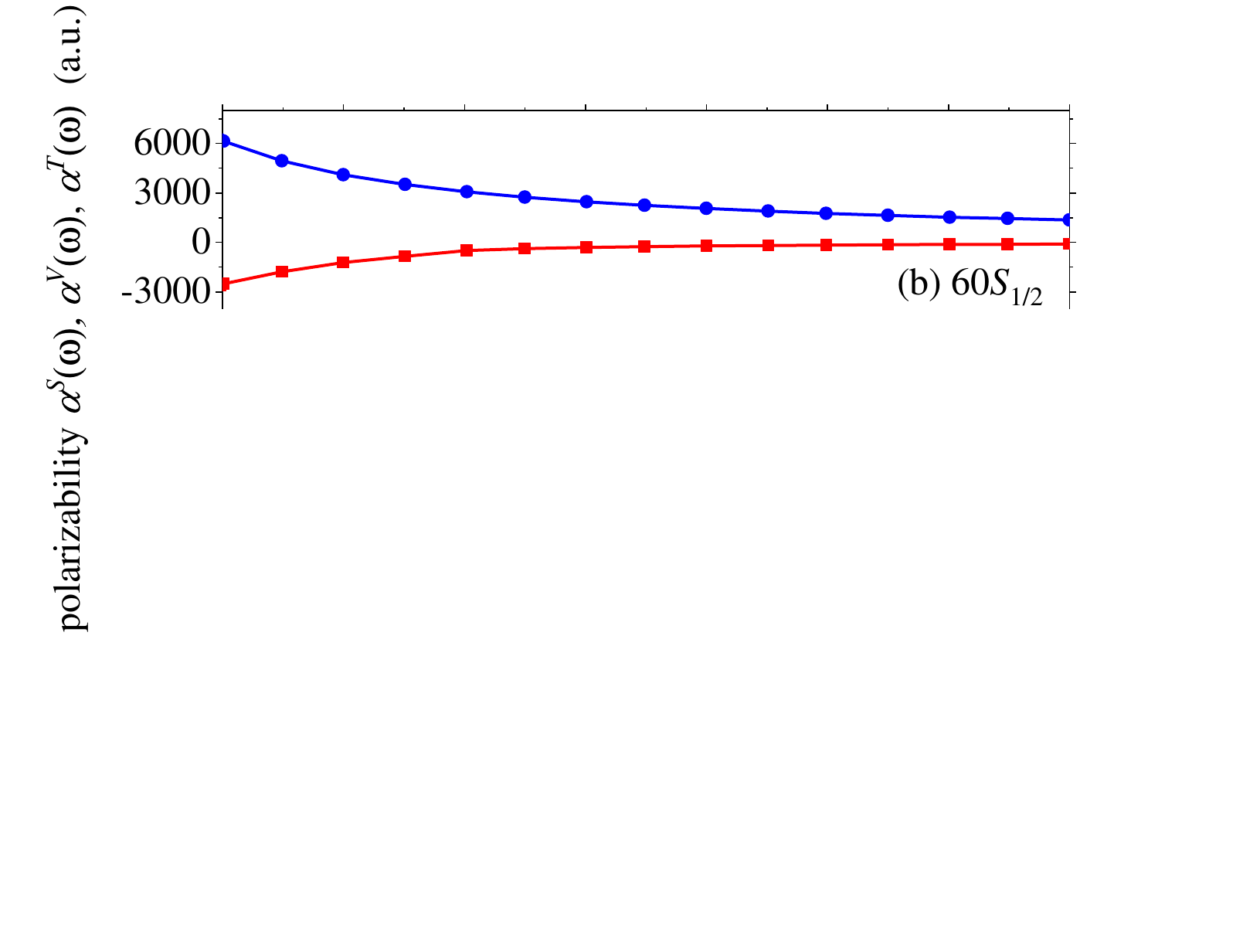}}\\
\vspace{-5.76cm}
{\includegraphics[ scale=.35]{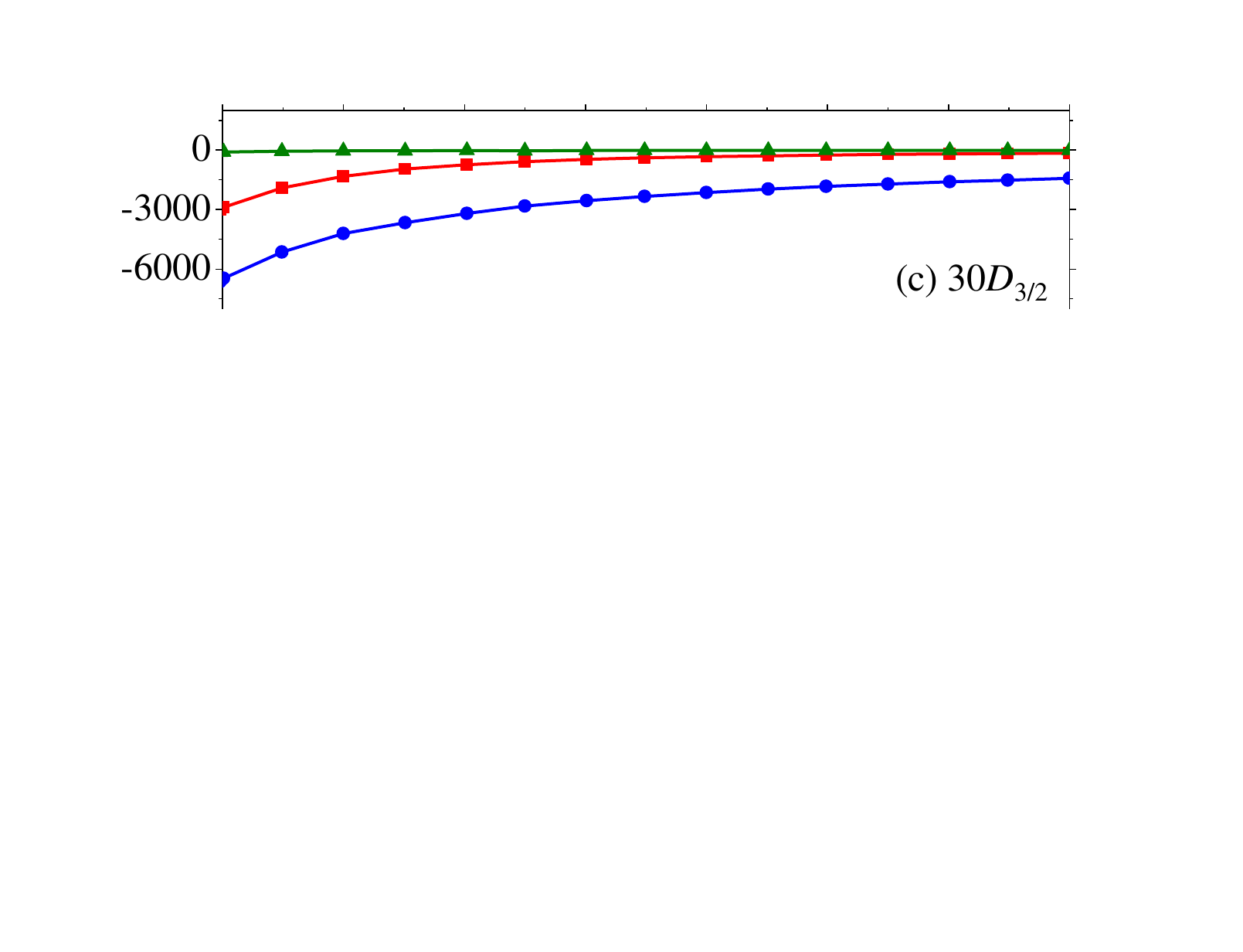}}\\
\vspace{-5.76cm}
{\includegraphics[ scale=.35]{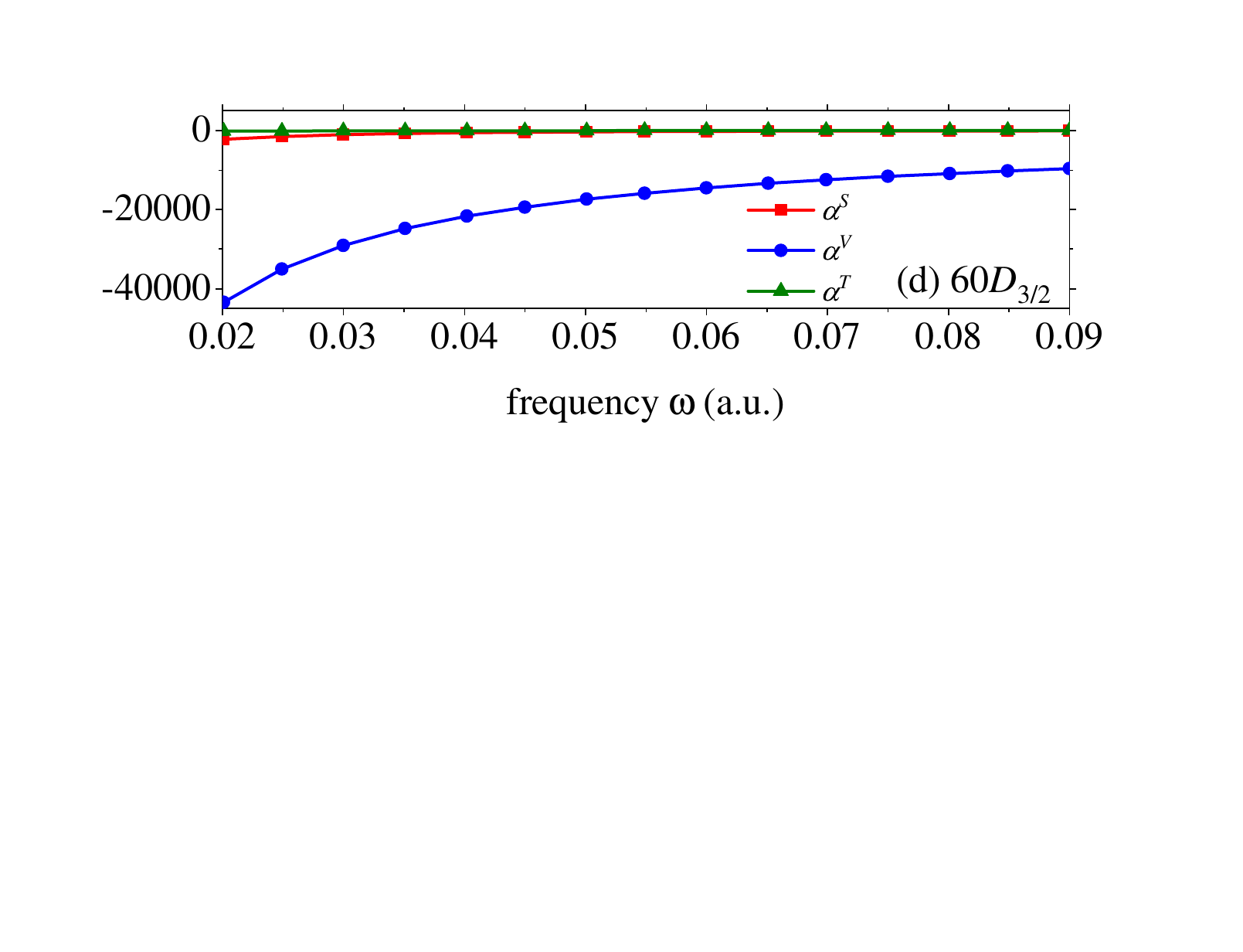}}\\
\vspace{-4.28cm}
\caption{Squares, circles, and triangles show the numerically obtained scalar, vector, and tensor polarizabilities  $\alpha^S(\omega)$, $\alpha^V(\omega)$,  and $\alpha^T(\omega)$, respectively, for rubidium as a function of the frequency $\omega$  ["a.u." stands for "atomic unit;" the  corresponding wavelengths $\lambda$ range from 500~nm to 2,200~nm, see the  top axis]. The one-parameter fits (lines) agree excellently with the symbols. Panels (a), (b), (c), and (d) are for the $30S_{1/2}$, $60S_{1/2}$, $30D_{3/2}$, and $60D_{3/2}$ states, respectively.}
\label{Fig1}
\end{figure}

   Figure~\ref{Fig1} shows  the scalar, vector, and tensor polarizabilities for selected rubidium  Rydberg states  as a function of $\omega$. The frequency range shown corresponds to experimentally accessible wavelengths between $500$ and $2,200$~nm (top axis).
   Figure~\ref{Fig1} considers the $nS_{1/2}$ and $nD_{3/2}$ states with  $n=30$ and $n=60$. The polarizabilities obtained from the sum-over-states  approach are shown by symbols while lines show one-parameter  fits (see below) to the numerical data.   Note that the tensor polarizability vanishes for  the $S_{1/2}$ states due to their spherical  shape of the $S$-orbital.   For  the $nD_{3/2}$ states with  $n=30$ and $n=60$,  $|\alpha^T(\omega)|$ is   smaller than  $|\alpha^S(\omega)|$   over the entire frequency range shown. 
   Comparison of the vector and scalar polarizabilities shows the following:
   For the $30S_{1/2}$ state, $|\alpha^V(\omega)|$
   is smaller than $|\alpha^S(\omega)|$.
For the $60S_{1/2}$, $30D_{3/2}$, and $60D_{3/2}$ states, in contrast, $|\alpha^V(\omega)|$
   is larger than $|\alpha^S(\omega)|$.
    This suggests that there exists, for each 
    $nL_{J}$ 
    series, a "reordering"  of the polarizability contributions from $|\alpha^V(\omega)|<|\alpha^S(\omega)|$ to $|\alpha^V(\omega)|>|\alpha^S(\omega)|$ as $n$ increases.
    For the $nD_{3/2}$ series, this reordering occurs for $n<30$.   We  confirmed that this reordering exists also for the $nP_{1/2}$, $nP_{3/2}$, and $nD_{5/2}$ series for both  rubidium and cesium.    The comparatively large magnitude of  $\alpha^V(\omega)$ suggests that  a positive total  polarizability can be realized by tuning $f^V(A,\theta_k, J, M_J)$.

\begin{figure}[t]
\vspace*{-0.65cm}
{\includegraphics[ scale=.35]{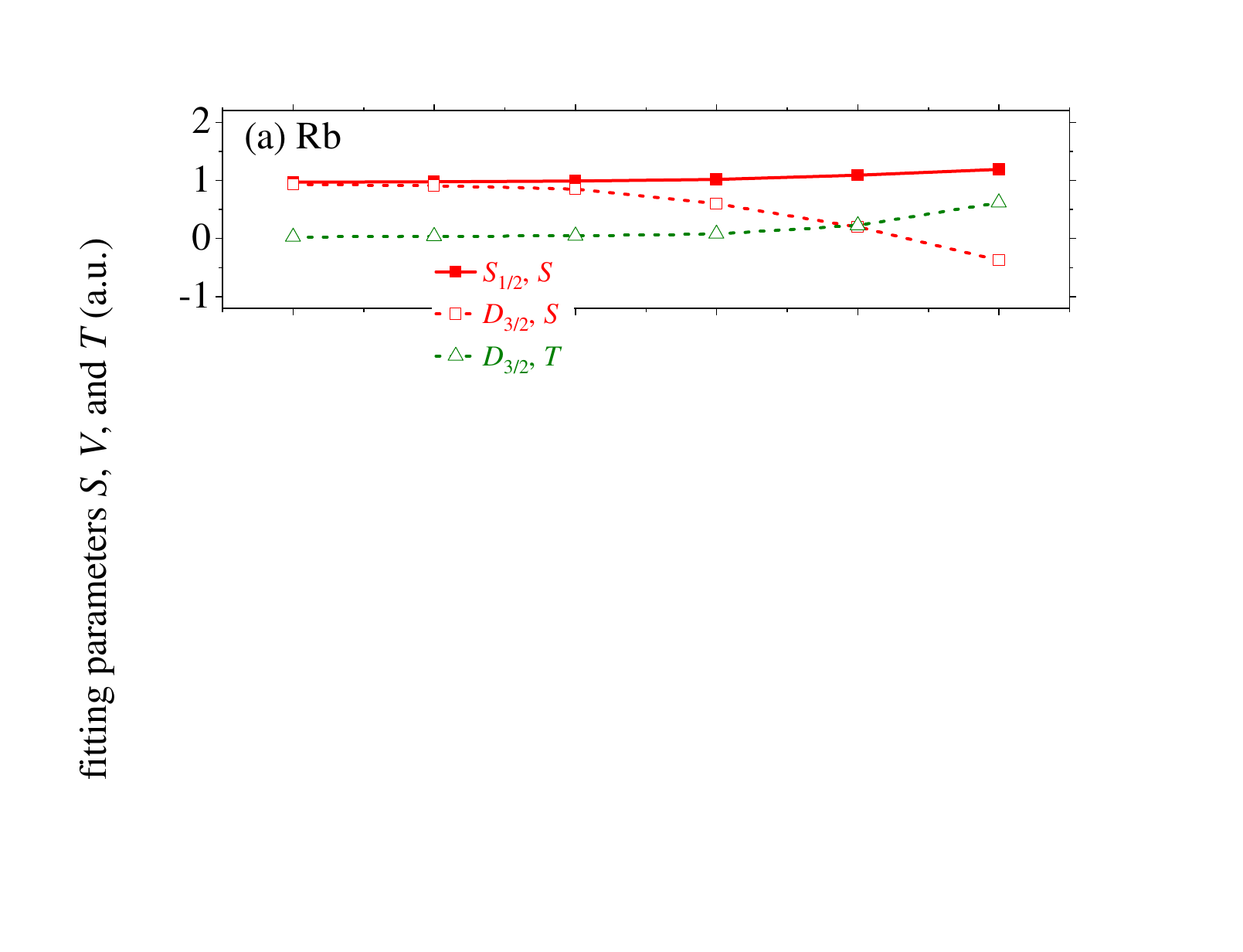}}\\
\vspace{-5.785cm}{\includegraphics[ scale=.35]{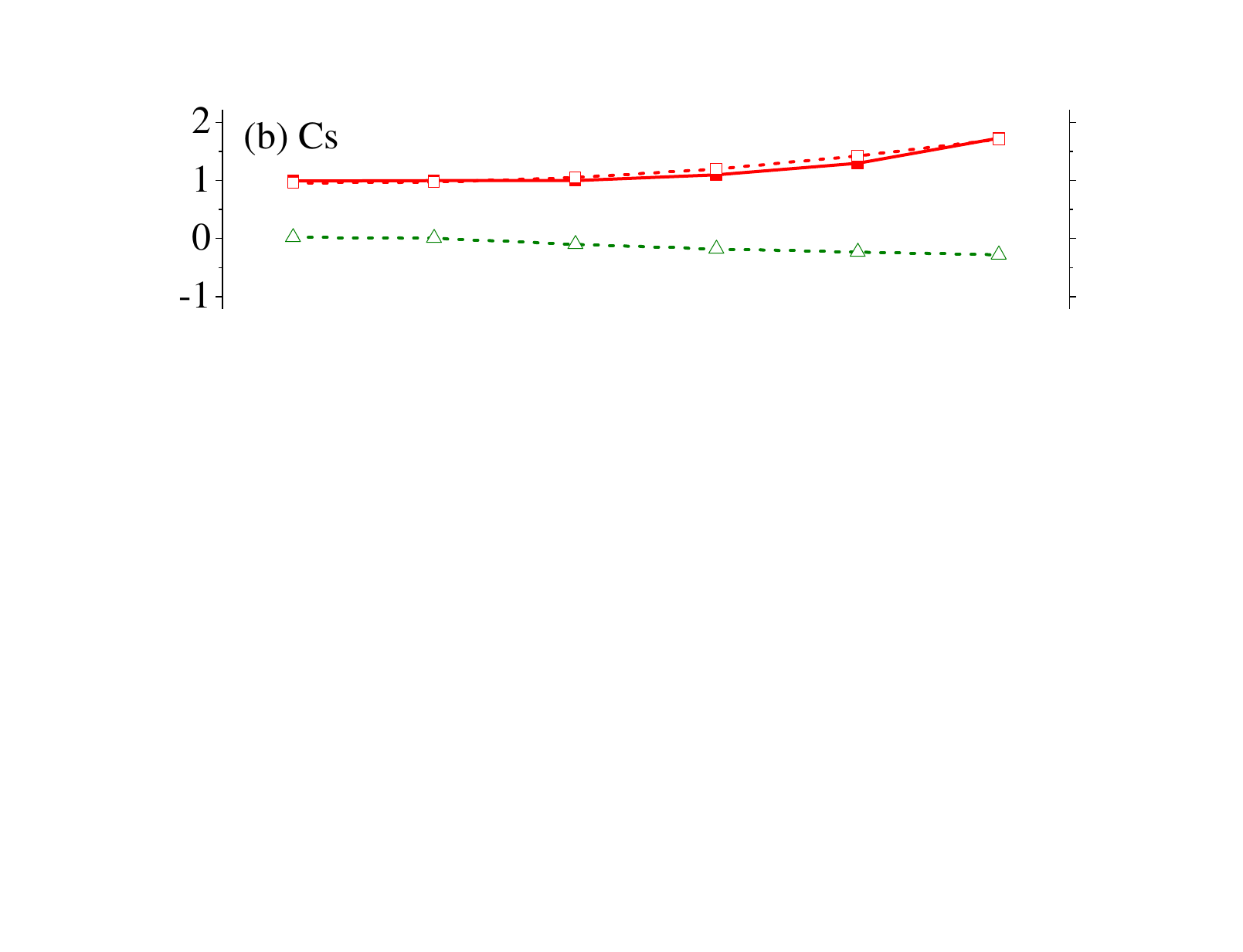}}\\
\vspace{-5.785cm}
{\includegraphics[ scale=.35]{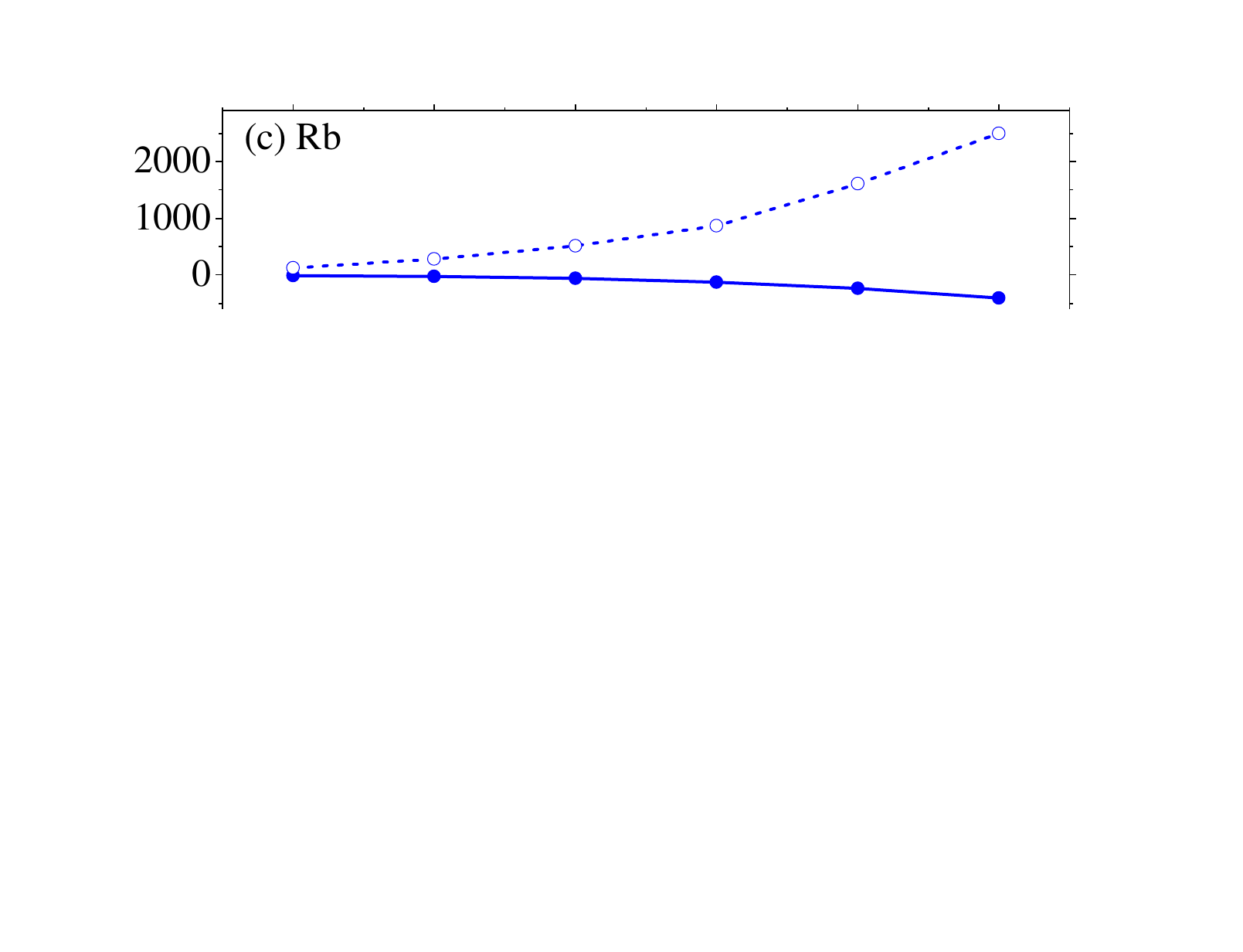}}\\
\vspace{-5.785cm}{\includegraphics[ scale=.35]{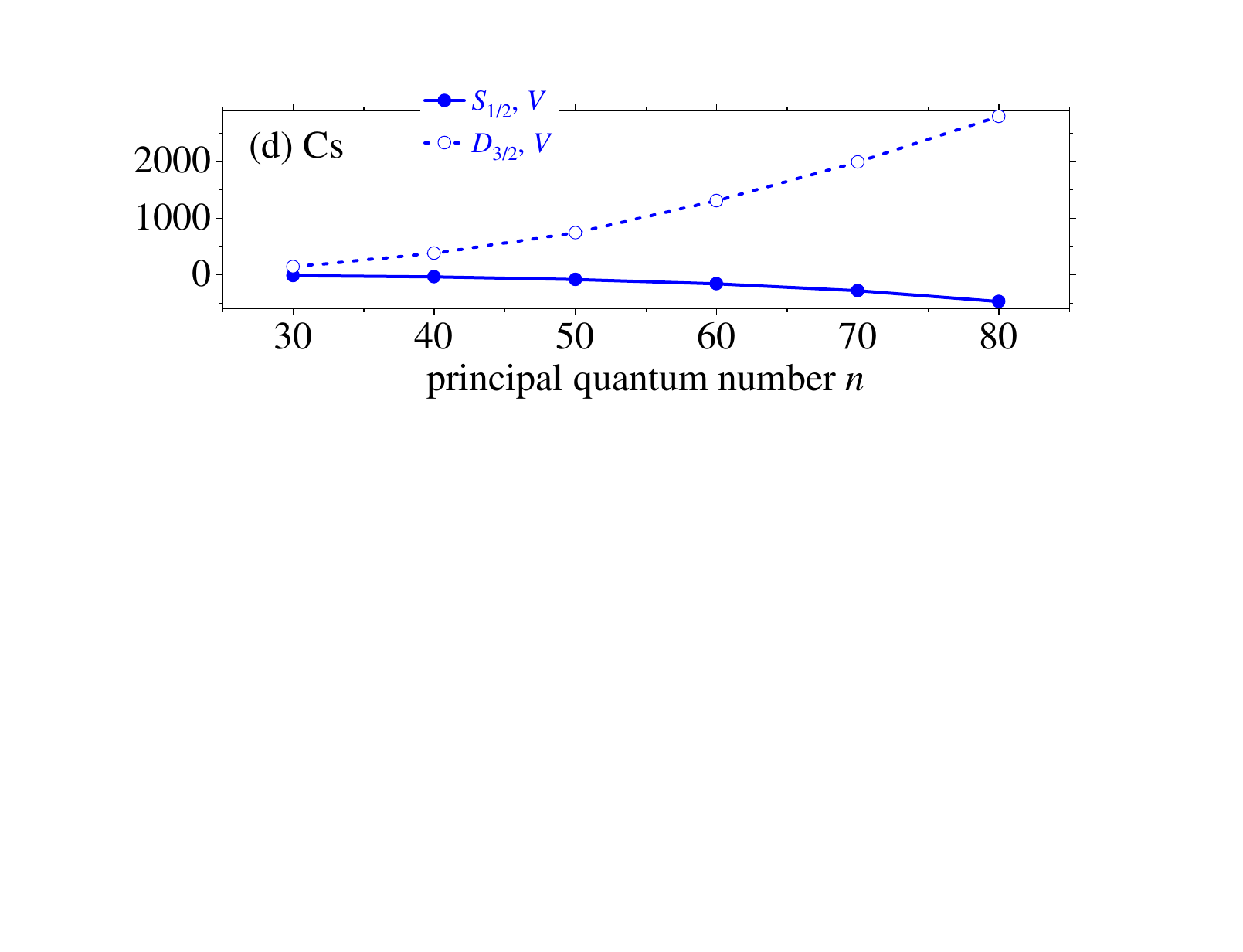}}\\
\vspace{-4.28cm}
\caption{ Symbols in (a)/(b) show the fitting parameters  $S$ [$\alpha^S(\omega)=-S\omega^{-2}$]
 and $T$ [$\alpha^T(\omega)=-T\omega^{-2}$]  while symbols in (c)/(d) show  
 $V$  [$\alpha^V(\omega)=-V\omega^{-1}$] as a function of  $n$ [to guide the eye, lines connect neighbouring data points]; (a) and (c) are for rubidium while (b) and (d) are for cesium.
In (a) and (b), filled squares, open squares, and open triangles show $S$ for the $S_{1/2}$ series, $S$ for the $D_{3/2}$ series, and $T$ for the $D_{3/2}$ series, respectively;
in (c) and (d),  filled circles and open circles show $V$ for the $S_{1/2}$ and  $D_{3/2}$ series, respectively. 
}
\label{Fig2}
\end{figure}

    In addition to the relative magnitudes of the contributions to the total polarizability, it is important to comment on the scaling of $\alpha^S(\omega)$, $\alpha^V(\omega)$, and $\alpha^T(\omega)$ with $\omega$.
    The lines in Fig.~\ref{Fig1} show  fits of the form  $\alpha^S(\omega)=-S\omega^{-2}$, $\alpha^V(\omega)=-V\omega^{-1}$, and $\alpha^T(\omega)=-T\omega^{-2}$ to the numerical data, where $S$, $V$, and $T$ are treated as fitting parameters.  Figure~\ref{Fig1} shows that the single-parameter fits (lines) describe  the numerical data (symbols) excellently.  This behavior should be contrasted with that for energetically low-lying states. Since the polarizability contributions for the ground state and low-lying excited states tend to feature broad resonances, they do not exhibit a  "pure"  $\omega^{-1}$ or $\omega^{-2}$ scaling. By comparison, resonances in  Rydberg states are very narrow \cite{Saffman2005, Bai2020_1, Bhowmik2024}; in fact, Fig.~\ref{Fig1}  "skips over" them by choosing an $\omega$-grid  that excludes  various extremely narrow resonances. The fact that the vector and scalar polarizabilities, which have a larger magnitude for Rydberg states than the tensor polarizability, scale differently with $\omega$ can be viewed as an  "indirect tuning knob" that may be leveraged to find parameter combinations for which the total polarizability of Rydberg states is quite flat over a wide range of frequencies.

 \begin{figure}[t]
 \vspace*{-0.55cm}
{\includegraphics[ scale=.35]{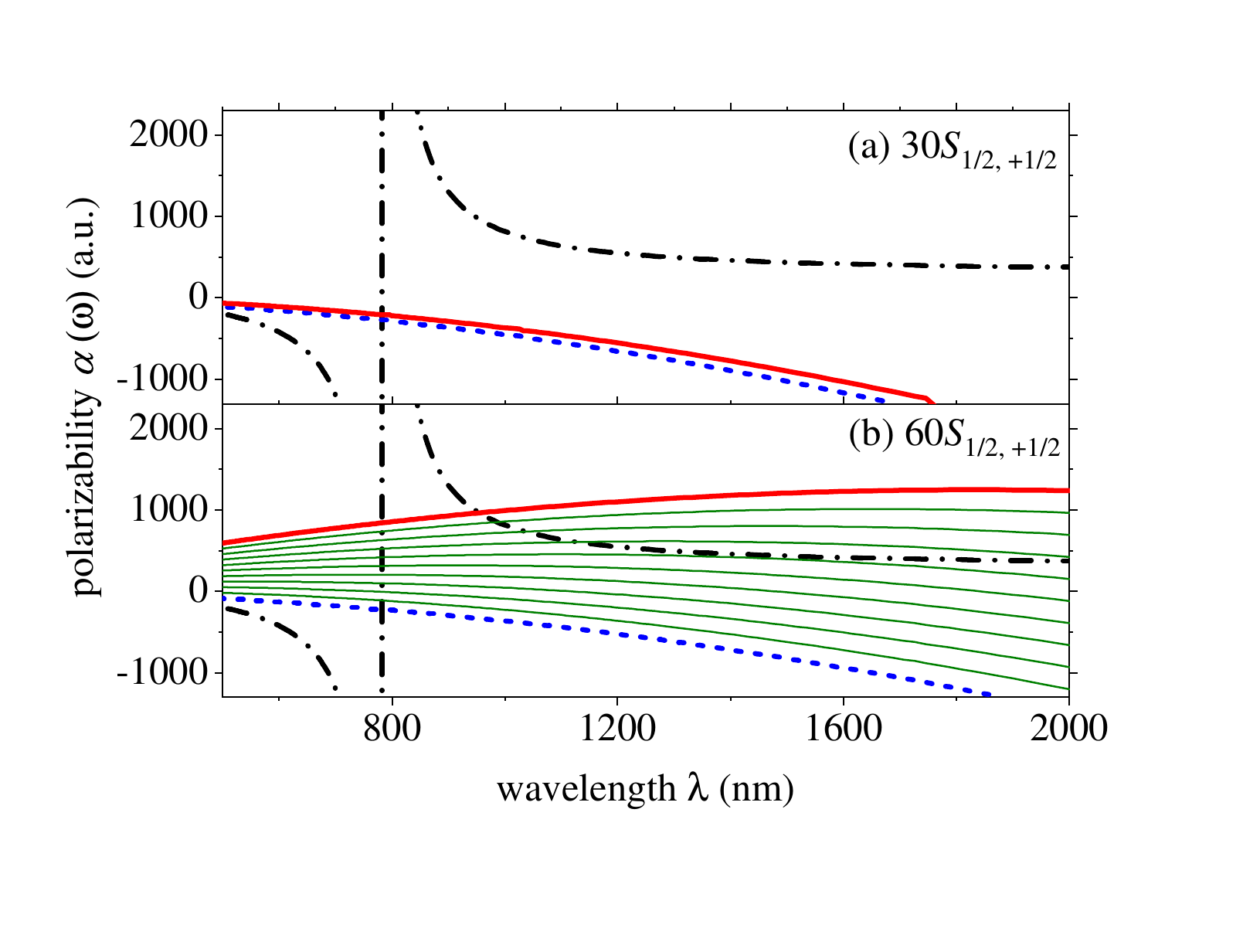}}\\
\vspace{-1.28cm}
\caption{ Polarizability of the ground and Rydberg states for Rb  as a function of wavelength. The Rydberg states are (a) $30S_{1/2,+1/2}$ and (b)  $60S_{1/2,+1/2}$.  The quantization axis is perpendicular to the  polarization vector, i.e.,  $\theta_p=\pi/2$.   The black dash-dotted lines show the polarizability of the ground state.  The blue dashed  and  red solid lines  show $\alpha(\omega)$ of the Rydberg state for  linearly and circularly polarized light, respectively.   In (b), the thin green lines show the polarizability of the Rydberg states for elliptically polarized light for various values of $A\cos\theta_k$ 
[$A\cos\theta_k=0.1$ (bottom-most curve) to $0.9$ (top-most curve)].  For the  $60S_{1/2,+1/2}$ state,  variation of the geometric factor $A \cos \theta_k$ affords  appreciable tunability of $U_{\text{stark}}(X, Y, Z, \omega)$. }
\label{Fig3}
\end{figure}

 To showcase how the frequency-dependent  scalar, vector, and  tensor polarizabilities change  with  increasing $n$,  symbols in Fig.~\ref{Fig2} show   $S$, $V$, and $T$ as a function of $n$ for the $nS_{1/2}$ and $nD_{3/2}$ series of rubidium and cesium. It can be seen that the magnitude of  $S$, $V$, and $T$ increases monotonically with increasing $n$.   The most prominent feature in Fig.~\ref{Fig2} is that the magnitude of $V$ changes by several orders of magnitude as $n$ changes from $30$ to $80$. Even though the units of $V$ and $S$ differ, it is meaningful to note that the  percentage   change of $|V|$ with $n$ is larger than the percentage  change of $|S|$ with $n$. This is consistent with the "reordering" of the importance of the scalar and vector polarizabilities discussed in the context of Fig.~\ref{Fig1}.

 \begin{figure}[t]
 \vspace*{-0.55cm}
{\includegraphics[ scale=.35]{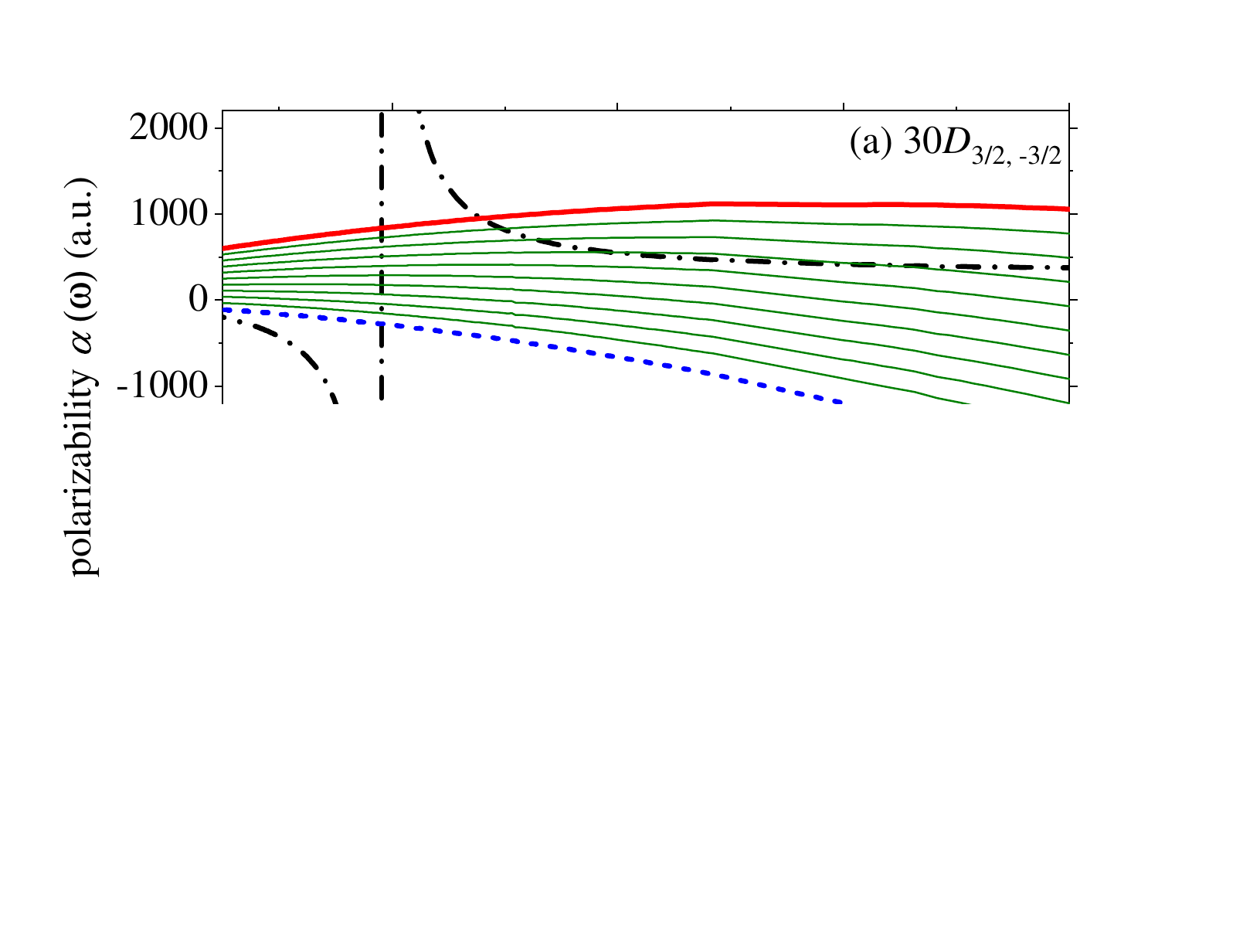}}\\
\vspace{-5.055cm}
{\includegraphics[ scale=.35]{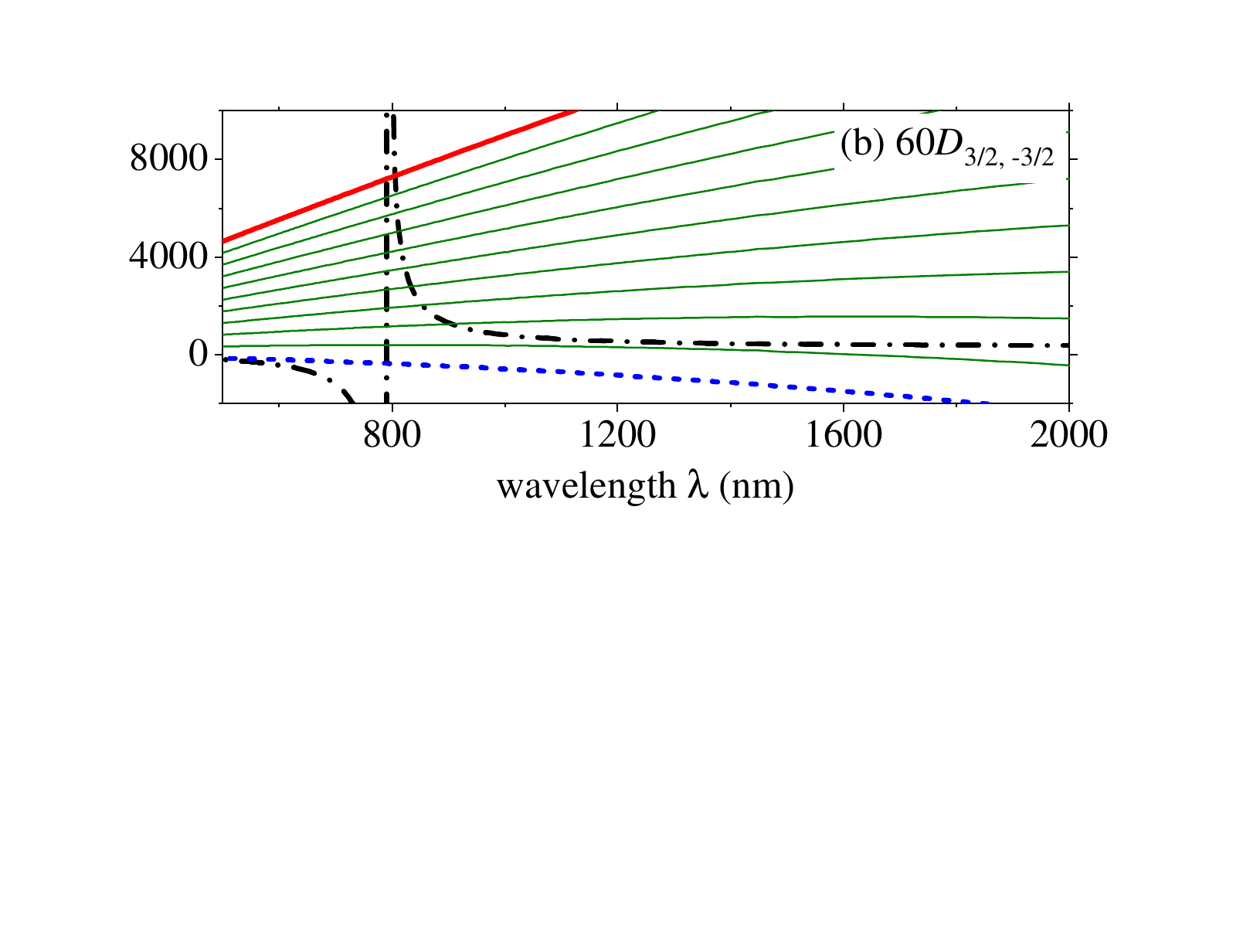}}\\
\vspace{-3.28cm}
\caption{Same as Fig.~\ref{Fig3} but for the $D_{3/2,-3/2}$ states as opposed to $S_{1/2,+1/2}$ states.  For both $D$-states shown,  variation of the geometric factor $A \cos \theta_k$ affords  appreciable tunability of $U_{\text{stark}}(X, Y, Z, \omega)$.}
\label{Fig4}
\end{figure}

We now demonstrate that the vector component affords a tremendous tunability of the total dynamic polarizability $\alpha(\omega)$ for a large number of alkali Rydberg states. 
 Figure~\ref{Fig3} depicts $\alpha(\omega)$ for the ground state of rubidium (black dash-dotted line) and the $nS_{1/2,+1/2}$ Rydberg states with (a) $n=30$  and (b) $n=60$.   The results are shown for the scenario in which the quantization axis is perpendicular to the polarization vector, i.e., for $\theta_p=\pi/2$.  Since the tensor contribution to the total polarizability is very small,  
 the choice $\theta_p=\pi/2$ provides maximal flexibility in selecting  the angle $\theta_k$, which is related to $\theta_p$ via the "geometric constraint" $\cos^2\theta_k+\cos^2\theta_p\leq 1$~\cite{Beloy2009,Jiang2017}.  In Figs.~\ref{Fig3}(a) and \ref{Fig3}(b), the blue dashed line  shows the polarizability of the Rydberg state for linearly polarized light ($A=0$). As expected~\cite{Bai2020_1,Bhowmik2024}, the Rydberg state polarizabilities are negative for the wavelengths considered. As the polarization of the light  changes from linear to elliptical, the polarizability of the Rydberg state with $M_J=+1/2$ shifts up  while that with $M_J=-1/2$ shifts down (here, we consider $A\cos\theta_k>0$; the situation for   $A\cos\theta_k<0$ is reversed).   For $n=30$ [Fig.~\ref{Fig3}(a)], the shift is very small and lies, for $M_J=+1/2$, between the blue dashed  line and the red solid line, which shows the result for the maximal up-shift, namely for $A \cos \theta_k=1$. For $n=60$ [Fig.~\ref{Fig3}(b)], in contrast, the up-shift for $M_J=+1/2$ is appreciable, with the maximal shift occurring  for $A \cos \theta_k=1$ (red solid line). The  green  lines in-between the blue dashed and  red solid lines show the polarizability for
 various $A \cos \theta_k$.  The fact that the total polarizability for $n=60$ is  positive for some  geometric factors is a consequence of the reordering of the  polarizability contribution 
 from  $|\alpha^V(\omega)|<|\alpha^S(\omega)|$ to  $|\alpha^V(\omega)|>|\alpha^S(\omega)|$.  This observation affords the  ``counteracting'' of the repulsive ponderomotive potential $U_{\text{pon}}(X,Y,Z,\omega)$ by a negative (and larger in magnitude)  Stark shift  $U_{\text{stark}}(X,Y,Z,\omega)$, as demonstrated in Fig.~\ref{Fig3_extra}.

Figure~\ref{Fig4} shows that the $n$ value for which the vector contribution affords appreciable tunability depends on the Rydberg series considered.
While the tunability for the rubidium  $nS_{1/2}$ series
requires $n$ of order $55$,  the tunability extends to $n$ values as low as $30$.  As an example, Fig.~\ref{Fig4} shows the $nD_{3/2,-3/2}$ series with $n=30$ and $60$. For this series, the upshift of $\alpha(\omega)$ for $A\cos\theta_k>0$ occurs for negative $M_J$, as opposed to for positive $M_J$  as in Fig.~\ref{Fig3}, since $\alpha^V(\omega)$ is negative. Use of $M_J=+3/2$ would, for $A\cos\theta_k>0$,  lead to a down-shift of $\alpha(\omega)$  of the Rydberg states  in Fig.~\ref{Fig4}. Another notable feature of the tunability via the vector contribution is that the polarizabilities display a relatively weak frequency dependence  giving rise to several 100~nm-wide frequency windows in which the ground state and Rydberg state polarizabilities change in a similar manner.

Figure~\ref{Fig2} shows that the fitting parameters $S$, $V$, and $T$ for the $nS_{1/2}$ and $nD_{3/2}$ series vary
roughly  in the same manner with $n$ for rubidium and for cesium. This suggests that the tunability afforded by the vector contribution does not only apply to rubidium   but also to cesium as well as other alkali Rydberg atoms. Indeed, calculations show that this reasoning is true. As  examples, Figs.~\ref{FigS2} and \ref{FigS3} 
from Appendix~\ref{sec_appendix_b} show $\alpha(\omega)$ for cesium;  
the resemblance between the results for cesium and rubidium is evident. Interestingly, though,  there exist Rydberg series for which different alkali atoms display distinct behaviors.  Figure~\ref{FigS4} from Appendix~\ref{sec_appendix_c} shows, e.g.,  
 that  rubidium possesses  sufficient tunability to simultaneously trap  the ground state and the $60P_{3/2,-3/2}$ Rydberg state, utilizing  a positive total polarizability,  while cesium does not.

Figures~\ref{Fig1}-\ref{Fig4}  illustrate that the Stark shift of the Rydberg atom can be tuned appreciably by  tuning the laser wavelength and the geometric factor $A\cos\theta_k$. This tunability is leveraged in Figs.~\ref{Fig3_extra}, \ref{Fig4_extra}, \ref{FigS1_extra}, and \ref{FigS2_extra} to  engineer the trap potential $U_{\text{ryd}}(X,Y,Z,\omega)$ so as to facilitate the  simultaneous trapping of the ground state and a Rydberg state, using parameter combinations that minimize the differential potential $\Delta U(X,Y,Z,\omega)$ for the $(X,Y,Z)$ values where the Rydberg atom has appreciable probability.

\section{Conclusion}
\label{sec_conclusion}

In conclusion,   we propose to employ  non-linearly polarized far off-resonant light to exert a force on an alkali  Rydberg atom in the direction of the  high intensity region of the  light by tuning the geometry of the experimental setup (i.e., by varying $\theta_k$ and $\theta_p$). The non-linearly polarized light induces a vector polarizability in the Rydberg atom, 
which is accompanied by a fictitious magnetic field. Working away from resonances, we demonstrated that the vector polarizability cannot only dominate the total polarizability but can also change its sign; for the $nS$, $nP$, and $nD$ series investigated, such  sign change
is not afforded by the
 scalar and tensor polarizabilities.  The resulting Stark shift was shown to ``counteract'' the repulsive ponderomotive potential, leading to an overall attractive trapping potential.  Our approach strongly suppresses unwanted losses, thereby enabling long coherence times.   Extensions to multiple Rydberg states is an exciting  topic for future study.  

The proposed approach for trapping Rydberg atoms in the high intensity region of light 
is expected to enable advances in metrology, such as precision measurements of fundamental constants~\cite{Ramos2017,Jentschura2008} and Rydberg atom based magnetometry~\cite{Dietsche2019},
as well as in quantum science, where essentially lossless trapping of Rydberg atoms enables 
the generation of entangled states~\cite{Adams2020,Wu2021}. Our proposal may also be applied in Rydberg atom based quantum simulations of spin models, which aim to probe the long time dynamics~\cite{Browaeys2020, Wu2021}.

\section{Acknowledgments} 
Fruitful discussions with the Biedermann group are gratefully acknowledged.
This work was supported by an award from the W.~M. Keck Foundation.

\appendix

\section{Numerical details and benchmarking}
\label{sec_appendix_a}
\begin{widetext}

\begin{table}[h]
\scriptsize
\caption {The static scalar polarizability $\alpha^{S}(0)$ and the static tensor polarizability $\alpha^{T}(0)$    are compared with theoretical and experimental results from the literature (for rubidium, $n=30$, $40$, $50$, $60$, $70$, and $80$ are considered; for cesium, $n=30$ and $80$ are considered).  The polarizabilities are reported in atomic units 
 (a.u.) in the format  $x[y]$, which stands for $x\times 10^y$ or in the format $x(u)[y]$, which indicates $x\times 10^y$ with uncertainty $u$ in the last digit of $x$. Entries marked by the superscripts $a$, $b$, $c$, and $d$  are taken from  Ref.~\cite{Yerokhin2016}  (theoretical data), Ref.~\cite{Lai2018}  (theoretical data), Ref.~\cite{OSullivan1986} (experimental data), and Ref.~\cite{OSullivan1985} (experimental data),
  respectively.} % title of Table
\centering % used for centering table
\begin{tabular}{c |c|c | c| c|c||c|c|c} % centered columns (4 columns)  
  \hline
\hline
&\multicolumn{5}{c||}{scalar polarizability $\alpha^{S}(0)$}&\multicolumn{3}{c}{tensor polarizability $\alpha^{T}(0)$}\\
\hline
$n$	&	 $S_{1/2}$	&	 $P_{1/2}$	&	 $P_{3/2}$	&	 $D_{3/2}$	&	 $D_{5/2}$	&	 $P_{3/2}$	&	 $D_{3/2}$	&	 $D_{5/2}$	\\
\hline
\multicolumn{9}{c}{Rb}\\
\hline
30	&	0.555[10]	&	0.269[11]	&	0.297[11]	&	0.101[11]	&	0.909[10]	&	$-0.277[10]$	&	0.791[10]	&	0.132[11]	\\
&	0.555[10]$^a$	&	0.269[11]$^a$	&0.297[11]$^a$		&	0.101[11]$^a$	&	0.909[10]$^a$	&$-0.277[10]^a$		&	0.791[10]$^a$	&	0.132[11]$^a$	\\
&	0.566[10]$^b$	&	0.275[11]$^b$	&	0.293[11]$^b$	&	0.098[11]$^b$	&	0.916[10]$^b$	&	$-0.256[10]^b$	&	0.824[10]$^b$	&	0.130[11]$^b$	\\

	&		&		&		&	0.104(4)[11]$^c$	&	0.936(8)[10]$^c$	&		&	0.784(20)[10]$^c$	&	0.129(4)[11]$^c$	\\
    &	0.559(6)[10]$^d$	&		&		&		&		&		&		&		\\

	&		&		&		&		&		&		&		&		\\
40	&	0.425[11]	&	0.231[11]	&	0.256[12]	&	0.718[11]	&	0.636[11]	&	$-0.230[11]$	&	0.668[11]	&	0.111[12]	\\

&	0.425[11]$^a$	&	0.231[11]$^a$	&	0.256[12]$^a$	&	0.718[11]$^a$	&	0.636[11]$^a$	&$-0.230[11]^a$		&	0.668[11]$^a$	&	0.111[12]$^a$	\\

	&	0.438[11]$^b$	&	0.237[11]$^b$	&0.252[12]$^b$		&	0.695[11]$^b$	&	0.643[11]$^b$	&	$-0.210[11]^b$	&	0.695[11]$^b$	&	0.109[12]$^b$	\\

	&		&		&		&	0.74(3)[11]$^c$	&	0.67(2)[11]$^c$	&		&	0.64(3)[11]$^c$	&	0.11(4)[12]$^c$	\\
    &	0.425(8)[11]$^d$	&		&		&		&		&		&		&		\\

	&		&		&		&		&		&		&		&		\\
50	&	0.203[12]	&	0.119[13]	&	0.132[13]	&	0.329[12]	&	0.288[12]	&	$-0.116[12]$	&	0.341[12]	&	0.566[12]	\\
&	0.203[12]$^a$	&	0.119[13]$^a$	&0.132[13]$^a$		&	0.329[12]$^a$	&	0.288[12]$^a$	&	$-0.116[12]^a$	&	0.341[12]$^a$	&	0.566[12]$^a$	\\
&	0.209[12]$^b$	&	0.122[13]$^b$	&0.130[13]$^b$		&	0.318[12]$^b$	&	0.291[12]$^b$	&	$-0.106[12]^b$	&	0.356[12]$^b$	&	0.555[12]$^b$	\\

	&		&		&		&	0.341(12)[12]$^c$	&	0.289(16)[12]$^c$	&		&	0.329(12)[12]$^c$	&	0.539(20)[12]$^c$	\\
    &	0.203(13)[12]$^d$	&		&		&		&		&		&		&		\\

	&		&		&		&		&		&		&		&		\\

60	&	0.698[12]	&	0.455[13]	&	0.498[13]	&	0.106[13]	&	0.106[13]	&	$-0.398[12]$	&	0.132[13]	&	0.207[13]	\\
	&	0.747[12]$^b$	&	0.462[13]$^b$	&	0.490[13]$^b$	&	0.110[13]$^b$	&	0.100[13]$^b$	&	$-0.393[12]^b$	&	0.133[13]$^b$	&	0.209[13]$^b$	\\
	&		0.690(3)[12]$^d$&		&		&		&		&		&		&		\\

	&		&		&		&		&		&		&		&		\\

70	&	0.210[13]	&	0.135[14]	&	0.155[14]	&	0.321[13]	&	0.285[13]	&	$-0.124[13]$	&	0.397[13]	&	0.638[13]	\\
	&0.219[13]$^b$		&	0.141[14]$^b$	&	0.150[14]$^b$	&	0.316[13]$^b$	&	0.286[13]$^b$	&	$-0.118[13]^b$	&	0.402[13]$^b$	&	0.631[13]$^b$	\\
	&	0.215(8)[13]$^d$	&		&		&		&		&		&		&		\\
	&		&		&		&		&		&		&		&		\\
80	&	0.520[13]	&	0.364[14]	&	0.398[14]	&	0.791[13]	&	0.709[13]	&	$-0.315[13]$	&	0.101[14]	&	0.156[14]	\\
	&0.550[13]$^b$		&	0.369[14]$^b$	&	0.393[14]$^b$	&	0.788[13]$^b$	&	0.711[13]$^b$	&	$-0.306[13]^b$	&	0.105[14]$^b$	&	0.165[14]$^b$	\\
	&	0.538(52)[13]$^d$	&		&		&		&		&		&		&		\\

	&		&		&		&		&		&		&		&		\\
\hline
\multicolumn{9}{c}{Cs}\\
\hline %inserts single line
30	&	0.510[10]	&	0.740[11]	&	0.968[11]	&	$-0.526[11]$	&	$-0.639[11]$	&	$-0.865[10]$	&	0.357[11]	&	0.714[11]	\\
	&	0.510[10]$^a$	&	0.740[11]$^a$	&	0.968[11]$^a$	&	$-0.526[11]^a$	&	$-0.639[11]^a$	&	$-0.865[10]^a$	&	0.357[11]$^a$	&	0.714[11]$^a$	\\
	&		&		&		&		&		&		&		&		\\

80	&	0.554[13]	&	0.119[15]	&	0.162[15]	&	$-0.821[14]$	&	$-0.891[14]$	&	$-0.134[14]$	&	0.547[14]	&	0.976[14]	\\
\hline
\end{tabular}
\label{table_1} % is used to refer this table in the text
\end{table}

 \end{widetext}

To determine the polarizability,  we need to determine the eigen states, eigen energies, and dipole matrix elements for the various Rydberg states. 
Following our earlier work~\cite{Bhowmik2024}, we combine quantum defect theory and the Coulomb approximation method~\cite{Seaton1983,Wijngaarden1994}. Since
Ref.~\cite{Bhowmik2024} provided a detailed account of the approach for cesium, including various tables that show excellent agreement with the literature, the discussion here is kept comparatively brief.

According to  quantum defect theory, the principal quantum  number $n$  and, correspondingly, the energy levels $\epsilon_{n,L,J}$ of single-valence Rydberg states, can be parameterized analogously as those  of  the hydrogen atom. Specifically, the energy $\epsilon_{n,L,J}$ of a Rydberg state with principal quantum number $n$, orbital angular momentum quantum number $L$, and total electronic quantum number $J$  is \begin{equation}\label{eq_A1}
\epsilon_{n,L,J}=-\dfrac{m_r\alpha^2 c^2}{2(n_{\text{eff}})^2},
\end{equation}
where $m_r$ denotes the reduced mass
of the atom, $\alpha$ the finestructure constant,  $c$ the speed of light in vacuum, and $n_{\text{eff}}$  the effective principal quantum number. 
The effective principal quantum number 
$n_{\text{eff}}$ is deduced from the   quantum defect parameters $\delta_{L,J,2i}$ by using 
\begin{equation}\label{eq_A2}
n_{\text{eff}}=n-\sum_{i=0}^\infty \dfrac{\delta_{L,J,2i}}{(n-\delta_{L,J,0})^{2i}},
\end{equation}
where $i$ is a dummy index. The experimental quantum defect parameters of Rb for the $nS_{1/2}$, $nP_{1/2}$, $nP_{3/2}$, $nD_{3/2}$, and $nD_{5/2}$ states  are extracted from Refs.~\cite{Mack2011, Li2003} and for $nF_{5/2}$ and $nF_{7/2}$  from Ref.~\cite{Han2006}. For Cs, the experimental  quantum defect parameters for the $nS_{1/2}$, $nP_{1/2}$, $nP_{3/2}$,  and $nD_{5/2}$ states are taken from  Ref.~\cite{Deiglmayr2016}, for the $nD_{3/2}$ state from Ref~\cite{Lorenzen1983}, and  for the $nF_{5/2}$ and $nF_{7/2}$ states from Ref.~\cite{Bai2023}.

Knowing $n_{\text{eff}}$,  the wavefunction of a Rydberg state can be  approximated using  modified Coulomb solutions~\cite{Bates1949}. This approach has been shown to produce accurate polarizabilities for both Rb and Cs~\cite{Wijngaarden1997, Yerokhin2016}.  For Rydberg states, the  solution to the radial part of the Schr\"odinger equation that is regular in the $r \rightarrow \infty$ limit  ($r$ denotes the distance of the valence electron from the core) and accounts for the orbital angular momentum barrier and Coulomb potential reads~\cite{Bates1949} 
\begin{eqnarray}\label{eq_A3}
R_{n_{\text{eff}},L,J}(r)=
&&
\dfrac{1}{[a_0 n_{\text{eff}}^2\Gamma(n_{\text{eff}}+L+1)\Gamma(n_{\text{eff}}-L)]^{1/2}}
\times \nonumber \\
&&
W_{n_{\text{eff}}, L+1/2}(y),
\end{eqnarray}
where  $W_{n_{\text{eff}},L+1/2}(y)$
with $y= 2r/(n_{\text{eff}} a_0)$ denotes the Whittaker function and $a_0$ the Bohr radius.  The radial wavefunction $R_{n_{\text{eff}},L,J}(r)$ reduces to the well known non-relativistic bound state wavefunction for integer quantum numbers, i.e., for $n_{\text{eff}} = n$~\cite{Yerokhin2016}. Using the radial solutions, the  radial transition dipole matrix element $R_{ki}$  is given by
\begin{eqnarray}\label{eq_A7}
R_{ki}=\int_{r_0}^\infty dr r R_{n_{\text{eff},k},L_k,J_k }(r) R_{n_{\text{eff},i},L_i,J_i}(r),
\end{eqnarray}
where the lower  integration limit $r_0$ needs to be chosen sufficiently small to ensure convergence of  the  matrix element.
 We find that  $
     r_0=\dfrac{sn_{\text{eff},k} n_{\text{eff},i} a_0}{n_{\text{eff},k}+n_{\text{eff},i}} $
     with $s$ around $1/100$ yields converged results. 
      Note that, in terms of the radial wave function $R_{n_{\text{eff}},L,J}(r)$, the wave function $\Psi^{(0)}_{n_{\text{eff}},L,J, M_J}(\vec{r})$ reads $\Psi^{(0)}_{n_{\text{eff}},L,J,M_J}(\vec{r})=\frac{1}{r}R_{n_{\text{eff}},L,J}(r) Y_{J,M_J}(\hat{\vec{r}})$ where $Y_{J,M_J}(\hat{\vec{r}})$ is spherical harmonic of degree $J$ and order $M_J$.

The total polarizability of an atomic state is composed of the core polarizability, valence-core polarizability, and valence polarizability~\cite{Flambaum2008}. The core and valence-core polarizabilities are essentially frequency independent~\cite{Bhowmik2024} and are calculated using second-order relativistic many-body perturbation theory.  The values of the core polarizability for Rb and Cs are 9.08 a.u. and 16.12 a.u., respectively.  Since the valence-core polarizability for all Rydberg states considered in the present work is essentially zero, it does not have a contribution to the total polarizability. 

To calculate  each part of the valence polarizability, i.e., the scalar, vector, and tensor polarizabilities \cite{Flambaum2008, Bhowmik2020, Zhang2024}, the sum-over-states method is used~\cite{Mitroy2010}. In the  sum-over-states method,  we include contributions from  the virtual excited bound states until convergence is reached. Typically, to obtain converged results for the polarizability of a Rydberg state with  principal quantum number $n$, we  include all states that are energetically lower-lying than the Rydberg state of interest and around 50 to 60 states that are energetically higher-lying than the Rydberg state of interest. In all cases, convergence is checked explicitly by increasing the number of states included in the sums.

 To benchmark our results, Table~\ref{table_1} compares the static (i.e., $\omega = 0$) scalar $\alpha^S(0)$ and tensor $\alpha^T(0)$ polarizabilities for rubidium and cesium Rydberg states for selected $n$ values with available theoretical and experimental data from the literature. Our results show good agreement with those of Ref.~\cite{Yerokhin2016}, which employed the same Coulomb approximation method as used in our work. Another recent study~\cite{Lai2018} calculated the polarizabilities for various rubidium Rydberg states using a model potential and the B-spline expansion technique. The difference between our results and those of Ref.~\cite{Lai2018} is less than 5~\% across all states considered. Most of our values fall within the experimental uncertainties reported in Refs.~\cite{OSullivan1985, OSullivan1986}, with the exception of the scalar polarizabilities for the rubidium 
$30D_{5/2}$, $40D_{5/2}$, 
and $60S_{1/2}$ Rydberg  states, and the tensor polarizability for the rubidium $50D_{5/2}$ Rydberg state,  
which deviate by $2\sigma$ to $4\sigma$. The overall agreement between our calculations and both theoretical and experimental results demonstrates the reliability of our approach. In addition, we reiterate that our previous work~\cite{Bhowmik2024} carefully benchmarked the cesium  polarizabilities for $n = 40$–$70$.

\section{Results for cesium}
\label{sec_appendix_b}
\begin{figure}[t]
\vspace{-1.8cm}
{\includegraphics[ trim=0cm 0cm 0cm 0cm, clip=true, totalheight=0.30\textheight, angle=0]{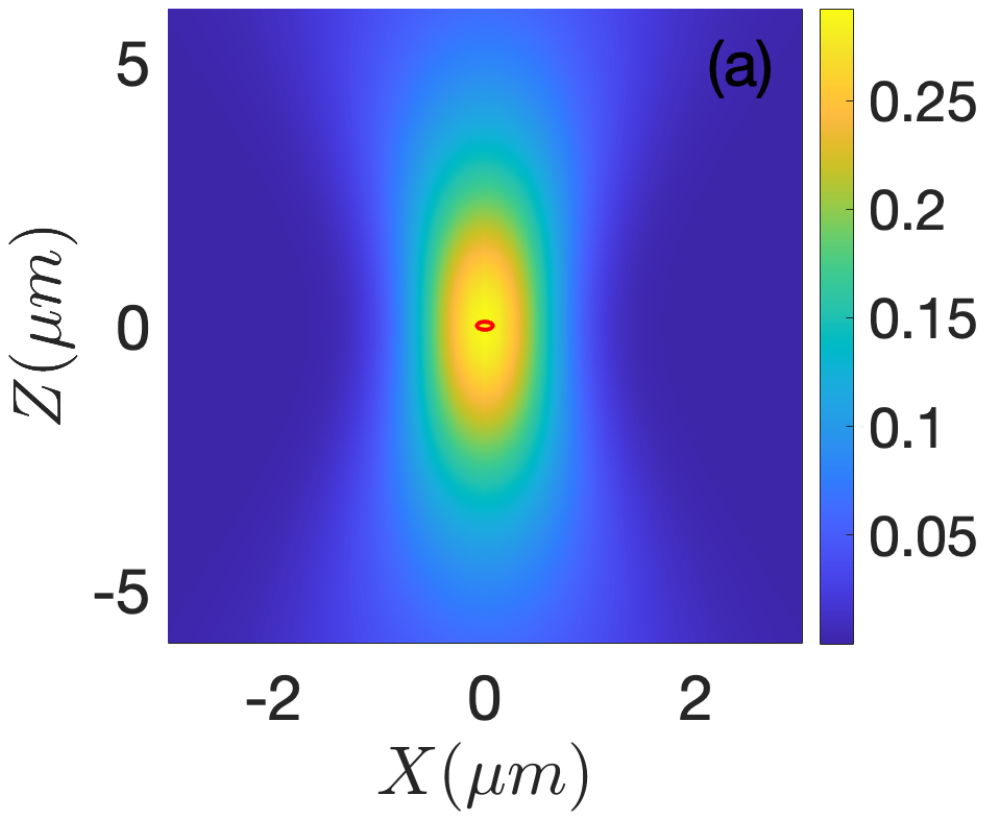}}
\hspace{-0.8cm}
{\includegraphics[trim=5.0cm 0cm 0cm 0cm, clip=true, totalheight=0.30\textheight, angle=0]{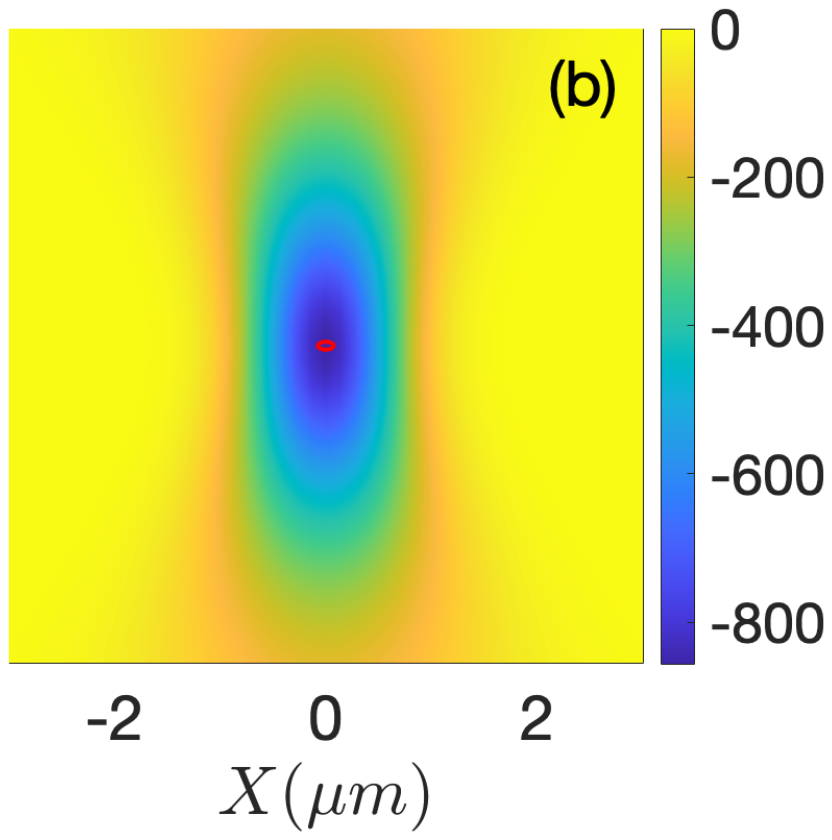}}\\
\vspace{-4.2cm}
{\includegraphics[ trim=0cm 0cm 0cm 0cm, clip=true, totalheight=0.30\textheight, angle=0]{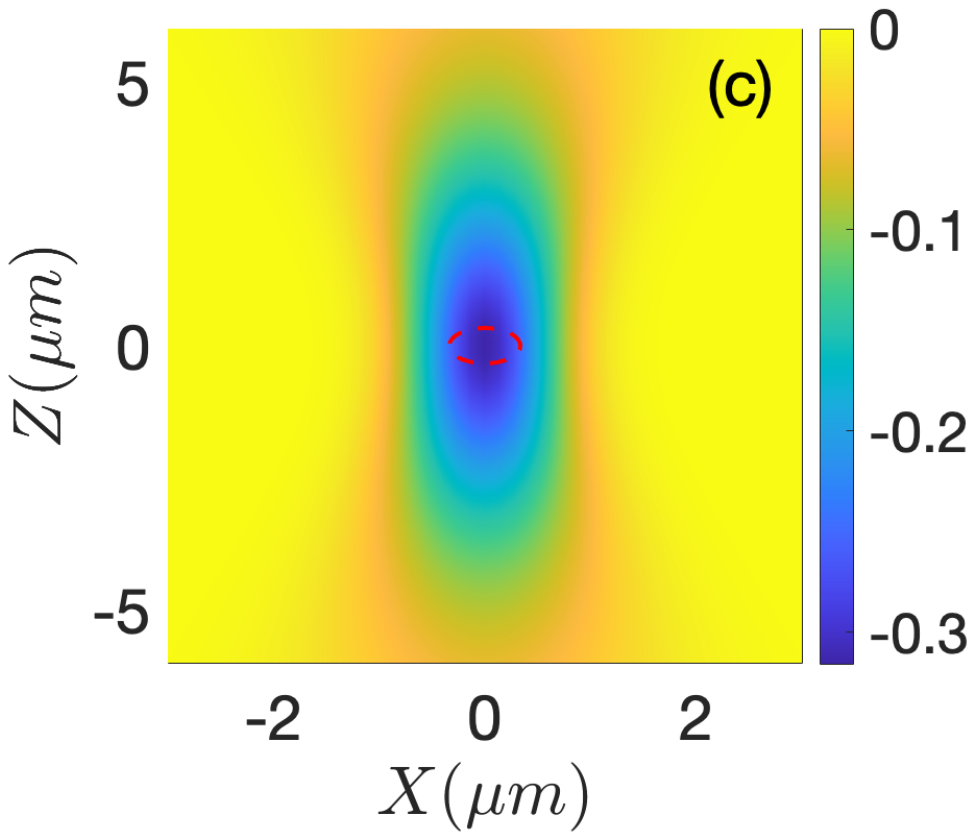}}
\hspace{-0.8cm}
{\includegraphics[trim=5.0cm 0cm 0cm 0cm, clip=true, totalheight=0.30\textheight, angle=0]{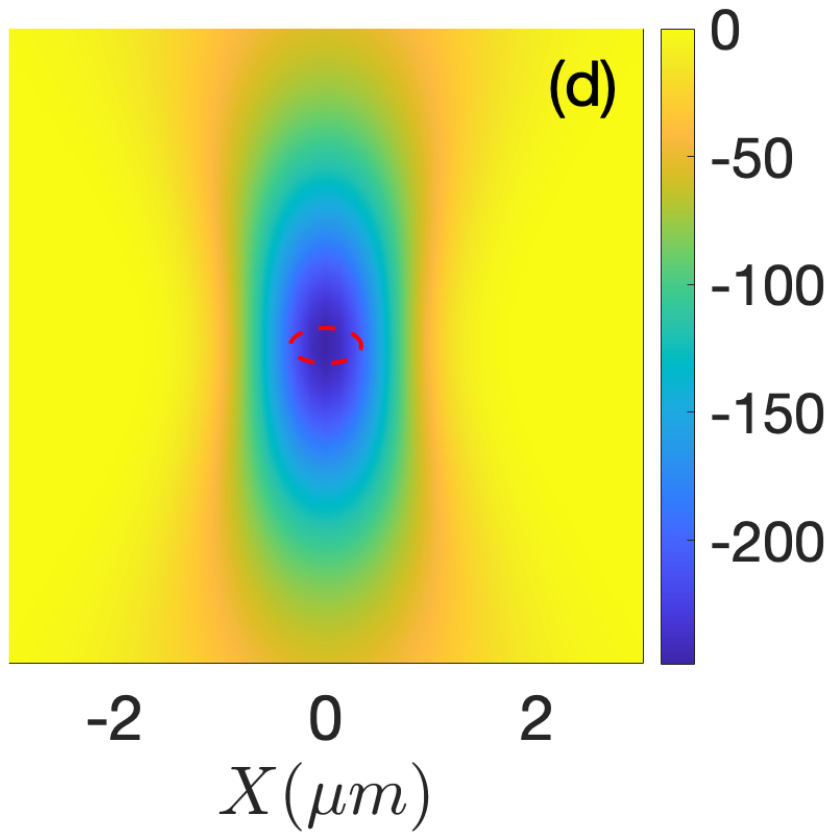}}\\
\vspace{-4.2cm}
{\includegraphics[ trim=0cm 0cm 0cm 0cm, clip=true, totalheight=0.30\textheight, angle=0]{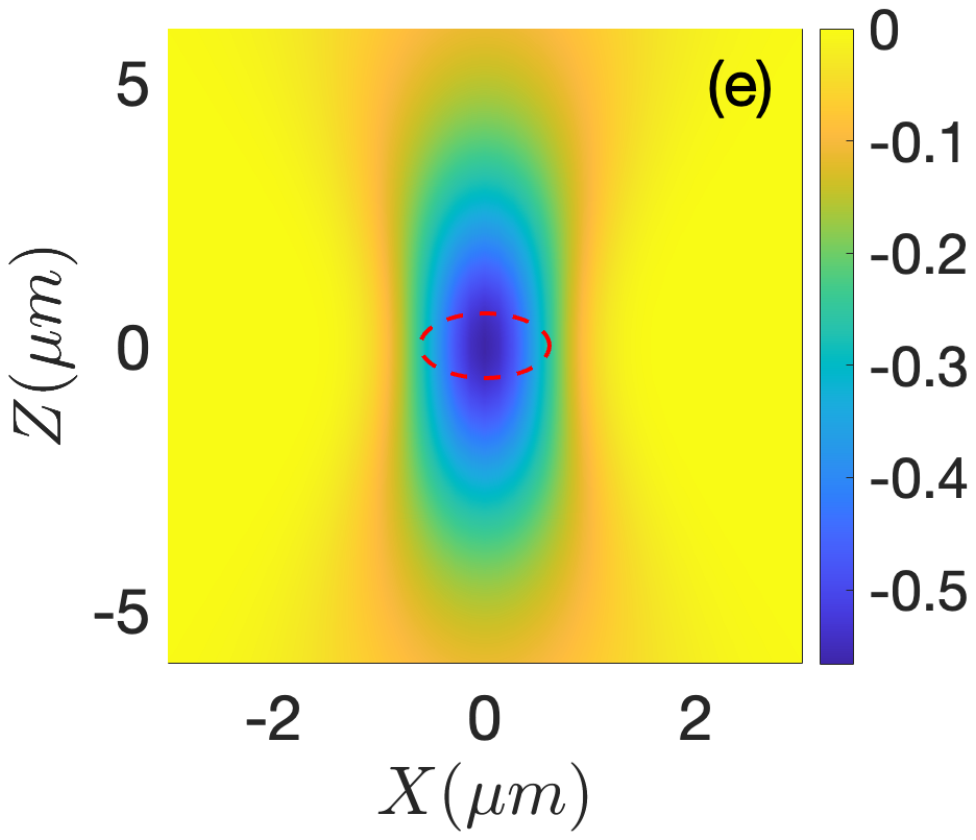}}
\hspace{-0.8cm}
{\includegraphics[trim=5.0cm 0cm 0cm 0cm, clip=true, totalheight=0.30\textheight, angle=0]{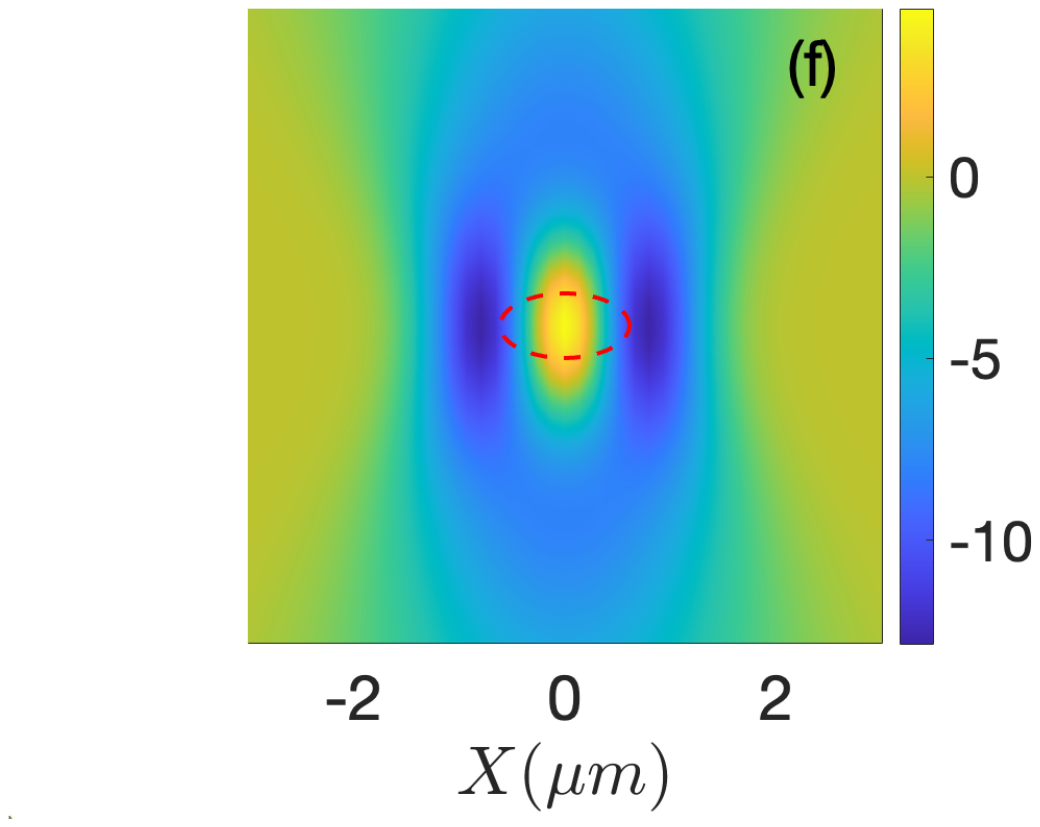}}\\
\vspace{-1.5cm}
\caption{ Trapping of $nS_{1/2,+1/2}$ series of the cesium Rydberg atom 
for $\lambda=1,000$~nm, $w_0=1$~$\mu$m,  $P=2.5$~mW, and $\theta_p=\pi/2$. 
Left column: Rydberg trapping potential $U_{\text{ryd}}(X,0,Z,\omega)$ as functions of $X$ and $Z$ for   (a) $n=30$ and $A \cos \theta_k=1$, (c) $n=60$ and $A \cos \theta_k=1$, and (e) $n=80$ and $A \cos \theta_k=0.5352$. The color bars indicate the values of $U_{\text{ryd}}(X,0,Z,\omega)$ in mK. 
 Right column: Differential trapping potential $\Delta U(X,0,Z,\omega)$  as functions of $X$ and $Z$. The $n$ and $A \cos \theta_k$ values in (b), (d), and (f) are the same as those used in (a), (c), and (e), respectively. The color bars indicate the values of $\Delta U(X,0,Z,\omega)$ in $\mu$K (it is important to note that the color bars in the left and right columns use different scales, namely mK and $\mu$K, respectively).
 The red dashed lines demarcate the $(X=x_*,Z=z_*)$ values for which 95~\% of the probability of the Rydberg electron, for $Y=0$, lies inside the red dashed line. }
\label{FigS1_extra}
\end{figure}

\begin{figure}[t]
\vspace{-1.8cm}
{\includegraphics[ trim=0cm 0cm 0cm 0cm, clip=true, totalheight=0.30\textheight, angle=0]{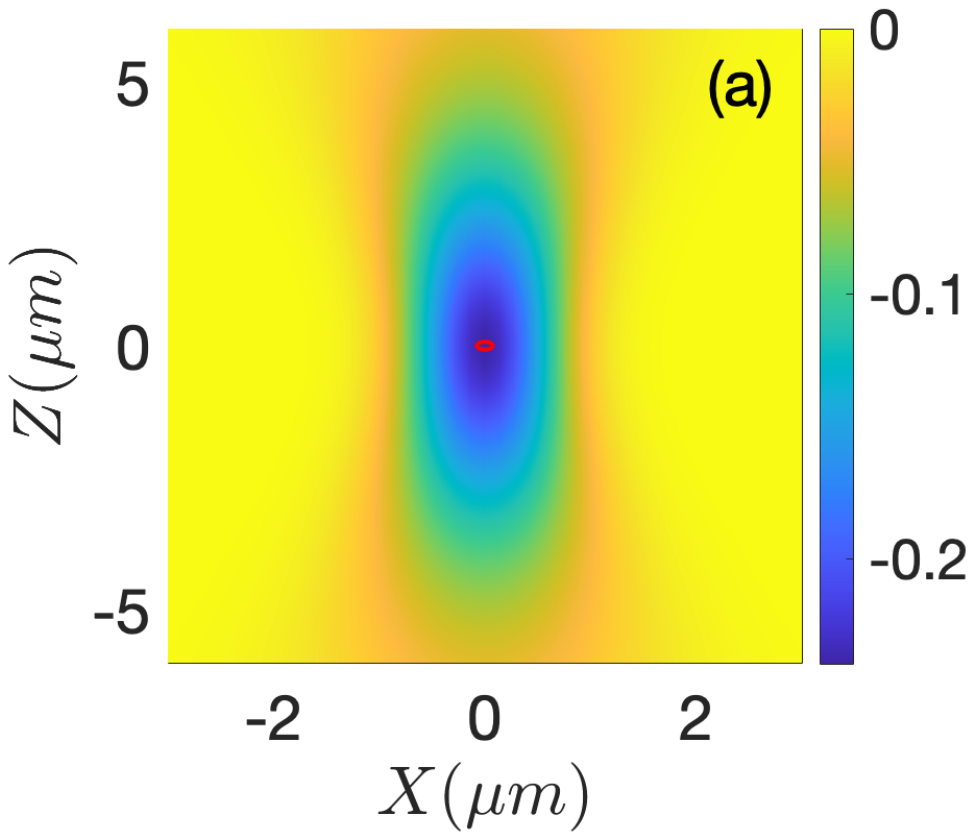}}
\hspace{-0.8cm}
{\includegraphics[trim=5.0cm 0cm 0cm 0cm, clip=true, totalheight=0.30\textheight, angle=0]{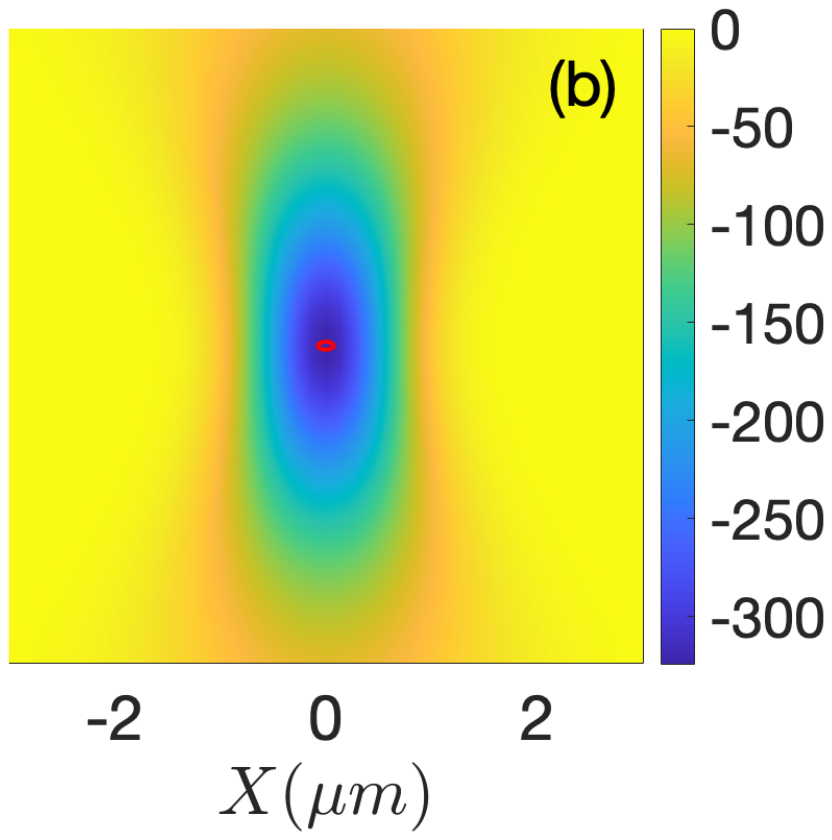}}\\
\vspace{-4.2cm}
{\includegraphics[ trim=0cm 0cm 0cm 0cm, clip=true, totalheight=0.30\textheight, angle=0]{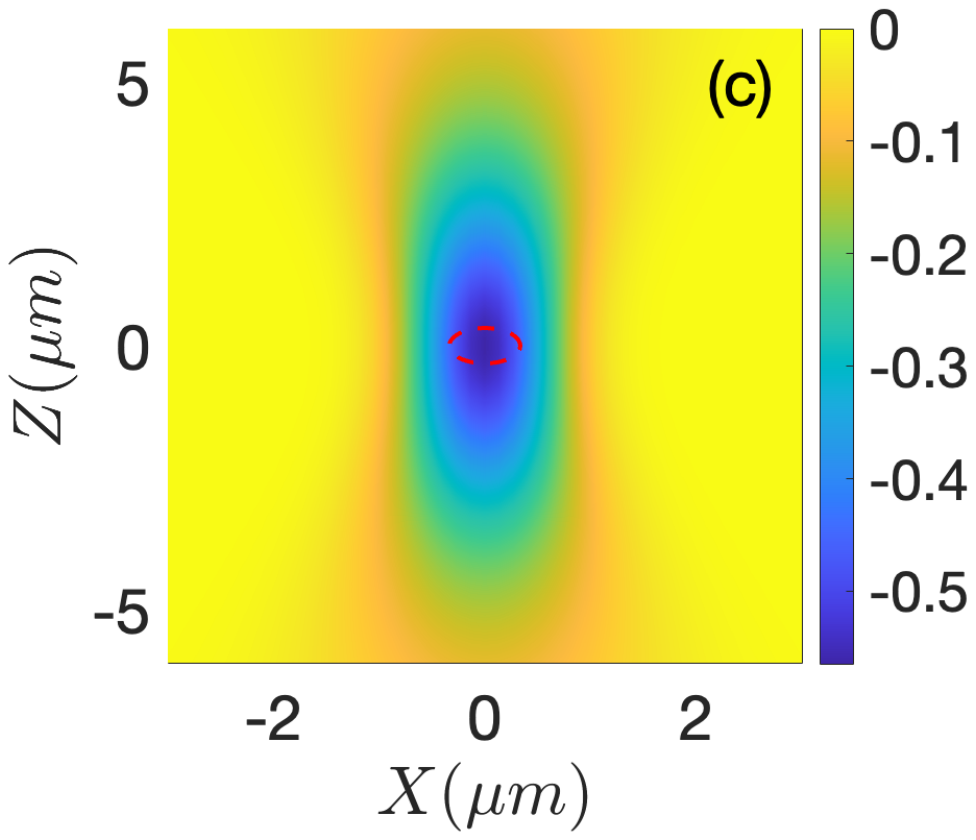}}
\hspace{-0.8cm}
{\includegraphics[trim=5.0cm 0cm 0cm 0cm, clip=true, totalheight=0.30\textheight, angle=0]{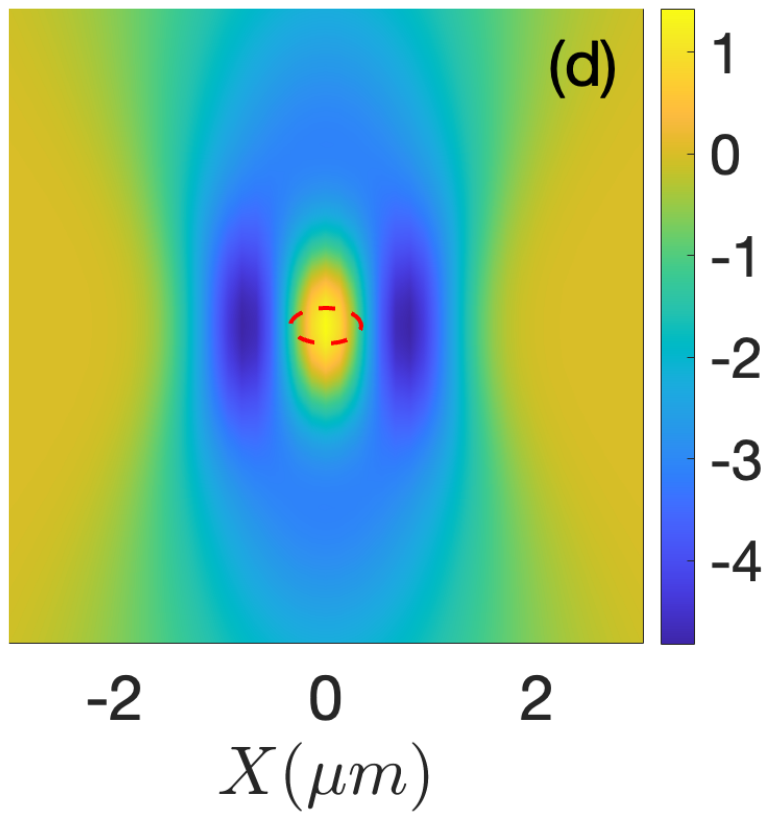}}\\
\vspace{-4.2cm}
{\includegraphics[ trim=0cm 0cm 0cm 0cm, clip=true, totalheight=0.30\textheight, angle=0]{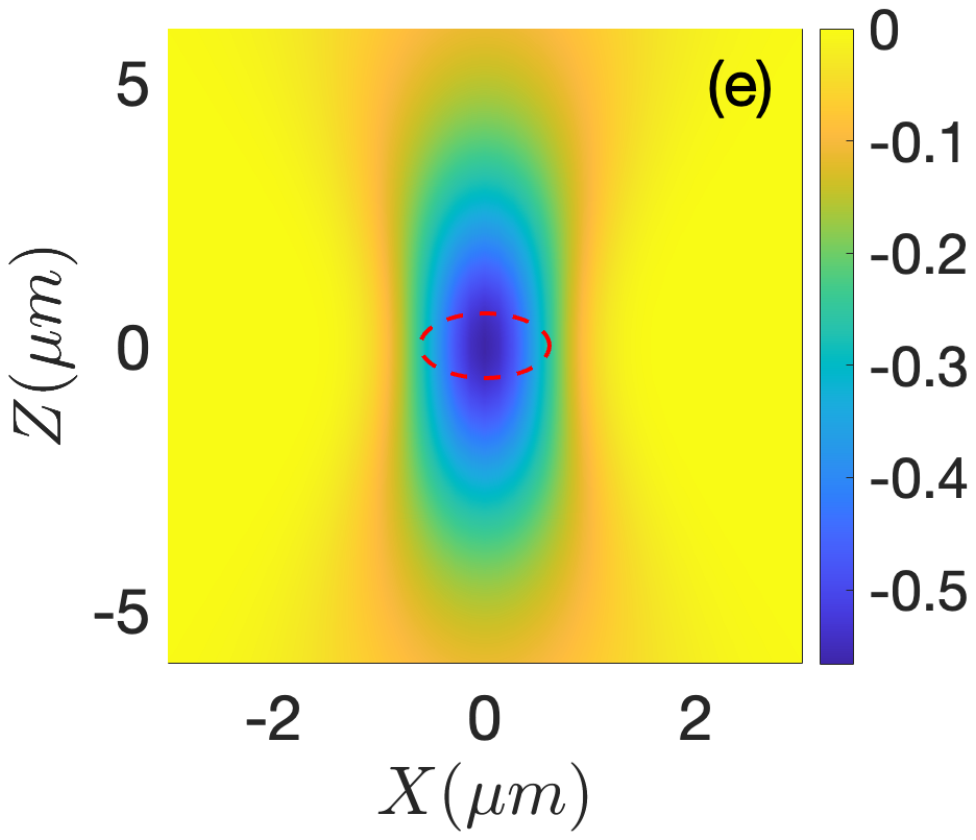}}
\hspace{-0.8cm}
{\includegraphics[trim=5.0cm 0cm 0cm 0cm, clip=true, totalheight=0.30\textheight, angle=0]{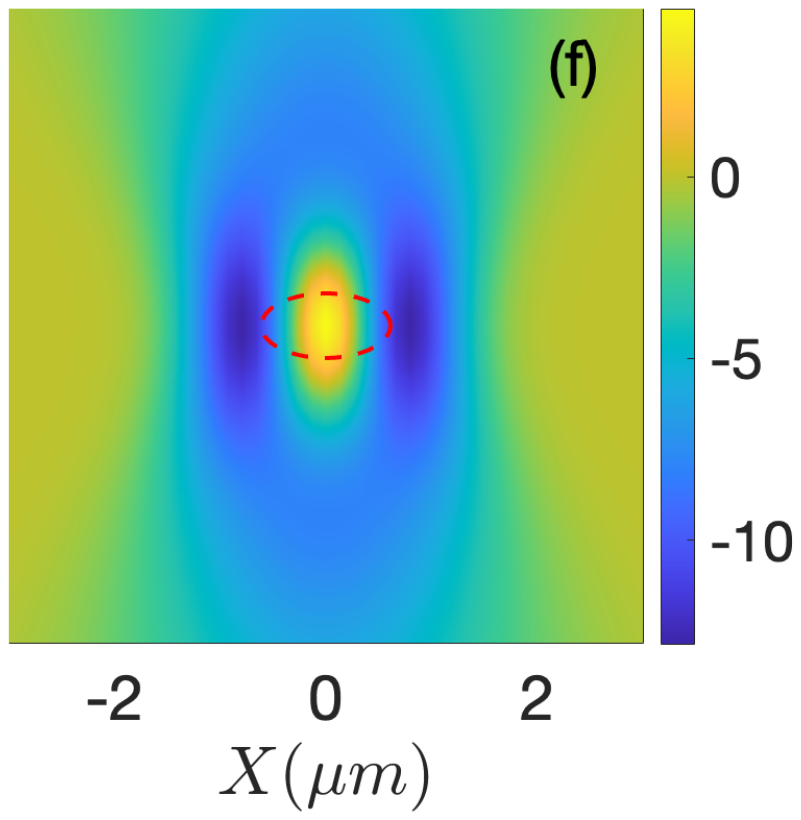}}\\
\vspace{-1.5cm}
\caption{ Trapping of $nD_{3/2,-3/2}$ series of the cesium  Rydberg atom 
for $\lambda=1,000$~nm, $w_0=1$~$\mu$m,  $P=2.5$~mW, and $\theta_p=\pi/2$. 
Left column: Rydberg trapping potential $U_{\text{ryd}}(X,0,Z,\omega)$ as functions of $X$ and $Z$ for   (a) $n=30$ and $A \cos \theta_k=1$, (c) $n=60$ and $A \cos \theta_k=0.1710$, and (e) $n=80$ and $A \cos \theta_k=0.0919$. The color bars indicate the values of $U_{\text{ryd}}(X,0,Z,\omega)$ in mK. 
 Right column: Differential trapping potential $\Delta U(X,0,Z,\omega)$  as functions of $X$ and $Z$. The $n$ and $A \cos \theta_k$ values in (b), (d), and (f) are the same as those used in (a), (c), and (e), respectively. The color bars indicate the values of $\Delta U(X,0,Z,\omega)$ in $\mu$K (it is important to note that the color bars in the left and right columns use different scales, namely mK and $\mu$K, respectively).
 The red dashed lines demarcate the $(X=x_*,Z=z_*)$ values for which 95~\% of the probability of the Rydberg electron, for $Y=0$, lies inside the red dashed line. }
\label{FigS2_extra}
\end{figure}

Figures~\ref{FigS1_extra} and \ref{FigS2_extra} are analogs of Figs.~\ref{Fig1_extra} and \ref{Fig2_extra} of the main text: the main text shows results for rubidium while this appendix presents results for cesium. As  in the main text, we use $P = 2.5~\text{mW}$,  $w_0 = 1~\mu\text{m}$, and $\lambda = 1{,}000~\text{nm}$.  For $\lambda=1,000$nm, the polarizability of the ground state of Cs is $1075$ a.u..  For these parameters, the ground state of cesium experiences a negative Stark shift, allowing for trapping by red-detuned light. The left columns of Figs.~\ref{FigS1_extra} and \ref{FigS2_extra} display the total trapping potential $U_{\text{ryd}}(X,0,Z,\omega)$ for selected cesium Rydberg states in the $nS_{1/2,+1/2}$ and $nD_{3/2,-3/2}$ series, respectively.   In Fig.~\ref{FigS1_extra}, the geometric factor $A\cos\theta_k$ is taken to be $1$ for the $30S_{1/2,+1/2}$ and $60S_{1/2,+1/2}$ states, and $0.5352$ for the $80S_{1/2,+1/2}$ state. The potential $U_{\text{ryd}}(X,0,Z,\omega)$ is positive (indicative of anti-trapping) for $n = 30$ but becomes negative (indicative of trapping) for $n = 60$ and $n = 80$.  The trapping potentials shown in Figs.~\ref{FigS1_extra}(c) and \ref{FigS1_extra}(e) are expected to be sufficient to trap the corresponding Rydberg states. In Fig.~\ref{FigS2_extra}, the geometric factor $A\cos\theta_k$ is set to $1$ for the $30D_{3/2,-3/2}$ state, to $0.1710$ for the $60D_{3/2,-3/2}$ state, and to $0.0919$ for the $80D_{3/2,-3/2}$ state. For these $D$-states, $U_{\text{ryd}}(X,0,Z,\omega)$ is negative for all three values of $n$, allowing for trapping of the corresponding Rydberg states.

Similar to what is observed for rubidium, the cesium Rydberg state $30S_{1/2,+1/2}$ shown in Fig.~\ref{FigS1_extra}(a) cannot be trapped using red-detuned light. In contrast, the cesium states $60S_{1/2,+1/2}$  [Fig.~\ref{FigS1_extra}(c)] and $30D_{3/2,-3/2}$ [Fig.~\ref{FigS2_extra}(a)]   can be trapped  using red-detuned light.  For the Rydberg states $80S_{1/2,+1/2}$, $60D_{3/2,-3/2}$, and $80D_{3/2,-3/2}$, the differential potential $\Delta U(X,0,Z,\omega)$ is on the order of a few $\mu$K within the region occupied by the Rydberg atom (as demarcated by the red dashed line in the figures). The, in magnitude, small differential potential indicates that, for the chosen parameters, the trapping potentials experienced by the ground and Rydberg states are nearly identical across the entire extent of the Rydberg electron wavefunction. As discussed previously,   we tune $A \cos \theta_k$  to minimize the differential potential, thereby optimizing the trapping conditions. A change of $A \cos \theta_k$ by approximately
$0.006$, $0.01$,  and $0.002$   for the $80S_{1/2,+1/2}$, $60D_{3/2,-3/2}$, and $80D_{3/2,-3/2}$ states, respectively,
results in a change by a factor of 2 in the metric defined in Eq.~(\ref{eq_metric}).

\begin{figure}[t]
{\includegraphics[ scale=.35]{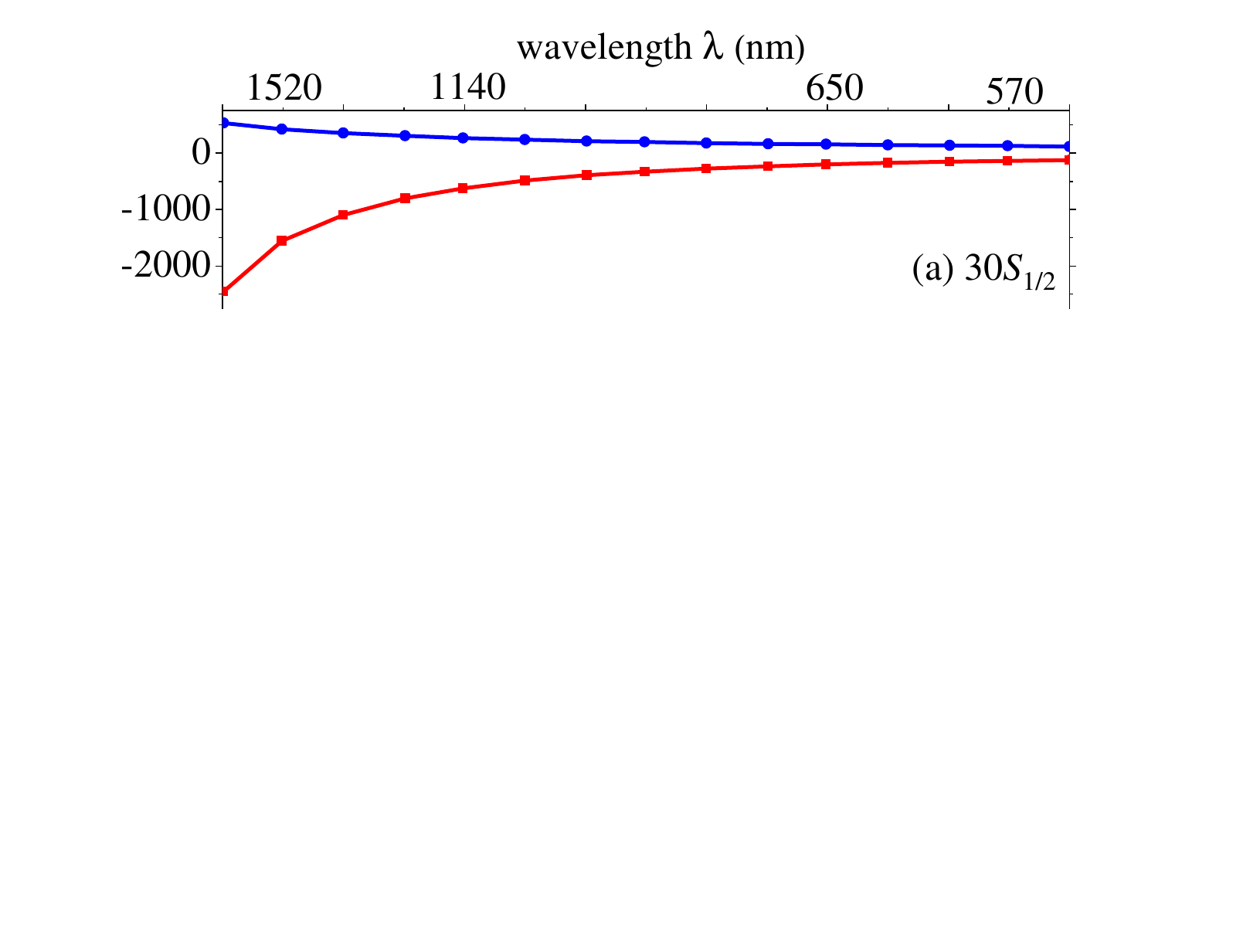}}\\
\vspace{-5.76cm}
{\includegraphics[ scale=.35]{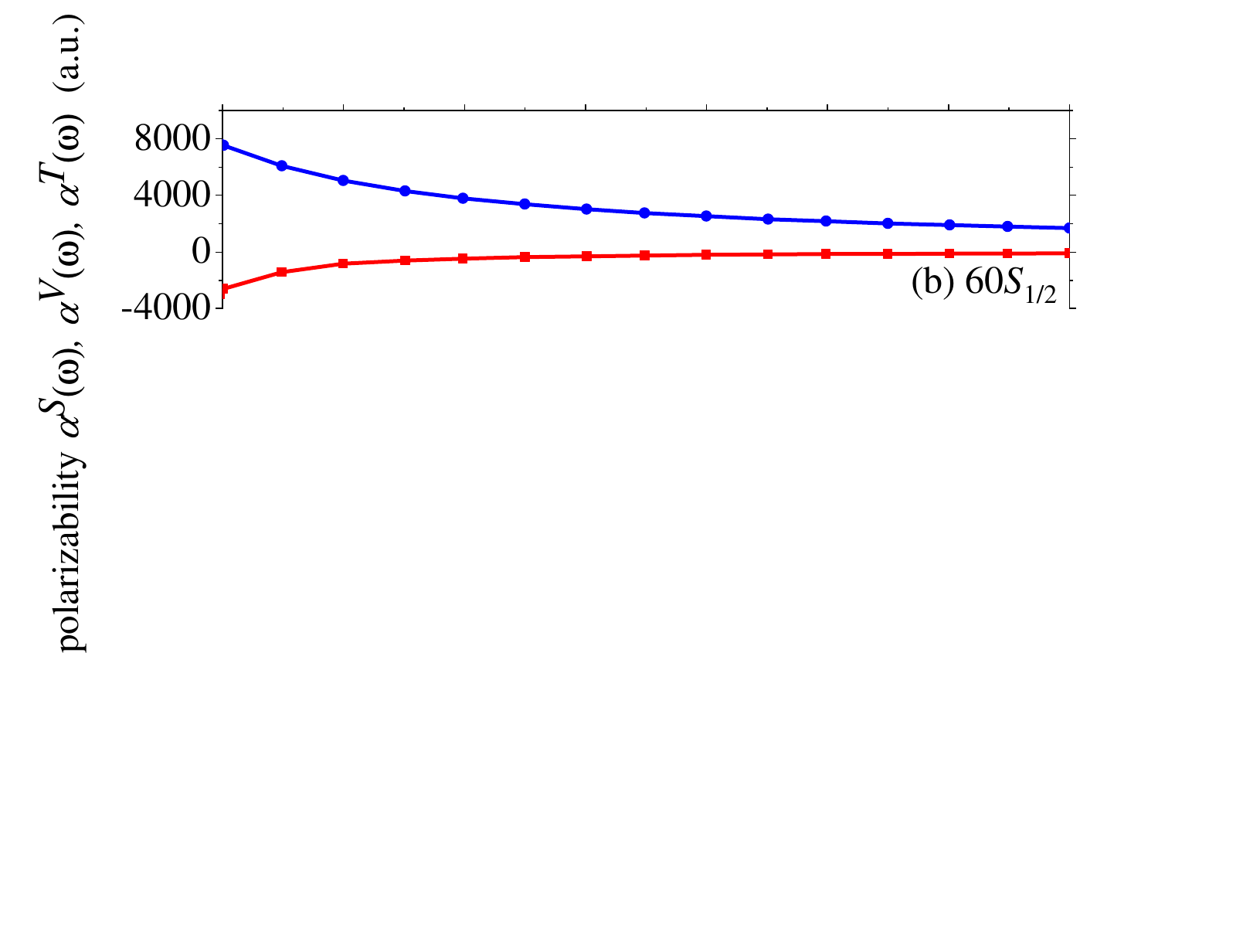}}\\
\vspace{-5.76cm}
{\includegraphics[ scale=.35]{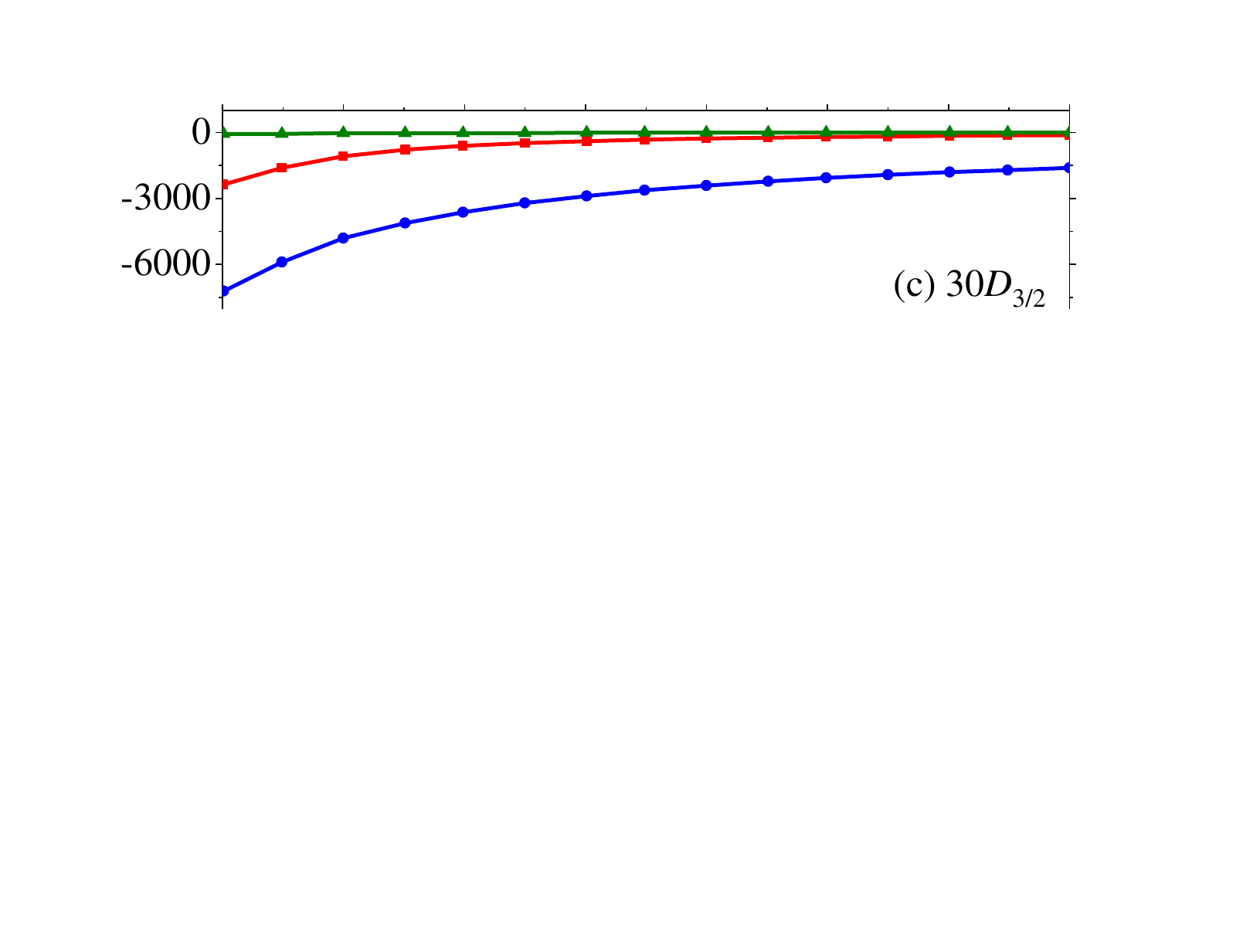}}\\
\vspace{-5.76cm}
{\includegraphics[ scale=.35]{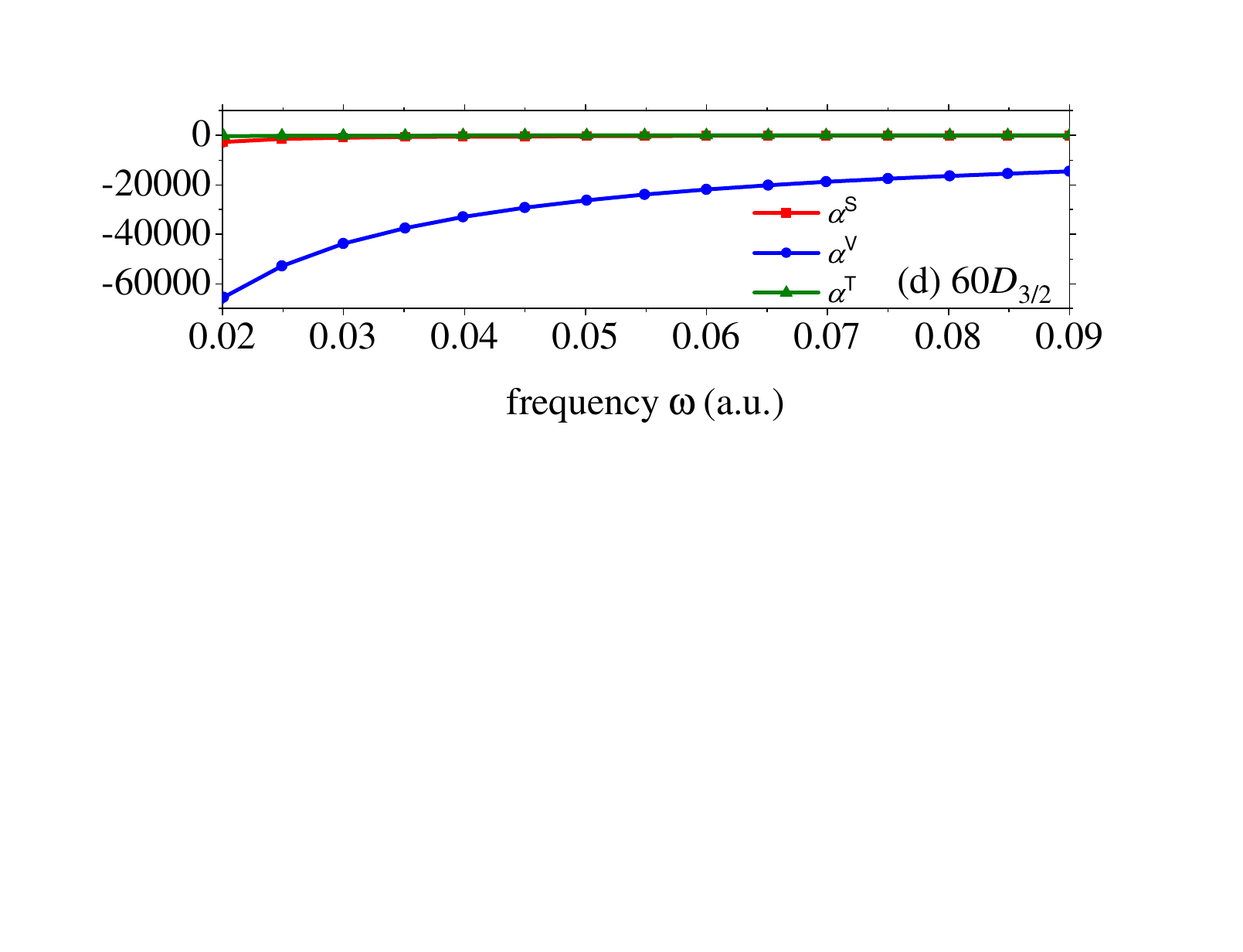}}\\
\vspace{-4.28cm}
\caption{Squares, circles, and triangles show the numerically obtained scalar, vector, and tensor polarizabilities  $\alpha^S(\omega)$, $\alpha^V(\omega)$,  and $\alpha^T(\omega)$, respectively, for cesium as a function of the frequency $\omega$ [the  corresponding wavelengths $\lambda$ range from 500~nm to 2,200~nm, see the  top axis]. The one-parameter fits (lines) agree excellently with the symbols. Panels (a), (b), (c), and (d) are for the $30S_{1/2}$, $60S_{1/2}$, $30D_{3/2}$, and $60D_{3/2}$ states, respectively.}
\label{FigS1}
\end{figure}

\begin{figure}[t]
{\includegraphics[ scale=.35]{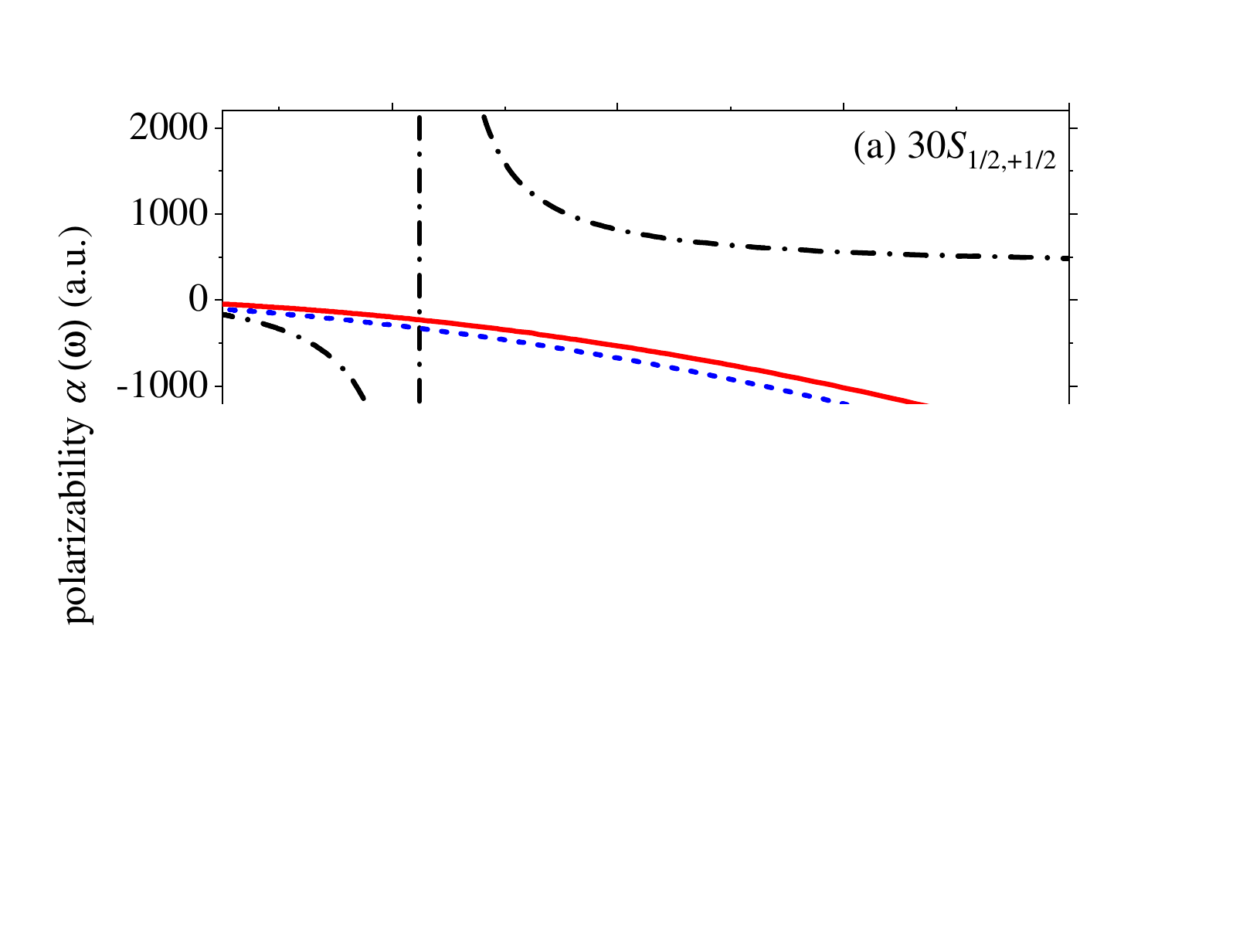}}\\
\vspace{-5.055cm}
{\includegraphics[ scale=.35]{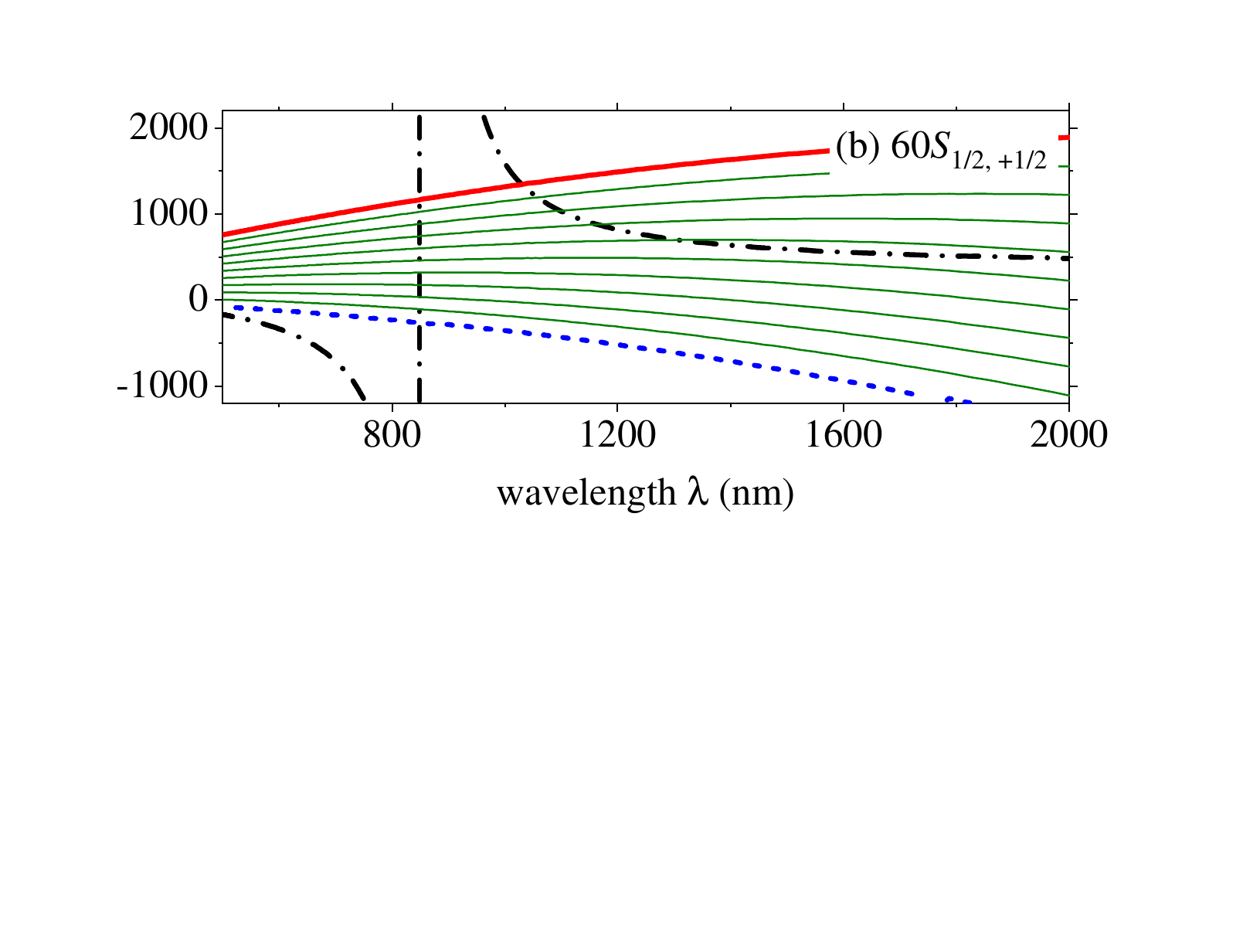}}\\
\vspace{-3.28cm}
\caption{Polarizability of the ground and Rydberg states for Cs  as a function of wavelength. The Rydberg states are (a) $30S_{1/2,+1/2}$ and (b)  $60S_{1/2,+1/2}$.  The quantization axis is perpendicular to the  polarization vector, i.e.,  $\theta_p=\pi/2$.   The black dash-dotted lines show the polarizability of the ground state.  The blue dashed  
and 
red solid lines 
show $\alpha(\omega)$ of the Rydberg state for  linearly and circularly polarized light, respectively.  In (b), the thin green lines show the polarizability of the Rydberg states for elliptically polarized light for various values of $A\cos\theta_k$ 
[$A\cos\theta_k=0.1$ (bottom-most curve) to $0.9$ (top-most curve)].  For the  $60S_{1/2,+1/2}$ state,  variation of the geometric factor $A \cos \theta_k$ affords  appreciable tunability of $U_{\text{stark}}(X, Y, Z, \omega)$.  }
\label{FigS2}
\end{figure}

\begin{figure}[t]
{\includegraphics[ scale=.35]{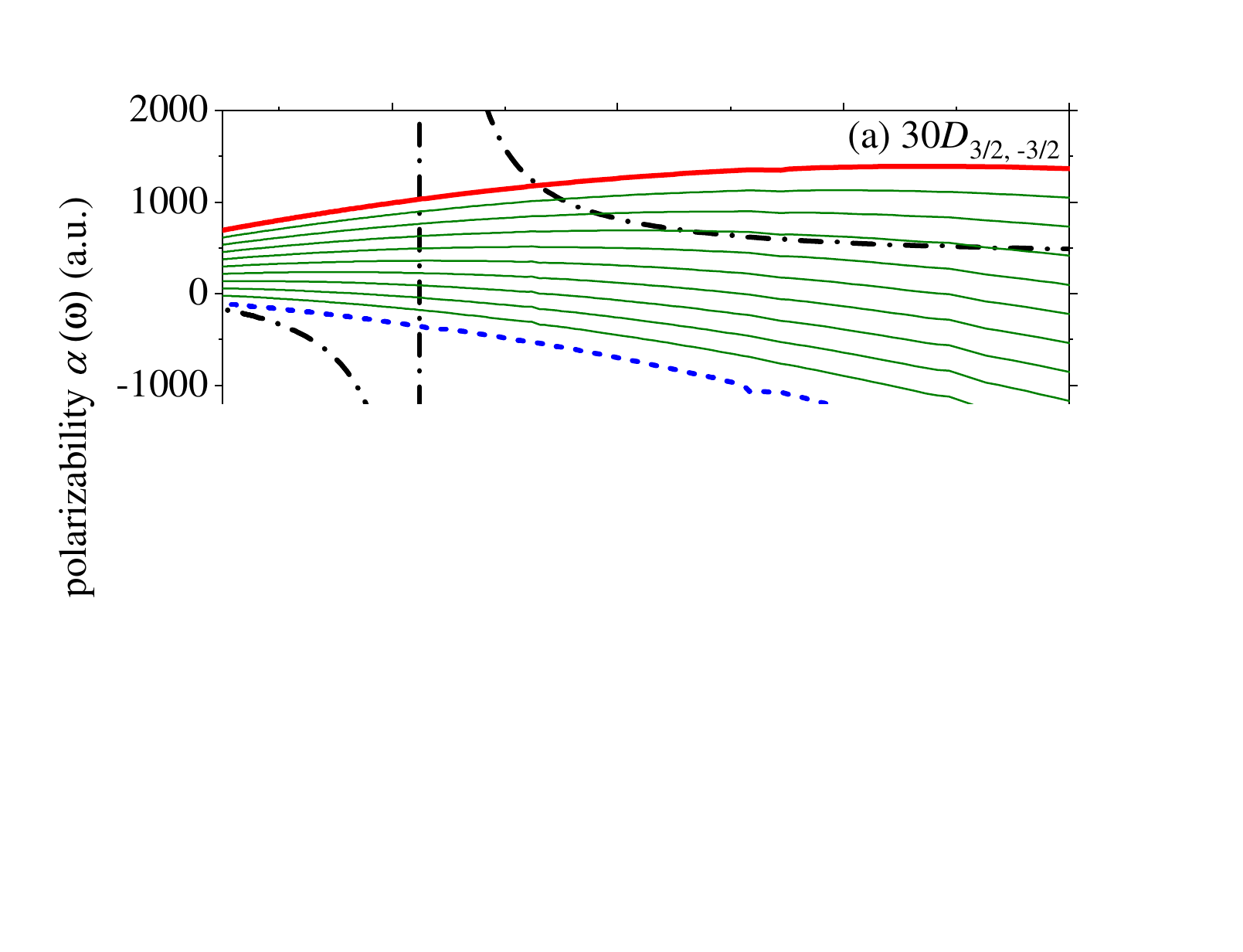}}\\
\vspace{-5.055cm}
{\includegraphics[ scale=.35]{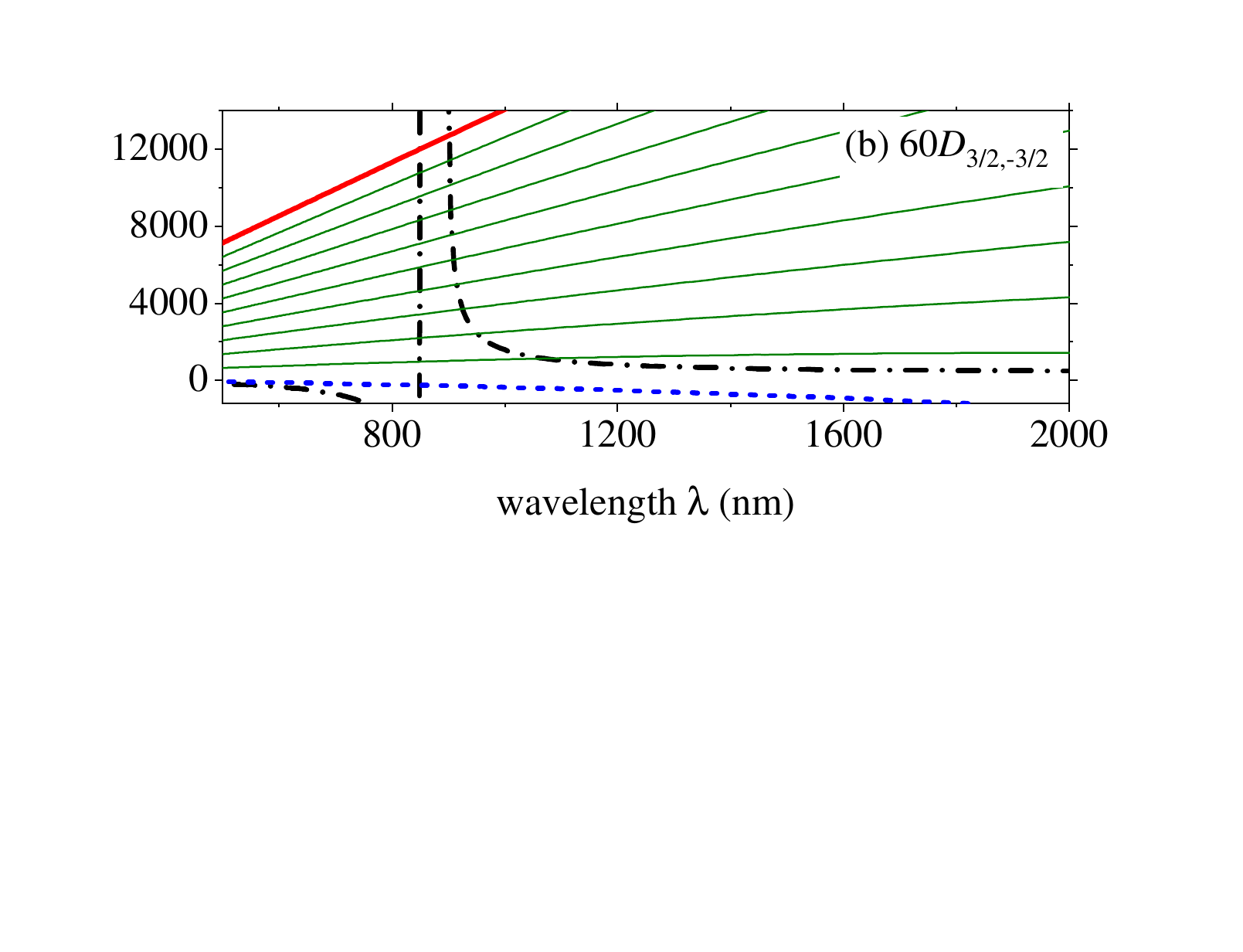}}\\
\vspace{-3.28cm}
\caption{Polarizability of the ground and Rydberg states for Cs  as a function of wavelength. The Rydberg states are (a) $30D_{3/2,-3/2}$ and (b)  $60D_{3/2,-3/2}$.  The quantization axis is perpendicular to the  polarization vector, i.e.,  $\theta_p=\pi/2$.   The black dash-dotted lines show the polarizability of the ground state.  The blue dashed  
and 
red solid lines 
show $\alpha(\omega)$ of the Rydberg state for  linearly and circularly polarized light, respectively.  The thin green lines show the polarizability of the Rydberg states for elliptically polarized light for various values of $A\cos\theta_k$ 
[$A\cos\theta_k=0.1$ (bottom-most curve) to $0.9$ (top-most curve)].   For both $D$-states,  variation of the geometric factor $A \cos \theta_k$ affords  appreciable tunability of $U_{\text{stark}}(X, Y, Z, \omega)$.}
\label{FigS3}
\end{figure}

Figures~\ref{FigS1}, \ref{FigS2}, and \ref{FigS3} are analogs of Figs.~\ref{Fig1}, \ref{Fig3}, and \ref{Fig4} of the main text: 
 the main text shows results for rubidium while this appendix presents results for cesium. The key take-away message is that the overall behavior of the polarizabilities of rubidium and cesium Rydberg states is similar for non-linearly polarized light.

Figure~\ref{FigS1} presents the scalar, vector, and tensor polarizabilities for the $nS_{1/2}$ and $nD_{3/2}$ Rydberg states ($n=30$  and $n=60$) of cesium as a function of $\omega$.  It can be seen that  for the $30S_{1/2}$ state, $|\alpha^V(\omega)|$
   is smaller than $|\alpha^S(\omega)|$. 
For the $60S_{1/2}$, $30D_{3/2}$, and $60D_{3/2}$ states, in contrast, $|\alpha^V(\omega)|$
   is larger than $|\alpha^S(\omega)|$. This behavior is consistent with what is observed for rubidium (see Fig.~\ref{Fig1}  of the main text). 
   Figure~\ref{FigS1}  shows that the single-parameter fits (lines) fall essentially on top of   the numerical data (symbols), using  the fitting functions $\alpha^S(\omega)=-S\omega^{-2}$, $\alpha^V(\omega)=-V\omega^{-1}$, and $\alpha^T(\omega)=-T\omega^{-2}$. The dependence of   $S$, $V$, and $T$ on $n$ is shown in Fig.~~\ref{Fig2}  of the main text.

Figure~\ref{FigS2} depicts the polarizabilities of the ground state (black dash-dotted line) and the $nS_{1/2,+1/2}$ Rydberg state of cesium for (a) $n=30$ and (b) $n=60$. The results are obtained, as  in Fig.~\ref{Fig3}   of the main text,  for $\theta_p=\pi/2$. The blue dashed and red solid lines show $\alpha(\omega)$ of  the Rydberg state for linearly polarized light ($A=0$) and circularly polarized light ($A\cos\theta_k=1$), respectively.  In  Fig.~\ref{FigS2}(b), the thin green (solid and dotted) lines show the polarizabilities for elliptically polarized light, with the bottom-most curve for $A\cos\theta_k=1/10$ and the top-most curve for $A\cos\theta_k=9/10$; these green curves lie between the polarizabilities for linearly and circularly polarized light.

 Figure~\ref{FigS3} depicts the polarizabilities of the  ground state and the $nD_{3/2,-3/2}$ Rydberg states  with $n=30$ and $n=60$ of cesium. The presentation is the same as that in Fig.~\ref{FigS2} and Figs.~\ref{Fig3}  and ~\ref{Fig4}  of the main text. As in the case for rubidium (see Fig.~\ref{Fig4}  of the main text),   we see that $nD_{3/2,-3/2}$ Rydberg states with $n$ as small as $30$ feature a positive polarizability if the  geometric factor $A\cos\theta_k$ is tuned appropriately. The positive total polarizability is tightly linked to  the fact that $|\alpha^V(\omega)|> |\alpha^S(\omega)|$ for the $nD_{3/2,-3/2}$ Rydberg state with $n=30$.

\section{Comparison of rubidium and cesium}
\label{sec_appendix_c}
\begin{figure}[t]
{\includegraphics[ scale=.35]{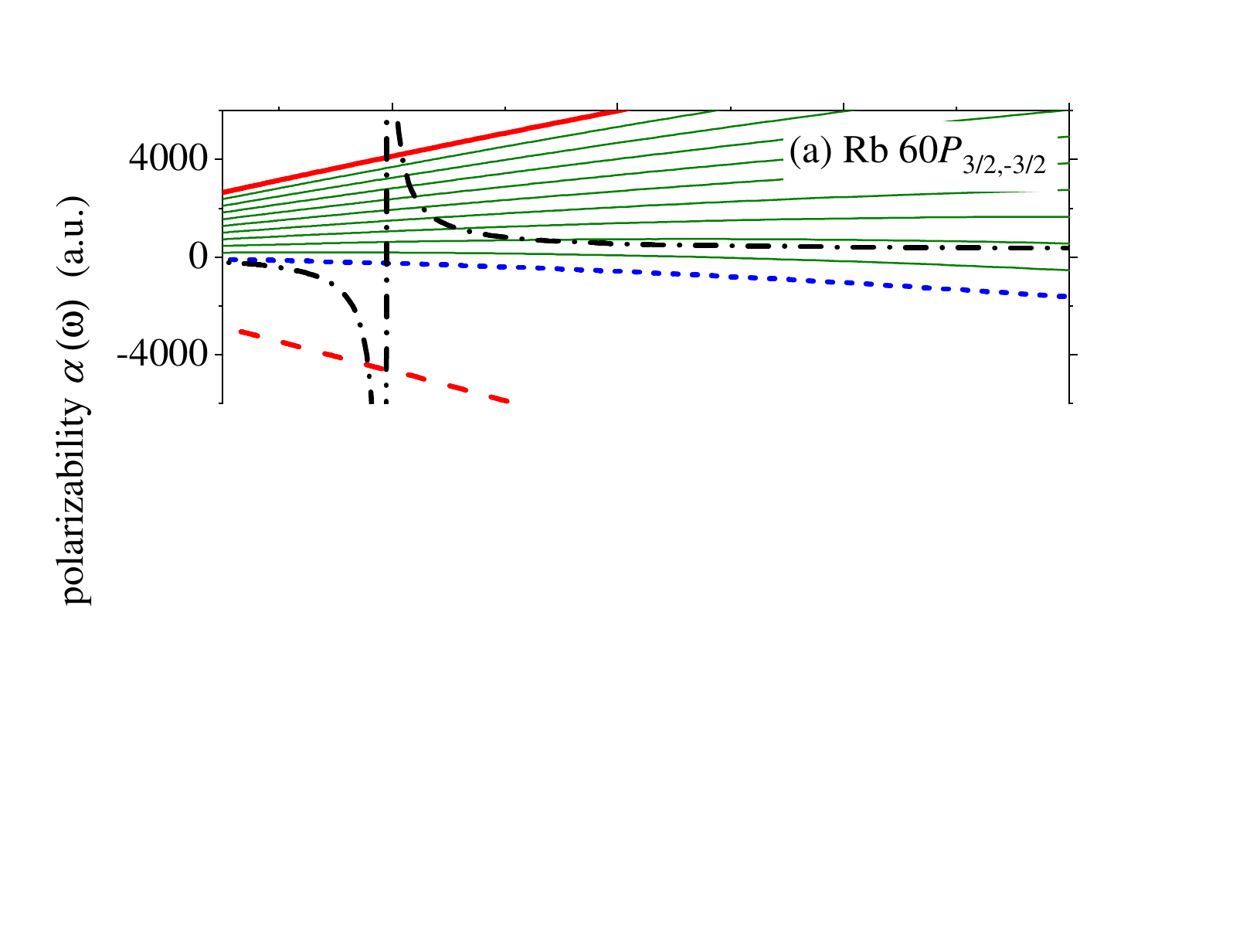}}\\
\vspace{-5.055cm}
{\includegraphics[ scale=.35]{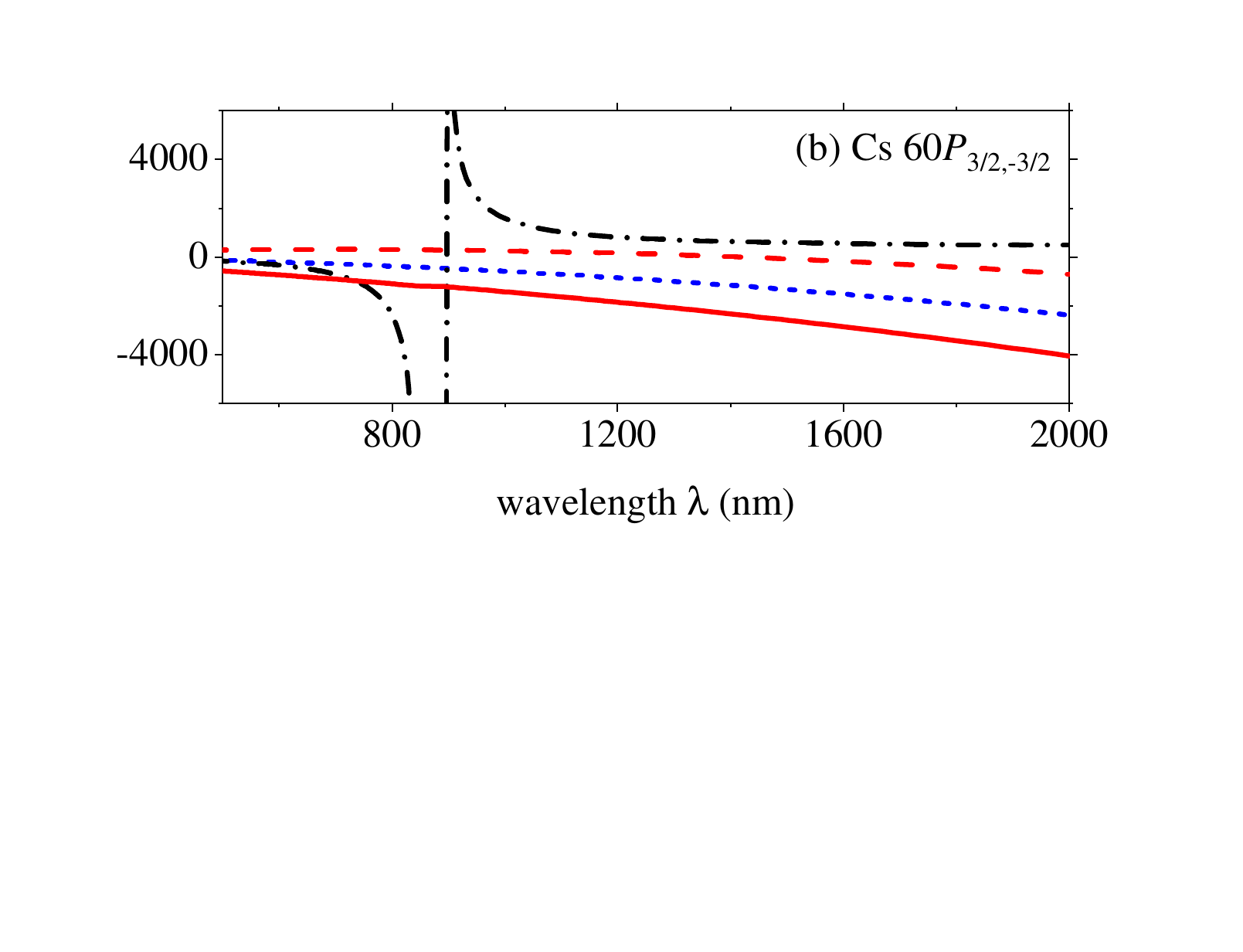}}\\
\vspace{-3.28cm}
\caption{Polarizability of the ground and Rydberg states for (a) Rb and (b) Cs  as a function of wavelength. The Rydberg state is   $60P_{3/2,-3/2}$.  The quantization axis is perpendicular to the  polarization vector, i.e.,  $\theta_p=\pi/2$.   The black dash-dotted lines show the polarizability of the ground state.  The blue small-dashed  lines show $\alpha(\omega)$ of the Rydberg state for  linearly polarized light. 
Red solid (dashed)  lines indicate  $\alpha(\omega)$ of the Rydberg state for   circularly polarized light with $A\cos\theta_k=+1(A\cos\theta_k=-1)$. In (a),  the thin green lines show the polarizability of the Rydberg states for elliptically polarized light for various values of $A\cos\theta_k$ 
[$A\cos\theta_k=0.1$ (bottom-most curve) to $0.9$ (top-most curve)].   For the  $60P_{3/2,-3/2}$ state of Rb,  variation of the geometric factor $A \cos \theta_k$ affords  appreciable tunability of $U_{\text{stark}}(X, Y, Z, \omega)$. }
\label{FigS4}
\end{figure}

While the qualitative behavior of the polarizabilities of rubidium and cesium is, for many Rydberg states, qualitatively similar, the main text pointed out that this is not always the case.
 Figure~\ref{FigS4} shows an example in support of this discussion.
  Figures~\ref{FigS4}(a) and \ref{FigS4}(b) show the polarizability of the $60P_{3/2,-3/2}$ state of rubidium  and cesium, respectively. One notices the following: (i)   $M_J=-3/2$ leads to an upshift for rubidium while $M_J=+3/2$ leads to an upshift for cesium [the ordering of the red solid and red dashed curves in Figs.~\ref{FigS4}(a) and \ref{FigS4}(b) is reversed]. (ii) The maximal shift is much larger for rubidium than  for cesium.

\bibliography{ref.bib}
\end{document}